\documentclass[amsfonts,amsmath,prd,preprint,nofootinbib,superscriptaddress]{revtex4}
\usepackage{amssymb,amsmath,amsfonts,amsbsy}
\usepackage{color}
\usepackage{enumerate}
\usepackage{graphicx}
\usepackage{pifont}
\usepackage{import}
\usepackage{float}
\usepackage{physics}
\RequirePackage[colorlinks,citecolor=red,urlcolor=red,linkcolor=red]{hyperref}
\usepackage{bm}
\usepackage[normalem]{ulem}
\usepackage{tikz}
\usetikzlibrary{arrows.meta,decorations.markings,calc,3d}
\usepackage{tikz-3dplot}
\usepackage{subcaption}
\usepackage{ragged2e}
\justifying
\newcommand{\beqa}{\begin{eqnarray}}
\newcommand{\eeqa}{\end{eqnarray}}
\newcommand{\bite}{\begin{itemize}}
\newcommand{\eite}{\end{itemize}}

\begin{document}
\title{Spherical orbits around a Newman–Janis-generated rotating Bronnikov black hole}
\author{F.~Subandi}
\email{ferdysubandi@postech.ac.kr}
\affiliation{\textit{Departemen Fisika, FMIPA, Universitas Indonesia, Depok, 16424, Indonesia}}
\affiliation{\textit{Department of Physics, Pohang University of Science and Technology (POSTECH), Pohang, 37673, Republic of Korea}}
\author{H.~S.~Ramadhan\footnote{Corresponding author.}}
\email{hramad@sci.ui.ac.id}
\affiliation{\textit{Departemen Fisika, FMIPA, Universitas Indonesia, Depok, 16424, Indonesia}}


\begin{abstract}
We investigate spherical geodesics in a Newman--Janis-generated rotating regular geometry constructed from the magnetically-charged Bronnikov mass profile. The
Kretschmann scalar is exponentially suppressed at the center, and is smooth across the disk and ring
within the single-sheeted domain $r\geq0$. We show that
all physically admissible spherical photon orbits outside
the event horizon are radially unstable in the black-hole
regime, whereas stable spherical photon orbits can exist in
horizonless configurations. The unstable photon orbits determine the shadow, which for an edge-on observer is smaller and distorted
compared with the Kerr shadow. For timelike motion, we identify
marginally stable spherical orbits and determine the innermost stable spherical orbit (ISSO). We further examine the null, weak, strong, and dominant
energy conditions, finding violations in the strong-field
region but satisfaction of all four conditions asymptotically
and throughout the exterior of every black-hole configuration.
Together with stable causality and the curvature bound, these results show that the metric satisfies the five geometric criteria
C1--C5 proposed by Maeda for physically reasonable
nonsingular black holes. 
\end{abstract}
\maketitle


\section{Introduction}

The motion of test particles and photons in the gravitational
field of a compact object provides an important means of
probing the underlying spacetime geometry. This connection
has become increasingly relevant in relativistic astrophysics,
where observations of black-hole shadows by the Event Horizon
Telescope (EHT)~\cite{EventHorizonTelescope:2019dse,EventHorizonTelescope:2022wkp}, precision measurements of stellar orbits around the Galactic-centre
black hole \cite{GRAVITY:2020gka}, and gravitational-wave signals from merging compact objects~\cite{LIGOScientific:2016aoc,LIGOScientific:2016lio} offer complementary probes of gravity in the strong-field
regime. 

In general relativity, rotation (as in Kerr and Kerr-Newman solutions~\cite{Kerr1963,Newman1965}) introduces features that are absent in static spacetimes. In particular, the loss of spherical symmetry allows particles and photons to follow non-planar trajectories, including spherical orbits that maintain a constant radial coordinate while undergoing
oscillations in the polar direction. Such orbits provide a
useful framework for investigating the dynamics of matter
and light in the strong-field region. In particular,
unstable spherical photon orbits determine the boundary of
the black-hole shadow as seen by a distant observer
\cite{Cunha:2017eoe,Tsukamoto:2017fxq}.

The study of spherical orbits has a long history, beginning
with the analysis of geodesic motion in the Kerr spacetime.
Following Carter's identification of an additional conserved
quantity associated with the separability of the
Hamilton--Jacobi equation~\cite{Carter1968}, Wilkins investigated bound
timelike geodesics with constant radial coordinates around
a Kerr black hole \cite{wilkins1972bound}. The
corresponding spherical photon orbits were subsequently
studied in the Kerr geometry and extended to the
Kerr--Newman spacetime
\cite{Stuchlik1981,Calvani1980}. These investigations
established the role of spherical photon orbits in
characterizing the strong-field structure of rotating black
holes. 

More recently, Teo provided a systematic analytical
description of spherical photon orbits in Kerr spacetime
\cite{Teo2003,Teo2021Spherical}. The use of Mino's parameter
further facilitates the separation and integration of the
radial and polar geodesic equations
\cite{mino2003perturbative}. These developments have
provided a useful framework for extending the analysis of
spherical orbits to a broader class of rotating
black-hole geometries. In this context, one of us (HSR) previously
examined spherical photon and timelike orbits in
Kerr--Newman and Ghosh black holes
\cite{Alam2024}.

An important extension of the Kerr paradigm is provided
by regular black-hole geometries, in which the central
curvature singularity is replaced by a nonsingular core.
The construction of such geometries dates back to the
regular black-hole proposal of Bardeen and subsequent
developments showing that regular solutions can arise
within general relativity coupled to nonlinear
electrodynamics (NLED)
\cite{Bardeen1968,AyonBeatoGarcia2000}. Among these models,
Bronnikov constructed a regular magnetically charged
solution supported by a nonlinear electrodynamic source
\cite{Bronnikov2000}. Its mass distribution is characterized
by a nonlinear radial dependence and approaches a finite
asymptotic mass at spatial infinity, while vanishing rapidly
near the center. This provides a useful setting for
investigating how a regular matter distribution modifies
the gravitational field and the motion of test particles.
Unlike regular black-hole geometries specified solely
through phenomenological mass functions, the static
Bronnikov solution has a definite origin in a coupled
gravitational and nonlinear electromagnetic field theory.
Rotating extensions of regular black-hole geometries have
subsequently been constructed using the Newman--Janis
algorithm and related prescriptions
\cite{Bambi2013,Toshmatov2014,Ghosh2015}.

Motivated by these developments, in this work we investigate spherical photon and timelike orbits in a
Newman--Janis-generated rotating regular geometry constructed from the magnetically charged Bronnikov mass profile. This paper is organized as follows.
In Sec.~\ref{sec:metric}, we construct the rotating regular
geometry from the static Bronnikov seed using the
Newman--Janis algorithm and the Giampieri prescription,
examine its curvature regularity, and derive the
corresponding geodesic equations. Section~\ref{sec:spo}
is devoted to spherical photon orbits, including their
analytical description, representative trajectories, and
the construction of the black-hole shadow. In
Sec.~\ref{sec:sto}, we investigate spherical timelike
orbits, focusing on their orbital properties, marginal
stability, and the existence of ISSOs. Section~\ref{sec:energy} is devoted to the energy conditions
in the static and rotating geometries, with particular
attention to their behavior at the event horizon and in
the asymptotic region, as well as the fulfillment of
Maeda's criteria. Finally, we summarize our results and discuss
possible directions for further investigation in
Sec.~\ref{sec:conc}.

\section{Metric Construction}
\label{sec:metric}

We consider the static, spherically symmetric line element
\begin{equation}
    ds^2=-f(r)\,dt^2+\frac{dr^2}{f(r)}+r^2d\Omega_2^2,
    \qquad f(r)=1-\frac{2m(r)}{r},
    \label{eq:static_seed_metric}
\end{equation}
where $d\Omega_2^2=d\theta^2+\sin^2\theta\,d\phi^2$. As a seed for the rotating construction, we adopt the Bronnikov's mass profile~\cite{Bronnikov2000}
\begin{equation}
    \label{eq:seed_mass_profile}
    m(r)=\frac{q^{3/2}}{2b}
    \left[1-\tanh\left(\frac{b\sqrt{q}}{r}\right)\right],
\end{equation}
associated with the static, magnetically charged NLED regular-black-hole geometry. 
At spatial infinity, the mass function approaches
\begin{equation}
    m(r\rightarrow\infty)\longrightarrow M\equiv\frac{q^{3/2}}{2b},
    \label{eq:ADM_mass}
\end{equation}
so that $M$ is the asymptotic mass parameter. Expanding the hyperbolic
tangent for large $r$ gives
\begin{equation}
    f(r)=1-\frac{2M}{r}+\frac{q^2}{r^2}
    +\mathcal{O}(r^{-4}).
    \label{eq:f_asymptotic}
\end{equation}
Near the origin, using $1-\tanh x\simeq 2e^{-2x}$ for $x\to+\infty$,
we obtain
\begin{equation}
    m(r\to0^+)\simeq \frac{q^{3/2}}{b}
    e^{-2b\sqrt{q}/r}.
    \label{eq:m_near_origin}
\end{equation}
The mass profile and all its radial derivatives vanish faster than any positive power of $r$ as
$r\to0^+$. Thus, the static lapse approaches $f(r)\to1$. These properties support the regularity
of the static seed geometry at the center.

\subsection{Newman--Janis-Generated Rotating Regular Geometry}

We generate a rotating counterpart of the static seed metric by using
the Newman--Janis algorithm (NJA) \cite{NewmanJanis1965}, implemented
through the simplified tensor-differential prescription of Giampieri
\cite{Giampieri1990}. This construction is a metric-generating
procedure.

We introduce an ingoing Eddington--Finkelstein coordinate $u$ through
\begin{equation}
    du=dt-\frac{dr}{f(r)},
\end{equation}
so that Eq.~\eqref{eq:static_seed_metric} becomes
\begin{equation}
    \label{eq:ef_seed}
    ds^2=-f(r)\,du^2-2\,du\,dr+r^2d\Omega_2^2.
\end{equation}
The complex coordinate transformation is
\begin{equation}
    \label{eq:complex_transformation}
    u\rightarrow u'+ia\cos\theta,\qquad r\rightarrow r'-ia\cos\theta,
\end{equation}
together with the Giampieri prescription
\begin{equation}
    i\,d\theta=\sin\theta\,d\phi.
\end{equation}
Here $a$ is the rotation parameter. The prescription provides a simplified implementation of the NJA without explicitly constructing and transforming a null tetrad.

For the complexification of radial functions, we use the symmetrized
prescription \cite{Erbin2015Simplifications, Erbin:2016lzq}
\begin{align}
    r&\longrightarrow\frac{r+\bar r}{2}=\operatorname{Re}(r),
    &\frac{1}{r}&\longrightarrow
    \frac12\left(\frac1r+\frac1{\bar r}\right)
    =\frac{\operatorname{Re}(r)}{|r|^2},
    \nonumber\\
    r^2&\longrightarrow |r|^2,
    &\frac{1}{r^2}&\longrightarrow\frac{1}{|r|^2}.
    \label{eq:complexification}
\end{align}
Applying this prescription to Eq.~\eqref{eq:ef_seed} gives
\begin{align}
    \label{eq:intermediate_metric}
    ds'^2={}&-\widetilde f
    \left(du'-a\sin^2\theta\,d\phi\right)^2
    \nonumber\\
    &-2\left(du'-a\sin^2\theta\,d\phi\right)
    \left(dr'+a\sin^2\theta\,d\phi\right)
    +\Sigma\,d\Omega_2^2,
\end{align}
where
\begin{equation}
    \Sigma\equiv r^2+a^2\cos^2\theta.
\end{equation}
We then introduce Boyer--Lindquist-like coordinates according to
\begin{equation}
    \label{eq:bl_transform}
    du'\equiv dt'-\frac{r^2+a^2}{\Delta}\,dr',
    \qquad d\phi\equiv d\phi'-\frac{a}{\Delta}\,dr',
    \qquad \Delta\equiv\widetilde f\,\Sigma+a^2\sin^2\theta.
\end{equation}
After dropping the primes, the resulting rotating metric is
\begin{align}
    \label{eq:rotating_regular_metric}
    ds^2={}&-\left(1-\frac{2m(r)r}{\Sigma}\right)dt^2
    +\frac{\Sigma}{\Delta}\,dr^2+\Sigma\,d\theta^2
    \nonumber\\
    &+\left(r^2+a^2+
    \frac{2m(r)ra^2\sin^2\theta}{\Sigma}\right)
    \sin^2\theta\,d\phi^2
    -\frac{4m(r)ra}{\Sigma}\sin^2\theta\,dt\,d\phi,
\end{align}
with
\begin{equation}
    {\widetilde f\equiv 1-\frac{2m(r)\,r}{\Sigma}},
    \qquad \Delta\equiv r^2+a^2-2m(r)r,
\end{equation}
and $m(r)$ given by Eq.~\eqref{eq:seed_mass_profile}. We refer to
Eq.~\eqref{eq:rotating_regular_metric} as the
\emph{Newman--Janis-generated rotating regular geometry}. It reduces
to the chosen static seed when $a\to0$. In the limit in which the
mass profile becomes constant, $m(r)\to M$, it reduces to the Kerr
metric. 

The metric~\eqref{eq:rotating_regular_metric} belongs to the
G\"urses--G\"ursey class of rotating geometries with a
radially dependent mass function, which can also be
constructed within the Kerr--Schild framework
\cite{GursesGursey1975}, where it provides a common
description of regular rotating and nonrotating
geometries with smooth matter sources~\cite{Burinskii2002}. In this representation,
\begin{equation}
    g_{\mu\nu}=\eta_{\mu\nu}+2H\,l_\mu l_\nu,
    \qquad H=\frac{m(r)\,r}{\Sigma},
    \label{eq:KS}
\end{equation}
where $l_\mu$ is the Kerr null vector
\cite{Visser2007}. We retain the radial dependence of
the Bronnikov mass profile in the rotating geometry;
complexifying the argument of the hyperbolic tangent
would instead introduce an angular dependence and
alter its behavior near $r=0$.

Throughout this work, we restrict the radial coordinate
to $r\geq0$. For $a\neq0$, the surface $r=0$ corresponds
to a disk $x^2+y^2\leq a^2$ in the plane $z=0$,
whose boundary is the ring $\Sigma=0$ at
$(r,\theta)=(0,\pi/2)$. For visualization, the orbits
in Secs.~\ref{sec:spo} and~\ref{sec:sto} are plotted
using $(r,\theta,\phi)$ as spherical polar coordinates;
hence, a spherical orbit is represented as lying on
a sphere of constant coordinate radius $r_{\rm SO}$.

\begin{figure}[H]
    \centering
    \begin{minipage}{0.5\textwidth}
        \includegraphics[width=\linewidth]{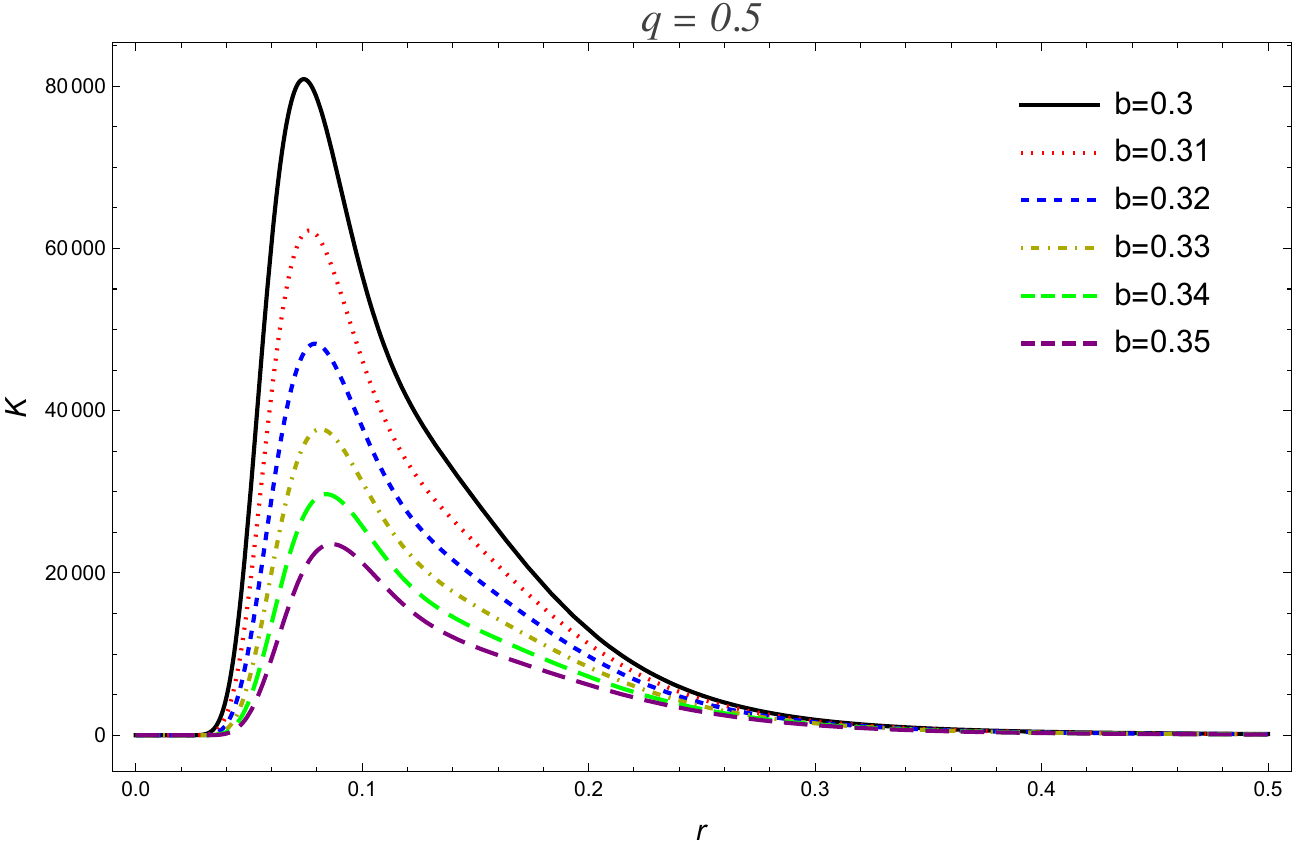}
    \end{minipage}\hfill
    \begin{minipage}{0.5\textwidth}
        \includegraphics[width=\linewidth]{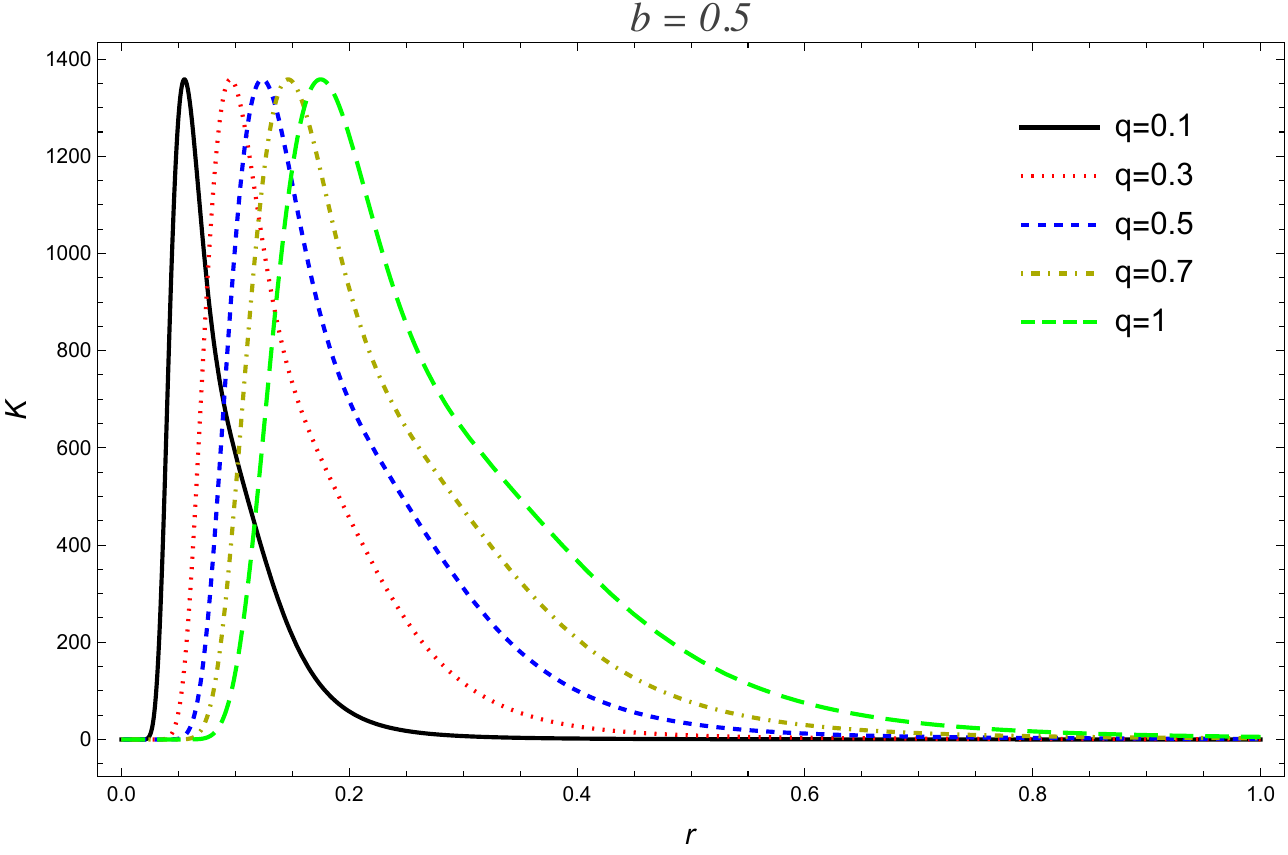}
    \end{minipage}
    \vspace{0.5cm}
    \begin{minipage}{0.5\textwidth}
        \includegraphics[width=\linewidth]{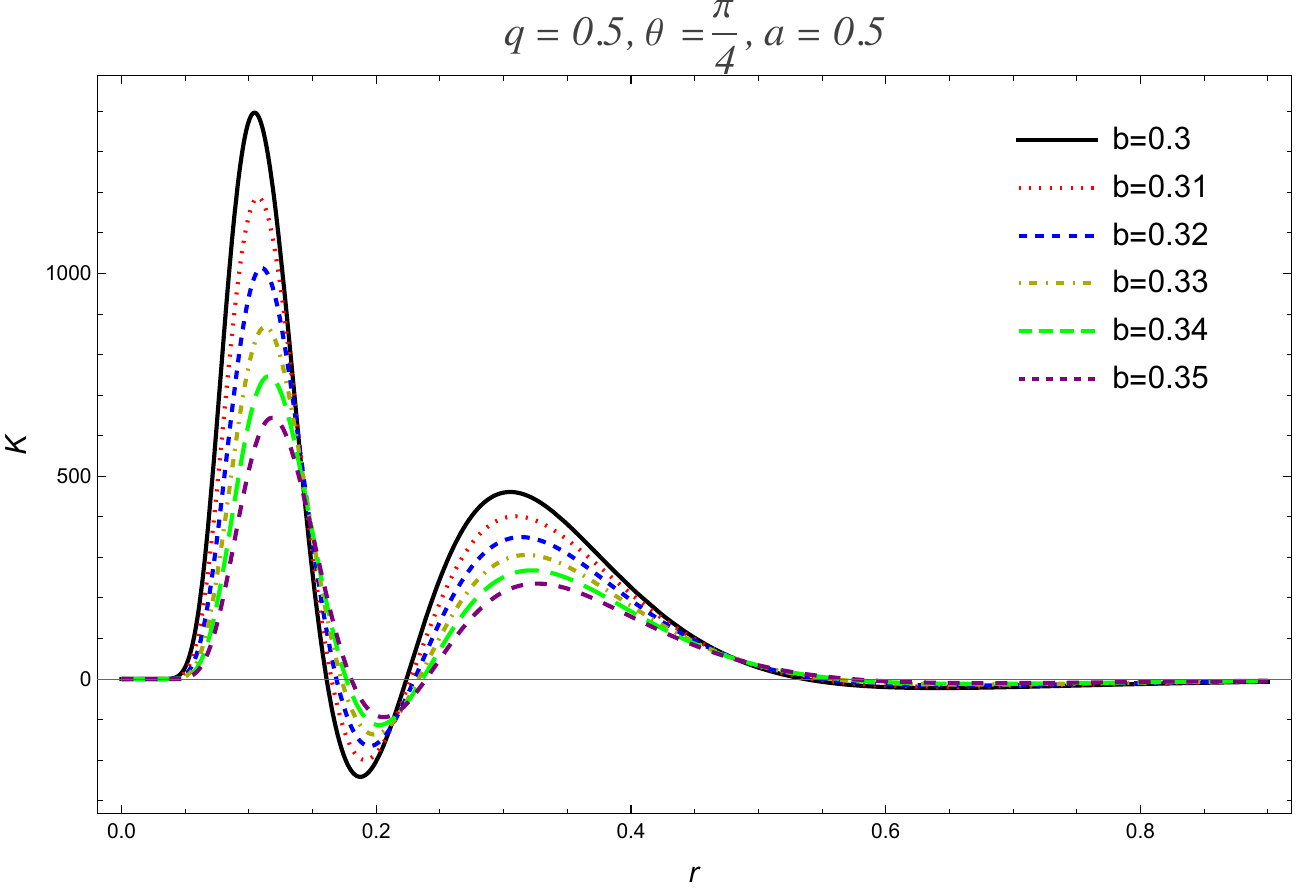}
    \end{minipage}\hfill
    \begin{minipage}{0.5\textwidth}
        \includegraphics[width=\linewidth]{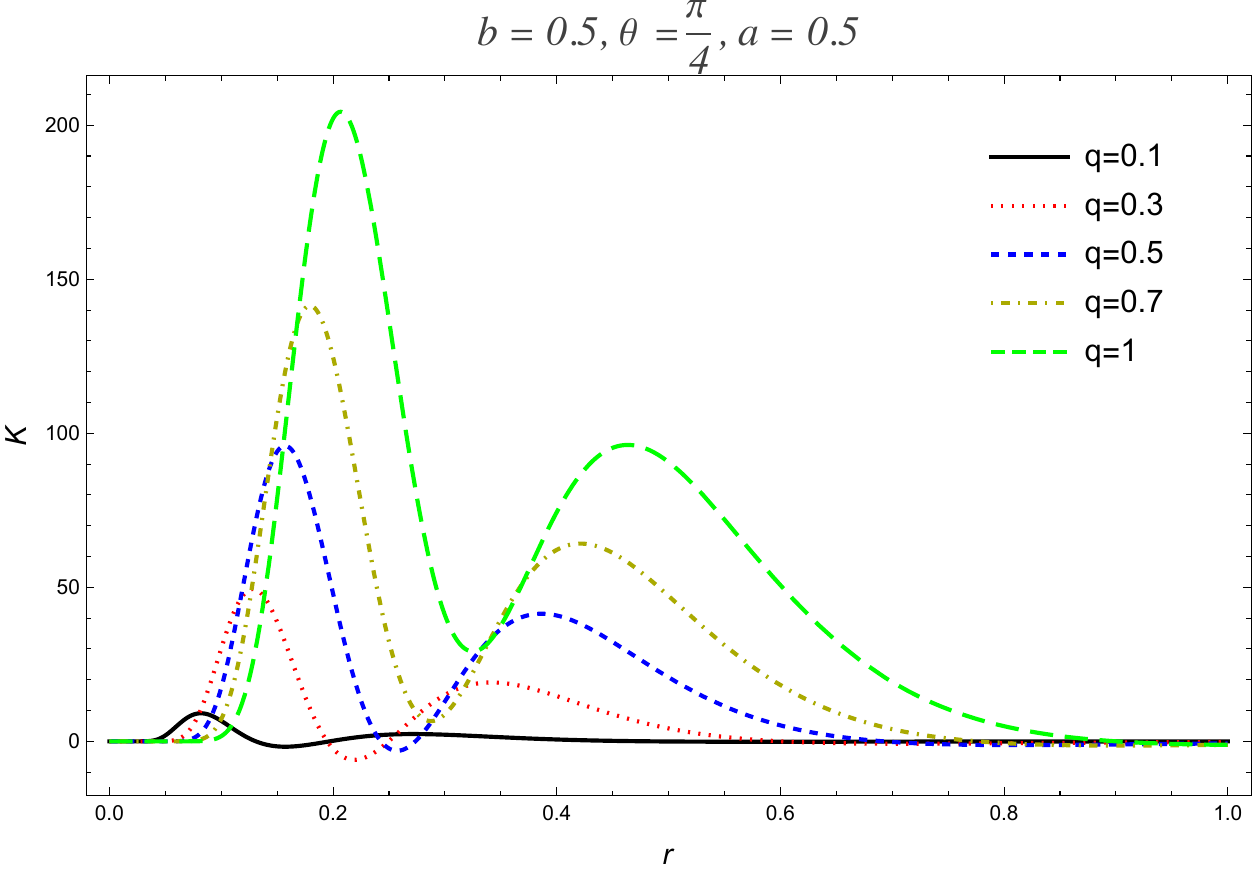}
    \end{minipage}
    \caption{
    Kretschmann scalar as a function of radius for the static (top row)
    and rotating (bottom row) Bronnikov black hole, for representative
    values of $b$ (left column) and $q$ (right column).
    }
    \label{fig:kretsch}
\end{figure}
To examine curvature regularity, we consider the Kretschmann scalar
\begin{equation}
    K\equiv R_{\mu\nu\rho\sigma}R^{\mu\nu\rho\sigma}.
\end{equation}
For the metric \eqref{eq:rotating_regular_metric}, it is
\begin{equation}
\begin{aligned}
K={}&\frac{2q^3}{b^2(r^3+a^2r\cos^2\theta)^6}
\Bigg\{
b^2q(r^2+a^2\cos^2\theta)^2
\operatorname{sech}^6\left(\frac{b\sqrt q}{r}\right)
\\
&\quad\times\Bigg[
-b^2qr^4+5r^6-2a^2b^2qr^2\cos^2\theta
-3a^2r^4\cos^2\theta-a^4b^2q\cos^4\theta
\\
&\qquad+
\left(
5r^6-3a^2r^4\cos^2\theta
+b^2q(r^2+a^2\cos^2\theta)^2
\right)
\cosh\left(\frac{2b\sqrt q}{r}\right)
\\
&\qquad
-4b\sqrt q\,r^3(r^2+a^2\cos^2\theta)
\sinh\left(\frac{2b\sqrt q}{r}\right)
\Bigg]
\\
&\quad+6r^6
\left(
r^6-15a^2r^4\cos^2\theta
+15a^4r^2\cos^4\theta-a^6\cos^6\theta
\right)
\\
&\qquad\times
\left[-1+\tanh\left(\frac{b\sqrt q}{r}\right)\right]^2
\\
&\quad+4b\sqrt q\,r^4(r^2+a^2\cos^2\theta)
\operatorname{sech}^2\left(\frac{b\sqrt q}{r}\right)
\\
&\qquad\times
\left[-1+\tanh\left(\frac{b\sqrt q}{r}\right)\right]
\\
&\qquad\times\Bigg[
3\left(r^5-6a^2r^3\cos^2\theta
+a^4r\cos^4\theta\right)
\\
&\qquad\quad+
b\sqrt q\left(-r^4+2a^2r^2\cos^2\theta
+3a^4\cos^4\theta\right)
\tanh\left(\frac{b\sqrt q}{r}\right)
\Bigg]
\Bigg\}.
\end{aligned}
\label{eq:kretschmann_rotating}
\end{equation}

In the non-rotating limit $a\to0$, Eq.~\eqref{eq:kretschmann_rotating}
reduces to the Kretschmann scalar of the static seed geometry. The
radial behavior of $K$ for representative parameter values is shown
in Fig.~\ref{fig:kretsch}, for both the static and rotating cases.

The hyperbolic functions in Eq.~\eqref{eq:kretschmann_rotating}
provide exponential suppression as $r\to0^+$, counteracting the explicit inverse powers of $r$. As demonstrated by the next subsection analysis, this behavior
persists throughout the parameter space considered, rather than
being restricted to the particular numerical profiles shown in
Fig.~\ref{fig:kretsch}. 

\subsection{Regularity and Maeda's Criteria}

The regularity of the geometry can be further examined using the criteria (C1-C7) proposed by Maeda~\cite{Maeda:2021jdc}
for physically reasonable nonsingular black holes. Among
these, C1, C2 and C5 are purely geometric. Criteria C3
and C4 concern the standard energy conditions in the
asymptotically flat region and on the event horizon of a
large black hole, respectively. The remaining criteria,
C6 and C7 , apply to black holes as solutions of a given theory. In this work, we examine C1, C2, C5, and C6, while C3 and C4 are
discussed in Sec.~\ref{sec:energy}.\\

\noindent\textbf{C1. Absence of non-coordinate singularities.}

For the static geometry, the absence of curvature
singularities follows directly from the near-origin
behavior, Eq.~\eqref{eq:m_near_origin},
which is locally Minkowskian. The Kretschmann scalar vanishes there, as illustrated in the top
row of Fig.~\ref{fig:kretsch}, and the tidal forces along radial timelike geodesics remain finite. For the rotating geometry~\eqref{eq:rotating_regular_metric}, the horizons $\Delta=0$ are coordinate singularities, as is evident from the Eddington--Finkelstein form
\eqref{eq:intermediate_metric}; likewise, $\sin\theta=0$ corresponds to the usual coordinate axis. The only
candidate for a genuine singularity is therefore the ring $\Sigma=0$. Since $m(0)=m'(0)=m''(0)=0$, all
second-order curvature invariants remain finite there~\cite{TorresFayos2017}. Moreover, the components of the Riemann tensor in the Carter frame are bounded, up to numerical factors independent of $a$ and
$\theta$, by combinations of $m/r^3$, $m'/r^2$, and $m''/r$. These quantities all vanish as $r\to0^+$, so the curvature tends to zero continuously along any path approaching the ring, not merely along directions with fixed $\theta$. This distinguishes the present profile from mass functions with a de Sitter core, $m(r)\propto r^3$ near the center, for which the curvature scalars remain finite but
take different limiting values at the ring depending
on the direction of approach; this discontinuity has
been associated with a string-like matter
distribution at the ring~\cite{BassoZanchin2025}.

The regularity can further be understood from the Kerr--Schild representation~\eqref{eq:KS}. The
components of $l_\mu$ are bounded, while Cartesian
derivatives of $r$, $\Sigma$, and $l_\mu$ introduce
at most inverse powers of $r$, since $\Sigma\geq r^2$.
These factors are dominated by the exponential
suppression in Eq.~\eqref{eq:m_near_origin}. Thus,
$H l_\mu l_\nu$ and all its Cartesian derivatives
vanish as $r\to0^+$. With the radial domain restricted
to $r\geq0$, crossing the disk $r=0$ leads back to
the same region, with $\theta\to\pi-\theta$. The
metric consequently admits a $C^\infty$ extension
across the disk, including at the ring, without a
curvature or conical singularity.

This behavior differs from the rotating models
examined in Ref.~\cite{Maeda:2021jdc}. Proposition $6$
of that work establishes a violation of C1 for their
rotating counterparts under a power-law assumption
$M(r)\simeq M_0r^{3+\alpha}$ near the center and an
extension through the disk into a region with $r<0$.
In that extension, the ring is at least a conical
singularity~\cite{Maeda:2021jdc,GibbonsVolkov2017}.
For the present mass profile~\eqref{eq:seed_mass_profile},
such a continuation is not naturally defined: the
profile is nonanalytic at $r=0$ and approaches $2M$
as $r\to0^-$, whereas a continuation that vanishes
identically for $r<0$ is also nonanalytic there.
The single-sheeted domain $r\geq0$ is therefore the
natural choice for the present geometry. A similar
exponential suppression occurs in other regular
mass profiles, including that of Ref.~\cite{Ghosh2015}.

\noindent\textbf{C2. Absence of closed causal curves.}

Since $m(r)>0$ for $r>0$,
\ proposition $5$ of Ref.~\cite{Maeda:2021jdc} then guarantees the existence of Doran coordinates~\cite{Doran2000}, in which the time coordinate $T$ satisfies $g^{TT}=-1$ everywhere,
including at and inside the horizons. Thus, $T$ is
a global time function and the spacetime is stably
causal~\cite{HawkingEllis}, excluding closed causal
curves. In particular, the closed timelike curves
of Kerr near the ring, which occur in the region
$r<0$ where $g_{\phi\phi}<0$, are absent because
that region is not part of the present spacetime.

\noindent\textbf{C5. Limiting curvature condition.}

In Bronnikov's nonlinear electrodynamics,
$\mathcal{L}(\mathcal{F})=\mathcal{F}\operatorname{sech}^2[b(\mathcal{F}/2)^{1/4}]$,with
$\mathcal{F}\equiv2q^2/r^4$~\cite{Bronnikov2000}, $b$ is a coupling
constant, whereas $q$ is an integration constant.
Regularity fixes the ADM mass through
$M=q^{3/2}/(2b)$. All static curvature quantities scale with the corresponding power of $b^{-1}$ and depend otherwise only on the combination $b\sqrt q/r$ that appears in the argument of the hyperbolic functions. In particular, $b^8K$ is a function of $b\sqrt q/r$ alone, which yields the bound $K\leq5.31/b^8$, where the maximum occurs at $b\sqrt q/r\simeq2.86$, independently of $q$.
This scaling is illustrated in the top-right panel
of Fig.~\ref{fig:kretsch}, where the curves with
$b=0.5$ attain the same peak value,
$K\simeq1358$. Varying $b$, as in the top-left
panel, changes the coupling constant and hence
the underlying theory, rather than merely changing
the solution.

The same scaling extends to the rotating case:
The Carter-frame components are bounded by combinations of $m/r^3$, $m'/r^2$,
and $m''/r$, with coefficients independent of $a$ and $\theta$. Consequently, every polynomial curvature invariant is bounded
uniformly in $q$, $a$, and $\theta$ for fixed $b$. Furthermore, the maximum of $K$ occurs on the equatorial plane and coincides with the static
bound above. The geometry therefore satisfies the limiting curvature condition, with
the fundamental length scale $L=b^2$. This result is tied to the regularity relation between $M$ and $q$, which also underlies the failure of C6, as
discussed below. A similar dependence on the coupling constant arises for the Dymnikova black hole when treated as a solution of nonlinear
electrodynamics. In Ref.~\cite{Maeda:2021jdc}, by contrast, its length scale is treated as a coupling constant while its mass is left free, leading to a violation of C5.

\noindent\textbf{C6. Regularity over a set of nonzero measure in the parameter space.}

The general static solution of general relativity coupled to Bronnikov's nonlinear electrodynamics with magnetic charge $q$ contains an additional integration constant $\bar M$.
Since the Kretschmann scalar behaves as $K\simeq48\bar M^2/r^6$ near the center, regularity
requires $\bar M=0$, equivalently the mass--charge relation~\eqref{eq:ADM_mass}. Thus, regular static black holes occupy a measure-zero subset of the parameter
space $(\bar M,q)$, and C6 is not satisfied, as is generally the case for nonsingular black holes in
general relativity coupled to nonlinear
electrodynamics~\cite{Maeda:2021jdc}.

For the rotating geometry~\eqref{eq:rotating_regular_metric}, the situation is more subtle. A consistent matter Lagrangian from which the geometry
can be derived as a solution of the Einstein field equations has not yet been identified (see
Sec.~\ref{sec:conc}). Thus, C6 cannot be assessed in the same sense.

\noindent\textbf{C7. Dynamical stability.}

The dynamical stability of the static solution and
of the rotating geometry is beyond the scope of
the present work.

\subsection{Geodesics}

We study the geodesic motion in the rotating geometry using the
Hamilton--Jacobi formalism \cite{Carter1968}. The stationary and axial symmetry enable the action to be separated as
\begin{equation}
    S=-\frac{1}{2}\mu^2\lambda-Et+L_z\phi
    +S_r(r)+S_\theta(\theta),
\end{equation}
where $\mu$ denotes the rest mass of the test particle, $\lambda$
is an affine parameter, and $E$ and $L_z$ are the conserved energy
and axial angular momentum, respectively. The conserved quantities are:
\begin{equation}
    p_t=-E,\qquad p_\phi=L_z,
\end{equation}
with the canonical momenta defined by
\begin{equation}
    p_\mu=g_{\mu\nu}\dot{x}^\nu.
\end{equation}

The Hamilton--Jacobi equation separates into radial and polar
equations, yielding the first-order geodesic equations
\begin{align}
    \Sigma\Delta\dot{t}
    &=
    \left[(r^2+a^2)^2-a^2\Delta\sin^2\theta\right]E
    -2am(r)rL_z,
    \label{timegeod}\\
    \Sigma\Delta\dot{\phi}
    &=
    \frac{\Delta-a^2\sin^2\theta}{\sin^2\theta}L_z
    +2am(r)rE,
    \label{phigeod}
\end{align}
together with
\begin{align}
    \Sigma^2\dot{r}^{\,2}&=R(r),
    \label{eq:radial_null_shadow}\\
    \Sigma^2\dot{\theta}^{\,2}&=\Theta(\theta),
    \label{eq:polar_null_shadow}
\end{align}
where the radial and polar potentials are given by
\begin{align}
    R(r)
    &\equiv
    \left[E(r^2+a^2)-aL_z\right]^2
    -\Delta\left[
    \mathcal C+\mu^2r^2+(L_z-aE)^2
    \right],
    \label{29}\\
    \Theta(\theta)
    &\equiv
    \mathcal C-\cos^2\theta
    \left[
    a^2(\mu^2-E^2)+\frac{L_z^2}{\sin^2\theta}
    \right],
    \label{30}
\end{align}
with $\mathcal C$ denoting Carter's separation constant. 
For later convenience, we introduce the dimensionless quantities
\begin{equation}
    C\equiv\frac{\mathcal C}{E^2},
    \qquad
    \Phi\equiv\frac{L_z}{E},
\end{equation}
and use $u=\cos\theta$ as the polar variable \cite{Wang2022}.
Eq.~\eqref{30} then takes the form
\begin{equation}
    \left(\frac{\Sigma}{E}\dot{u}\right)^2
    =
    C-
    \left[
    C+\Phi^2+\frac{a^2}{E^2}(\mu^2-E^2)
    \right]u^2
    +\frac{a^2}{E^2}(\mu^2-E^2)u^4.
    \label{31}
\end{equation}

\section{Spherical Photon Orbits}
\label{sec:spo}

For null geodesics, we set $\mu=0$ in Eqs.~\eqref{29} and \eqref{30}. The physical
range of the polar motion is determined by the condition
$\Theta(\theta)\geq0$. Introducing $u\equiv\cos\theta$, the turning
points of the latitudinal motion are obtained by setting
$\dot{u}=0$ in Eq.~\eqref{31}. The resulting physical solutions
are
\begin{eqnarray}
    &\underline{C>0:} \nonumber \\
    && u_0^2 = \frac{\left(a^2 - C - \Phi^2 \right) + \sqrt{\left(a^2 - C - \Phi^2 \right)^2+4a^2C}}{2a^2}, \\
    & \underline{C<0:} \nonumber \\
    \label{33}
    && \left(a^2 - C - \Phi^2 \right)>0.
\end{eqnarray}
Here, $u_0$ denotes the maximum value of $|u|$ along the
trajectory, corresponding to the turning point of the polar
motion.

Spherical photon orbits are characterized by a constant radial
coordinate, $r=r_{\rm SO}$. Consequently, the radial potential
must satisfy
\begin{equation}
    \left.R(r)\right|_{r=r_{\rm SO}}
    =
    \left.\frac{dR}{dr}\right|_{r=r_{\rm SO}}=0.
    \label{SOcond}
\end{equation}
Solving these conditions simultaneously yields two branches.
The first is
\begin{equation}
    \Phi=\frac{r_{\rm SO}^2+a^2}{a},
    \qquad
    C=-\frac{r_{\rm SO}^4}{a^2}.
    \label{34}
\end{equation}
The second branch gives
\begin{align}
    \Phi(r_{\rm SO})
    &=
    \frac{1}{a}
    \left[
    r_{\rm SO}^2+a^2
    -\frac{4r_{\rm SO}\Delta(r_{\rm SO})}
    {\Delta'(r_{\rm SO})}
    \right],
    \nonumber\\
    C(r_{\rm SO})
    &=
    \frac{
    \left[r_{\rm SO}^2+a^2-a\Phi(r_{\rm SO})\right]^2
    }{\Delta(r_{\rm SO})}
    -
    \left[\Phi(r_{\rm SO})-a\right]^2,
    \label{35}
\end{align}
where $\Delta'(r_{\rm SO})
    \equiv d\Delta/dr\bigg|_{r=r_{\rm SO}}.$

The first branch, Eq.~\eqref{34}, is excluded because it has $C<0$ and does not satisfy the
corresponding admissibility condition in Eq.~\eqref{33}.
We therefore focus on the second branch, Eq.~\eqref{35}, where the orbits are selected by requiring $C>0$. With the
convention $\Sigma^2\dot r^2=R(r)$, stability requires
\begin{equation}
\label{unstable}
    \left.
    \frac{d^2R}{dr^2}
    \right|_{r=r_{\rm SO}}<0.
\end{equation}

\begin{figure}[H]
    \centering
    \begin{minipage}{0.5\textwidth}
        \includegraphics[width=\linewidth]{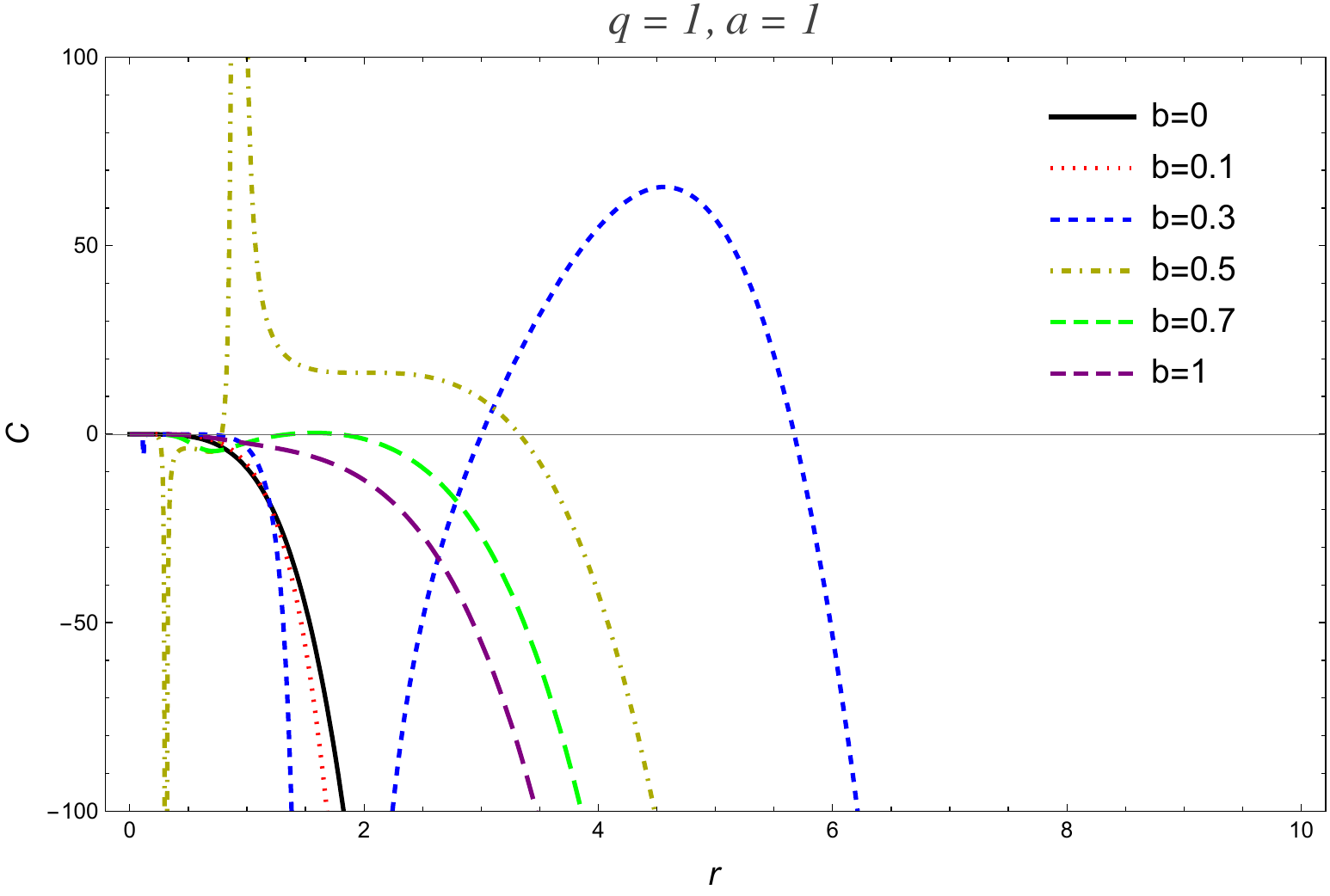}
    \end{minipage}\hfill
    \begin{minipage}{0.5\textwidth}
        \includegraphics[width=\linewidth]{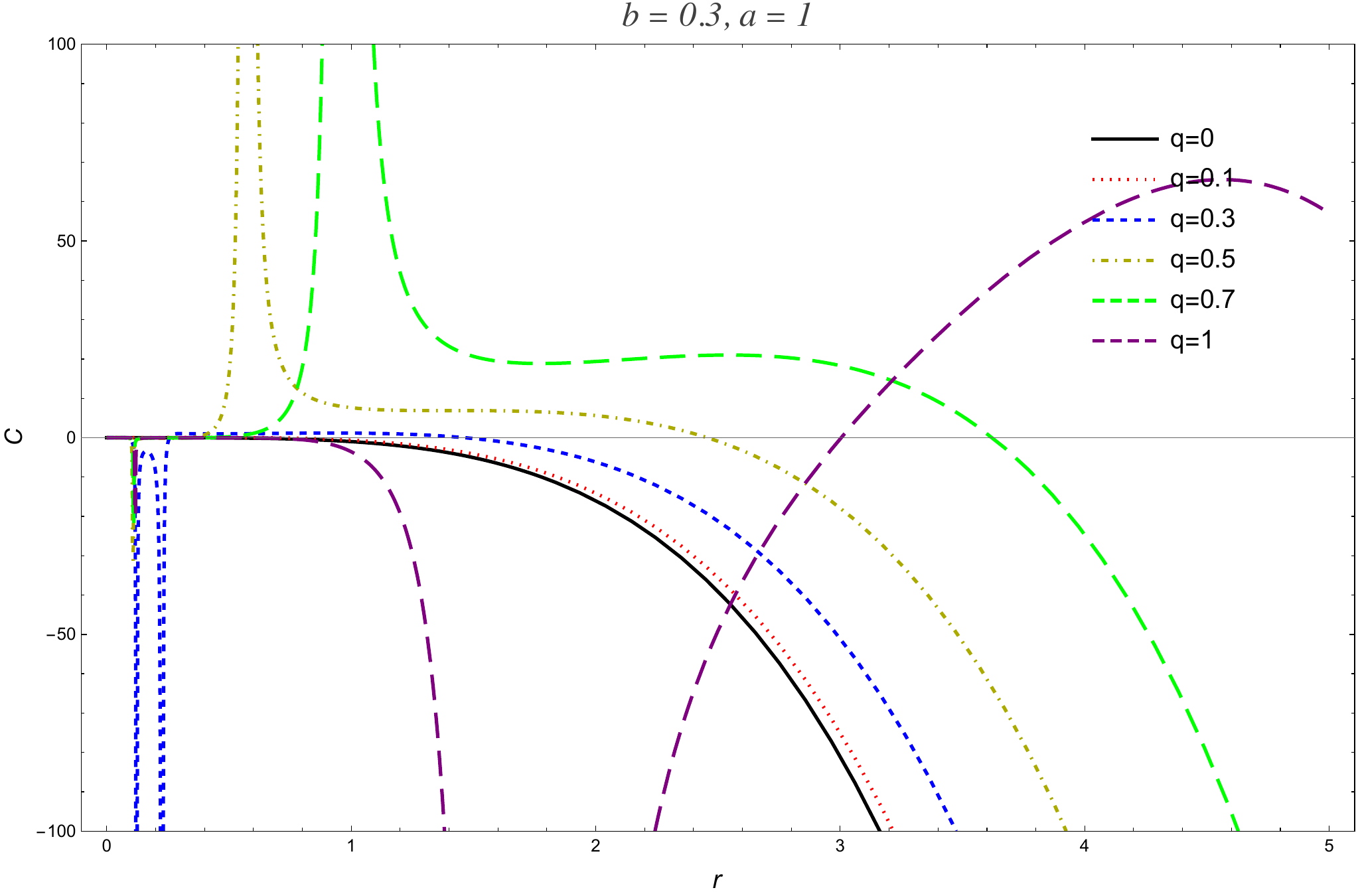}
    \end{minipage}
    \vspace{0.5cm}
    \begin{minipage}{0.5\textwidth}
        \includegraphics[width=\linewidth]{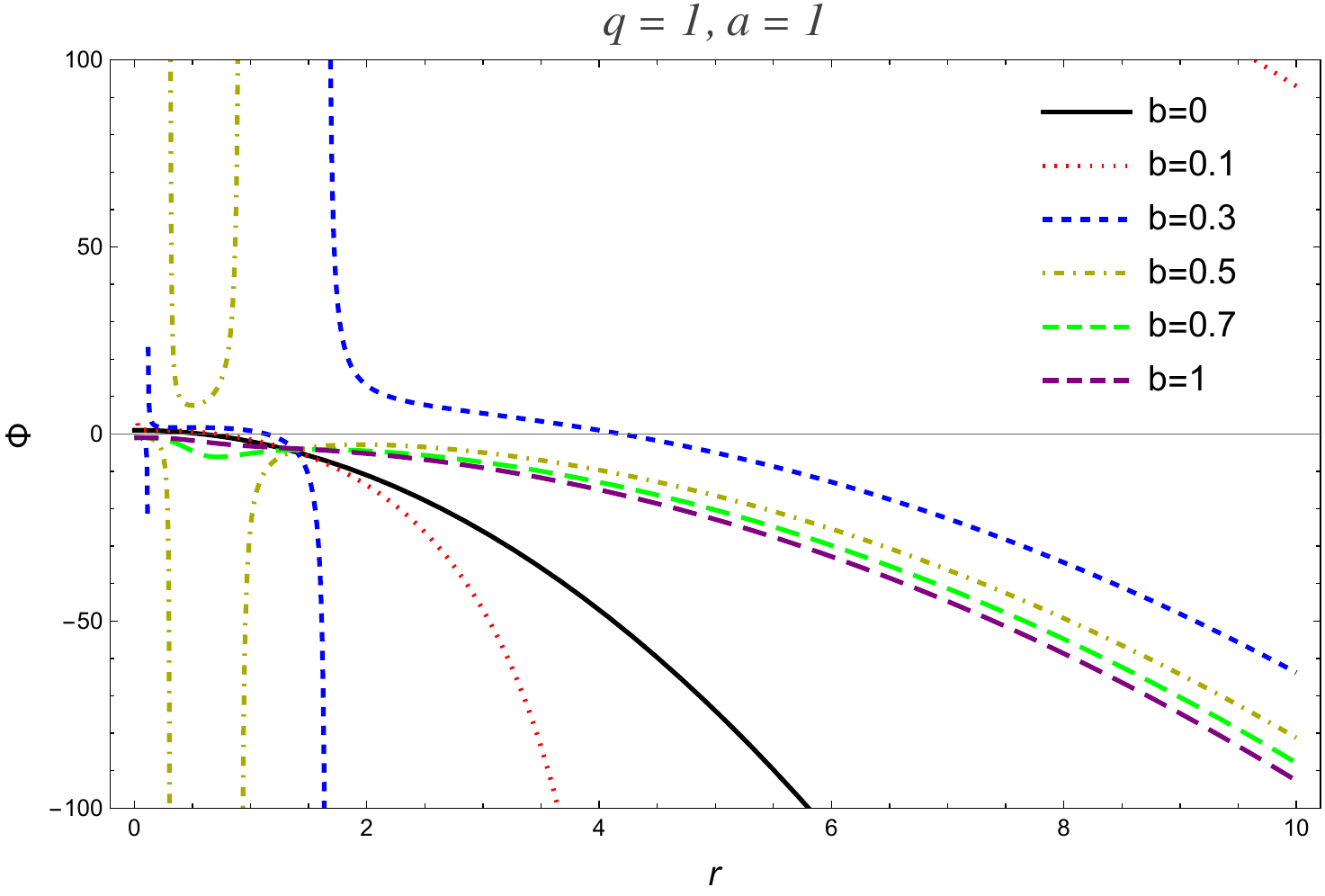}
    \end{minipage}\hfill
    \begin{minipage}{0.5\textwidth}
        \includegraphics[width=\linewidth]{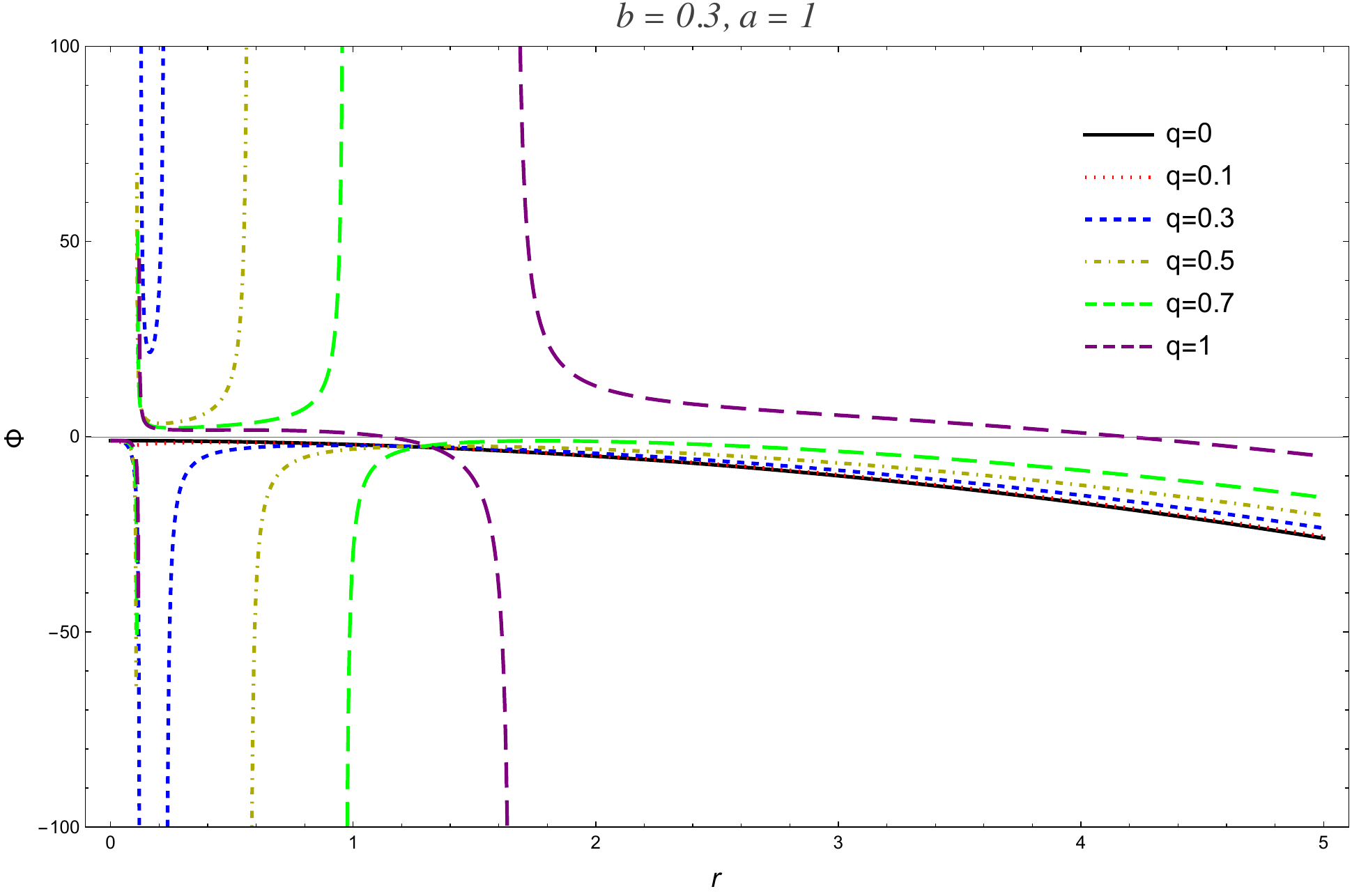}
    \end{minipage}
    \caption{Carter parameter $C$ (top row) and normalized
    azimuthal angular momentum $\Phi$ (bottom row) as functions
    of the spherical-orbit radius for different values of $b$
    and $q$.}
    \label{fig:CandPhi}
\end{figure}

To examine how the spherical photon orbits depend on the
parameters of the rotating geometry, we plot the corresponding
Carter parameter $C$ and normalized azimuthal angular momentum
$\Phi$ as functions of $r_{\rm SO}$ for different values of
$b$ and $q$. The results are displayed in Fig.~\ref{fig:CandPhi}.

\begin{figure}[H]
    \centering
    \begin{minipage}{0.5\textwidth}
        \includegraphics[width=\linewidth]{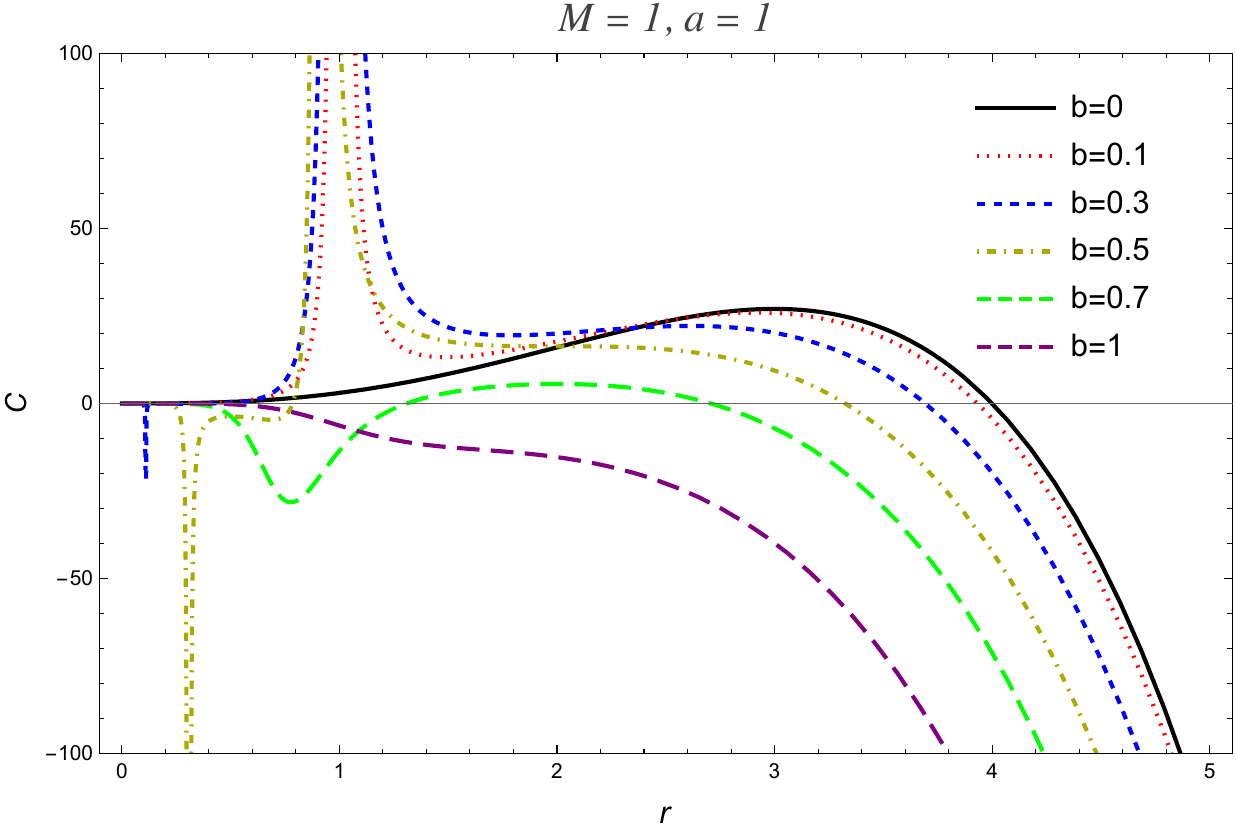}
    \end{minipage}\hfill
    \begin{minipage}{0.5\textwidth}
        \includegraphics[width=\linewidth]{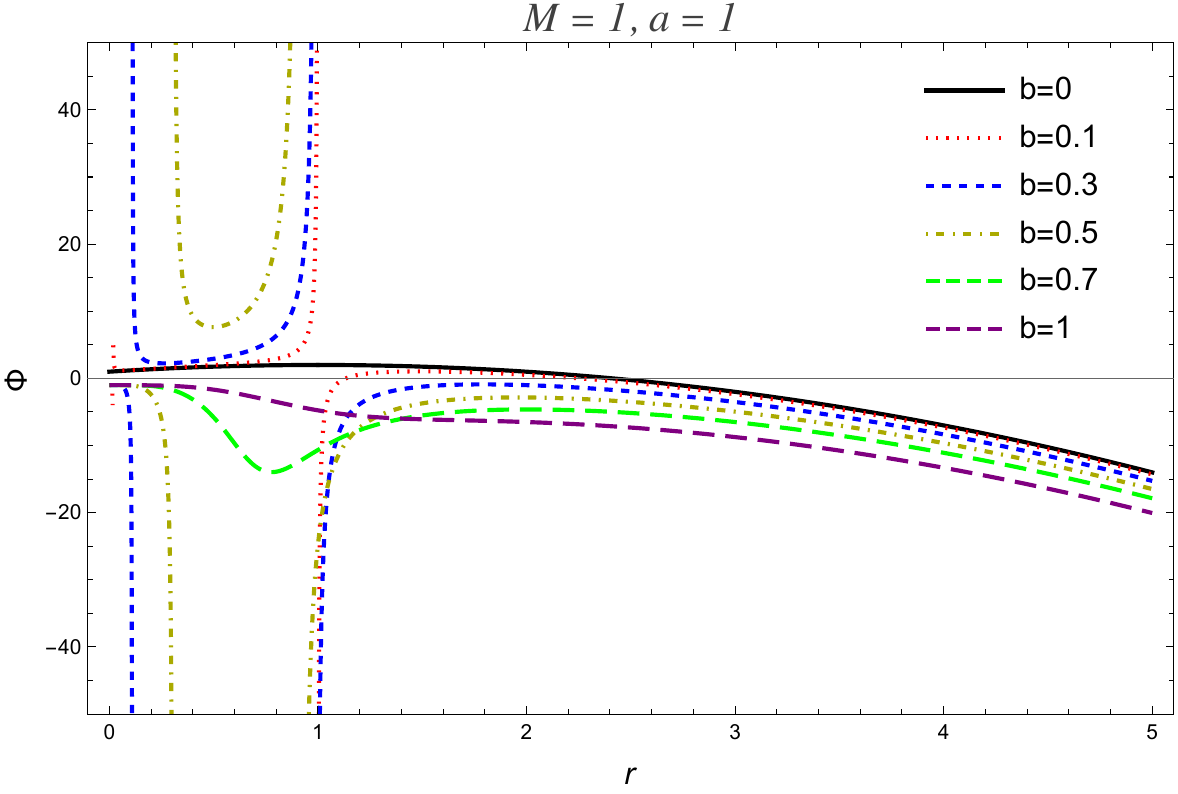}
    \end{minipage}
    \caption{Carter parameter $C$ (left) and normalized
    azimuthal angular momentum $\Phi$ (right) as functions of
    the spherical-orbit radius for different values of $b$,
    with the asymptotic mass held fixed.}
    \label{fig:CandPhiKerr}
\end{figure}

The plots reveal a nonlinear dependence of $C$ and $\Phi$ on $b$ and $q$. The functions may exhibit sharp peaks and extrema over restricted ranges of $r_{\rm SO}$. To compare the rotating geometry with the Kerr spacetime, we use
\begin{equation}
    q=(2bM)^{2/3}.
    \label{qsubs}
\end{equation}
The resulting profiles of $C$ and $\Phi$ are shown in Fig.~\ref{fig:CandPhiKerr}.

As $b\to0$ at fixed $M$, the mass profile approaches
$m(r)\to M$, and the rotating metric reduces to the Kerr
geometry. Accordingly, the profiles of $C$ and $\Phi$ approach
their Kerr counterparts. For finite $b$, deviations from the
Kerr behavior arise from the nontrivial radial dependence of
the mass function. The sharp variations in
$C$ and $\Phi$ indicate regions where the spherical photon
orbit parameters are particularly sensitive to the radial
structure of the geometry.

The Carter parameter $C$ also determines the allowed
latitudinal motion of the photon. In particular, its value
controls the maximum polar excursion $u_0=\lvert\cos\theta\rvert_{\rm max}$.
Thus, for a given rotation parameter, larger positive values
of $C$ generally permit a greater excursion from the
equatorial plane, whereas smaller values confine the motion
closer to that plane. 

\subsection{Analytical Solutions}

To obtain the analytical solutions of the polar and azimuthal
motions, we introduce Mino's parameter $\lambda$
\cite{mino2003perturbative}, such that
\begin{equation}
    \frac{d\tau}{d\lambda}=\Sigma,
    \qquad
    \frac{dx^\mu}{d\lambda}
    =\Sigma\frac{dx^\mu}{d\tau}.
    \label{Minoinitial}
\end{equation}
We adopt the normalization $E=1$ in what follows. Introducing
$w\equiv u^2$, Eq.~\eqref{31} can then be
written as
\begin{equation}
    \frac{dw}{d\lambda}=\pm 2Y(w),
    \qquad
    Y(w)^2=a^2w(w_1-w)(w-w_2),
    \label{MinoY}
\end{equation}
where $w_1$ and $w_2$ are the roots of the quadratic polynomial in $w$:
\begin{equation}
    w_{1,2}
    =
    \frac{
    a^2-C-\Phi^2
    \pm\sqrt{(C+\Phi^2-a^2)^2+4a^2C}
    }{2a^2}.
\end{equation}
For the physical polar motion considered here, the roots satisfy
$w_1\geq w\geq0>w_2$. We choose the positive branch of
Eq.~\eqref{MinoY} and impose the initial condition
$\lambda=0$ at $w=0$. Integration gives
\begin{equation}
    \lambda
    =\frac12\int_0^w\frac{dw}{Y(w)}
    =\frac{1}{a\sqrt{w_1-w_2}}F(\psi,k),
    \label{4.21}
\end{equation}
where $F(\psi,k)$ is the incomplete elliptic integral of the
first kind, with
\begin{equation}
    \psi
    =
    \arcsin\sqrt{
    \frac{w(w_1-w_2)}{w_1(w-w_2)}
    },
    \qquad
    k=\sqrt{\frac{w_1}{w_1-w_2}}.
\end{equation}
Here and below, the notation for elliptic integrals follows
Ref.~\cite{Gradshteyn1994Table}.

As $w$ increases from zero to its turning point $w_1$, the
corresponding Mino parameter reaches
\begin{equation}
    \lambda_0
    =\frac{K(k)}{a\sqrt{w_1-w_2}},
\end{equation}
where $K(k)$ is the complete elliptic integral of the first
kind. This value corresponds to one quarter of a complete
latitudinal oscillation. Inverting Eq.~\eqref{4.21}, we obtain
\begin{equation}
    \sin\psi
    =
    \operatorname{sn}
    \left(a\sqrt{w_1-w_2}\,\lambda,k\right),
    \label{4.24}
\end{equation}
where $\operatorname{sn}$ is the Jacobi elliptic sine function.
Consequently, the polar motion can be expressed as
\begin{equation}
    u(\lambda)
    =
    \sqrt{-w_2}\,k\,
    \operatorname{sd}
    \left(a\sqrt{w_1-w_2}\,\lambda,k\right),
    \label{4.26}
\end{equation}
where $\operatorname{sd}$ is defined in terms of the Jacobi
elliptic functions $\operatorname{sn}$ and $\operatorname{dn}$ as
\begin{equation}
    \operatorname{sd}(x,k)
    \equiv
    \frac{\operatorname{sn}(x,k)}
         {\operatorname{dn}(x,k)},
    \qquad
    \operatorname{dn}(x,k)
    \equiv
    \sqrt{1-k^2\operatorname{sn}^2(x,k)}.
\end{equation}

We next determine the azimuthal motion. Using
Eq.~\eqref{phigeod} together with the definition of Mino's
parameter, the azimuthal equation becomes
\begin{equation}
    \frac{d\phi}{d\lambda}
    =
    \frac{a\left[2m(r)r-a\Phi\right]}{\Delta}
    +\frac{\Phi}{1-w}.
    \label{Minophi}
\end{equation}
Using Eq.~\eqref{MinoY}, we can rewrite the azimuthal equation as
\begin{equation}
    \frac{d\phi}{dw}
    =
    \frac{d\phi}{d\lambda}\frac{d\lambda}{dw}
    =
    \pm\frac{1}{2Y(w)}
    \left[
        \frac{a\left[2m(r)r-a\Phi\right]}{\Delta}
        +\frac{\Phi}{1-w}
    \right].
    \label{Minophiw}
\end{equation}
For a spherical photon orbit, $r=r_{\rm SO}$ is constant,
so the first term on the right-hand side is constant along
the orbit. Combining Eqs.~\eqref{MinoY} and~\eqref{Minophi},
and integrating with respect to $\lambda$, we obtain
\begin{equation}
\begin{split}
\phi(\lambda)
&=
\frac{a\left[2m(r_{\rm SO})r_{\rm SO}-a\Phi\right]}
{\Delta(r_{\rm SO})}\lambda
\\
&\quad+
\frac{\Phi}
{a(1-w_2)\sqrt{w_1-w_2}}
\left[
F(\psi,k)
-w_2\Pi\left(\psi,k^2(1-w_2),k\right)
\right],
\end{split}
\label{4.28}
\end{equation}
where $\Pi(\psi,k^2(1-w_2),k)$ denotes the incomplete
elliptic integral of the third kind.

Unlike the polar coordinate, the azimuthal coordinate is not
periodic along the photon trajectory: $\phi$ accumulates
continuously as the photon orbits the black hole. Nevertheless,
the azimuthal change over one complete latitudinal oscillation
can be obtained by evaluating Eq.~\eqref{4.28} over the
corresponding Mino-parameter interval. Since a complete
latitudinal oscillation consists of four successive quarter
oscillations, the total azimuthal change is
\begin{equation}
\begin{split}
    \Delta\phi
    &=
    \frac{4}{\sqrt{w_1-w_2}}
    \Bigg[
    \frac{\Phi}{a(1-w_2)}
    \left(
    K(k)-w_2\Pi(k^2(1-w_2),k)
    \right)
    \\
    &\qquad\qquad
    +\frac{2m(r_{\rm SO})r_{\rm SO}-a\Phi}
    {\Delta(r_{\rm SO})}K(k)
    \Bigg],
\end{split}
\end{equation}
where $K(k)$ and $\Pi(k^2(1-w_2),k)$ are the complete
elliptic integrals of the first and third kinds, respectively.


\subsection{Representative Orbits}

The analytical expressions obtained in the preceding subsection allow us to reconstruct the spatial trajectories of spherical photon orbits (SPOs). In particular, the polar and azimuthal motions determine how a photon moves along a spherical orbit and how its spatial trajectory is shaped by the parameters of the rotating geometry. We use these solutions to examine representative SPOs in three different spacetime configurations: extremal, non-extremal, and horizonless (naked) cases.

Since the non-extremal configurations exhibit qualitative orbital features similar to those found in the extremal case, we restrict our discussion to the extremal and horizonless configurations. The corresponding orbit parameters are obtained by selecting the nonlinear parameter $b$ and the impact parameter $\Phi$, and then solving the SPO conditions discussed in the preceding section. The resulting parameters, together with the polar turning point $u_0$ and the accumulated azimuthal angle $\Delta\phi$, are listed in the tables below.

\subsubsection{Extremal Case}

The horizons of the rotating geometry are determined by
\begin{equation}
    \Delta(r_\pm)
    =
    r_\pm^2+a^2
    -\frac{q^{3/2}r_\pm}{b}
    \left(
        1-\tanh\frac{b\sqrt{q}}{r_\pm}
    \right)
    =0.
\end{equation}
An extremal configuration corresponds to a degenerate horizon, given by
\begin{equation}
    \Delta(r_{\rm ext})=0,
    \qquad
    \Delta'(r_{\rm ext})=0.
\end{equation}
For a fixed ADM mass $M=1$, we vary the nonlinear parameter $b$ and select the corresponding values of the spin parameter $a$ and the impact parameter $\Phi$ to obtain representative extremal SPOs.

\begin{table}[H]
    \centering
    \caption{Parameters of representative SPOs around an extremal rotating regular geometry with $M=1$.}
    \label{tab:Extreme}
    \vspace{0.2cm}
    \begin{tabular}{|c|c|c|c|c|c|c|c|}
        \hline
        \makebox[2cm]{Orbit} & \makebox[1.5cm]{$b$} & \makebox[1.5cm]{$\Phi$} & \makebox[1.5cm]{$a$} & \makebox[2cm]{$r_{\rm SO}$} & \makebox[2cm]{$C$} & \makebox[2cm]{$u_0$} & \makebox[2cm]{$\Delta\phi$} \\
        \hline
        (a) & 0   & 0     & 1 & 2.414211 & 22.3137  & 1 & 3.17612 \\
        (b) & 0.1   & -1     & 0.94 & 2.6798535 & 24.9681  & 0.981179 & 2.5123 \\
        (c) & 0.2   & 1     & 0.84 & 1.931538 & 14.8841  & 0.969312 & 4.14498  \\
        (d) & 0.3   & -6     & 0.71 & 3.395791 & 1.16507  & 0.178229 & 1.26812  \\
        (e) & 0.4   & 2     & 0.539 & 1.36701 & 5.65262  & 0.769981 & 6.8483  \\
        (f) & 0.5   & -2     & 0.286 & 2.204191 & 14.8294  & 0.887858 & 1.08638  \\
        \hline
    \end{tabular}
\end{table}

\begin{figure}[H]
    \centering
    \begin{minipage}{0.32\textwidth}
        \centering
        \includegraphics[width=\linewidth]{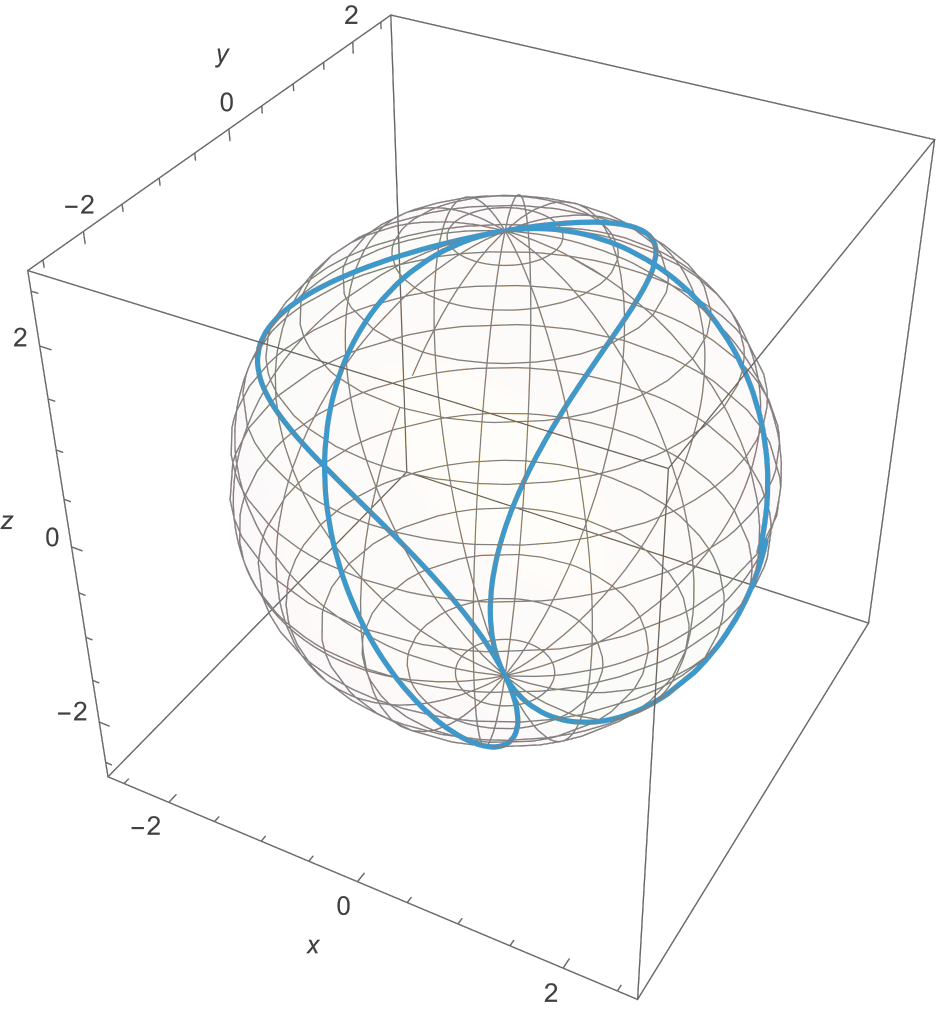}\vspace{0.2cm}
        \includegraphics[width=\linewidth]{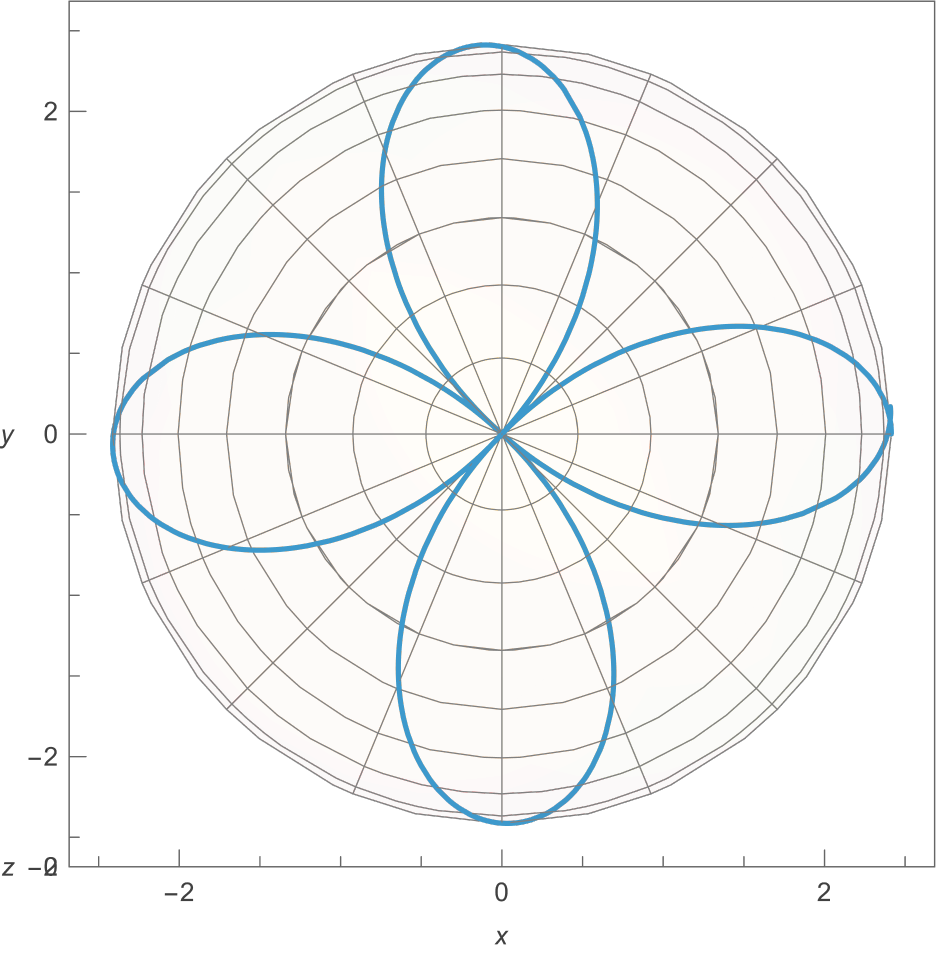}\vspace{0.2cm}
        \includegraphics[width=\linewidth]{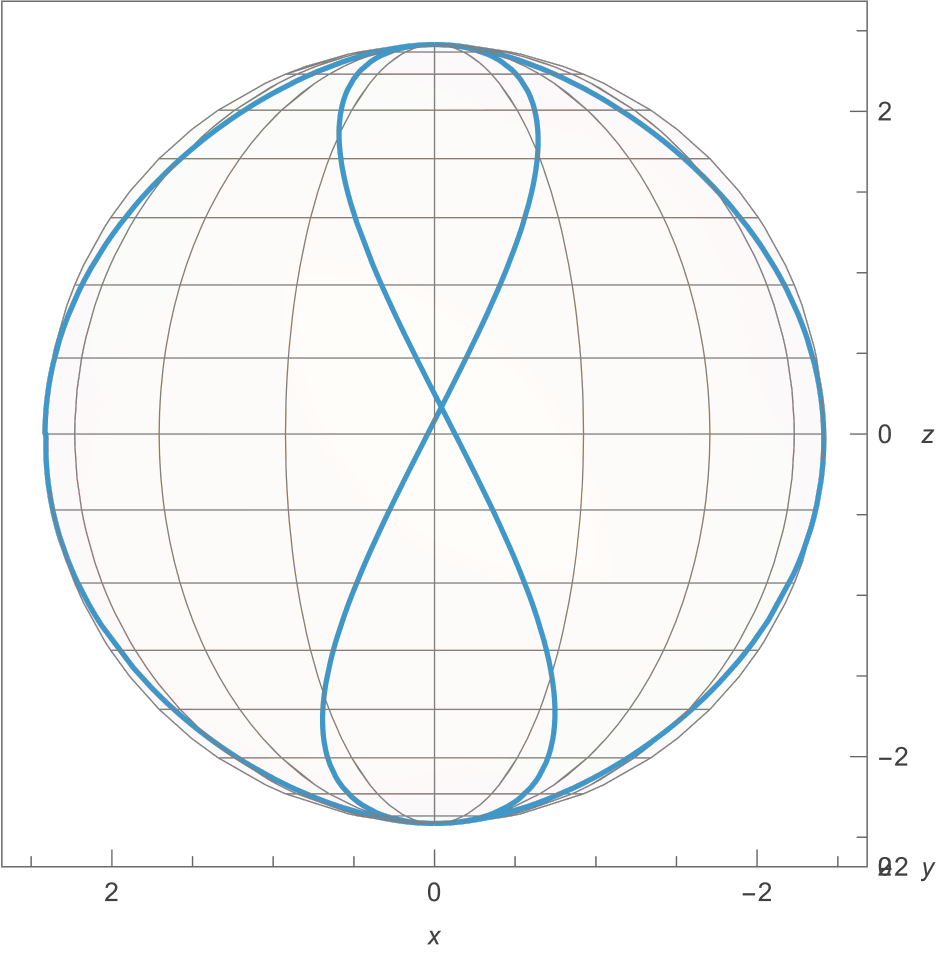}\vspace{0.2cm}
        (a)
    \end{minipage}\hfill
    \begin{minipage}{0.32\textwidth}
        \centering
        \includegraphics[width=\linewidth]{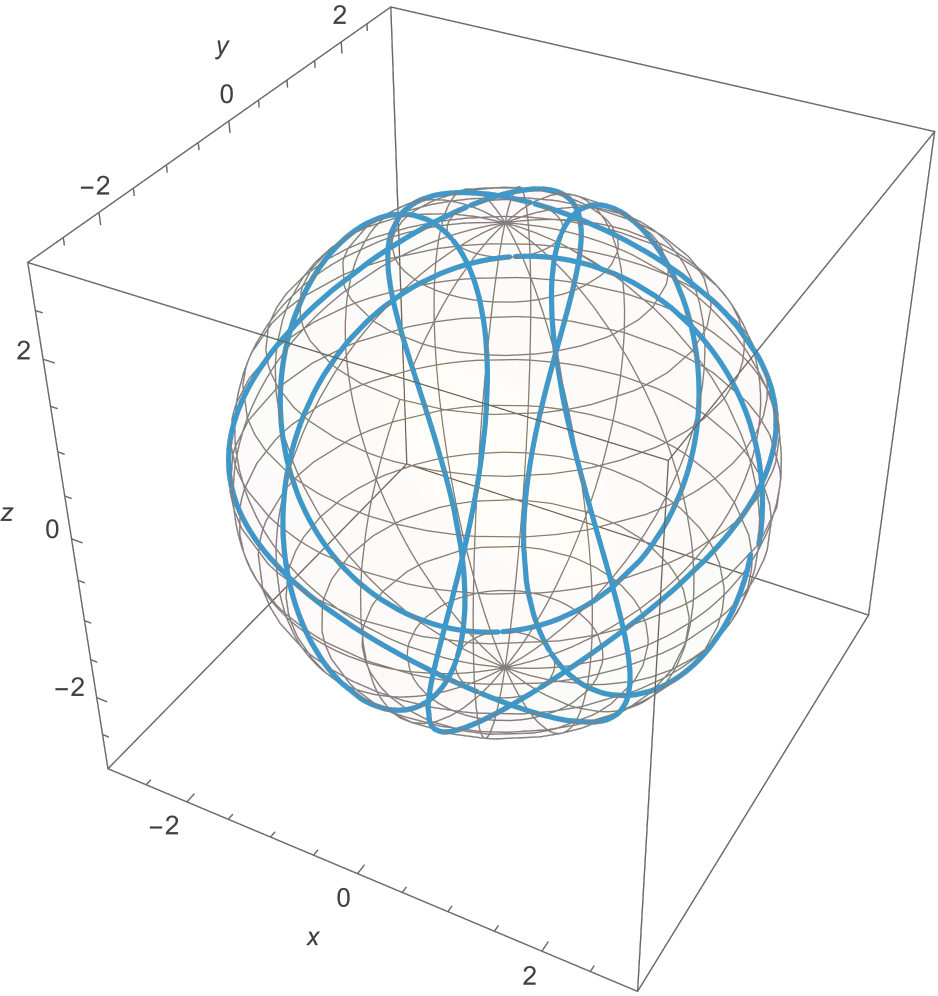}\vspace{0.2cm}
        \includegraphics[width=\linewidth]{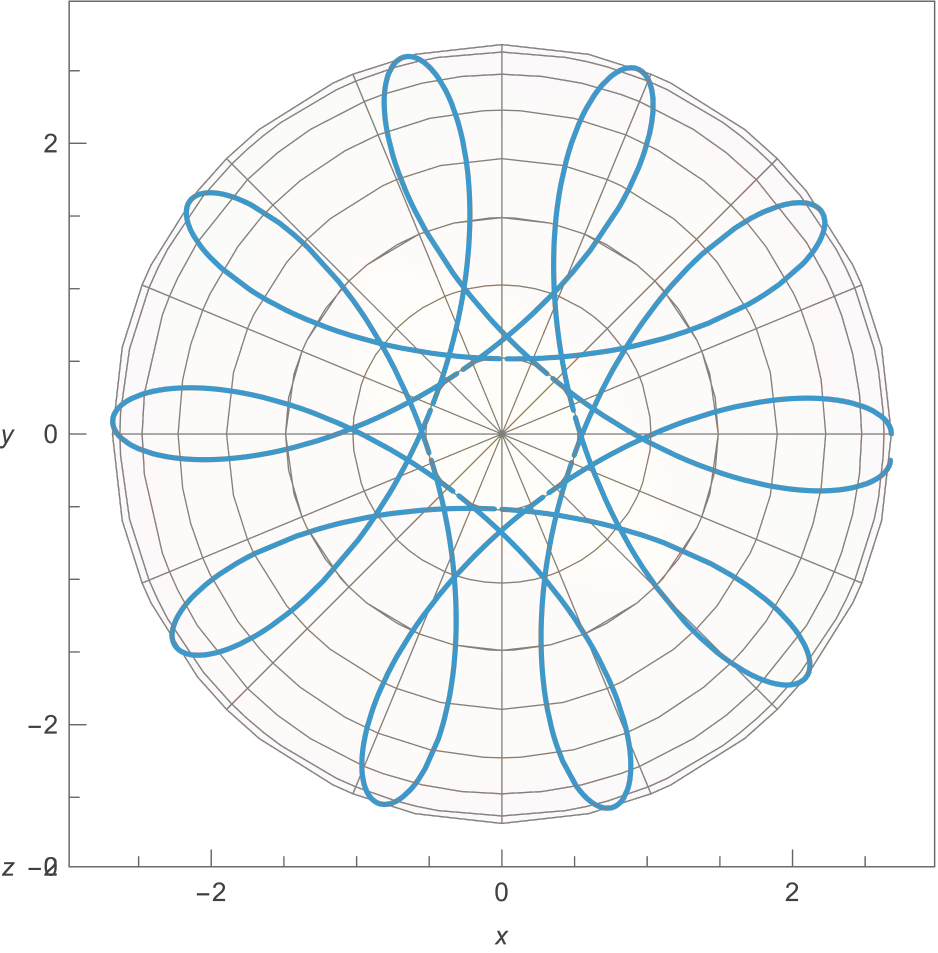}\vspace{0.2cm}
        \includegraphics[width=\linewidth]{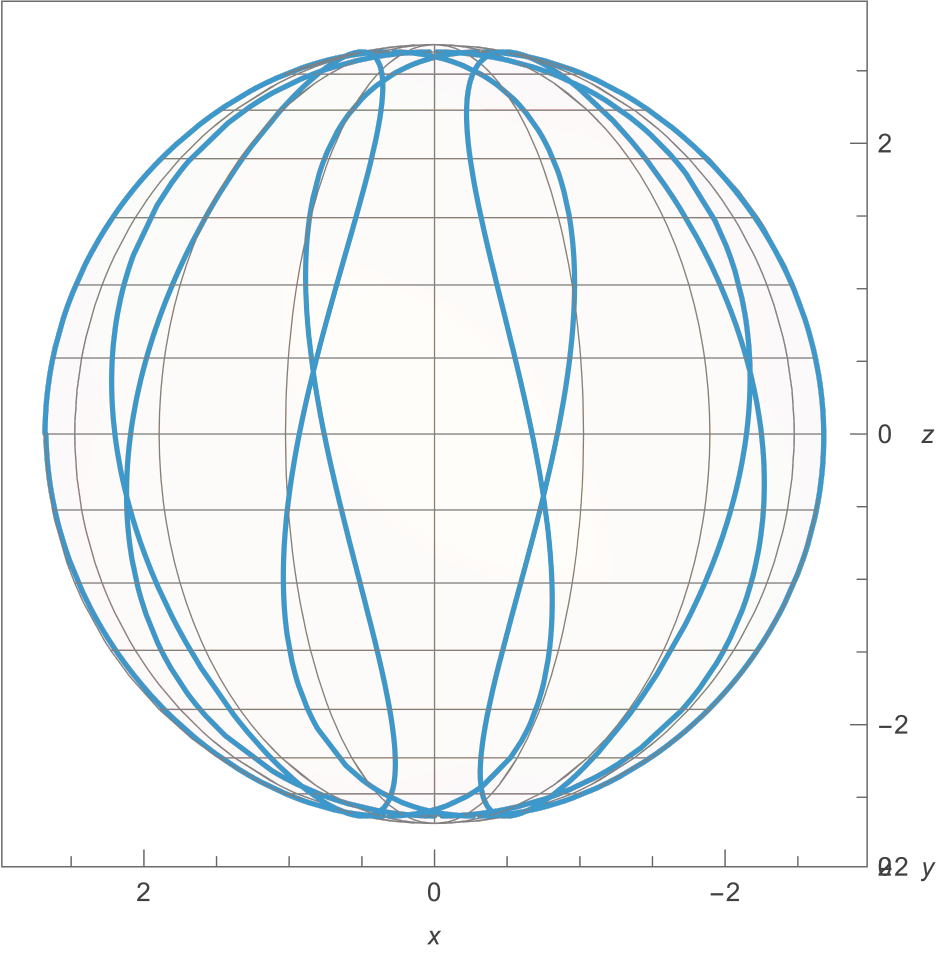}\vspace{0.2cm}
        (b)
    \end{minipage}\hfill
    \begin{minipage}{0.32\textwidth}
        \centering
        \includegraphics[width=\linewidth]{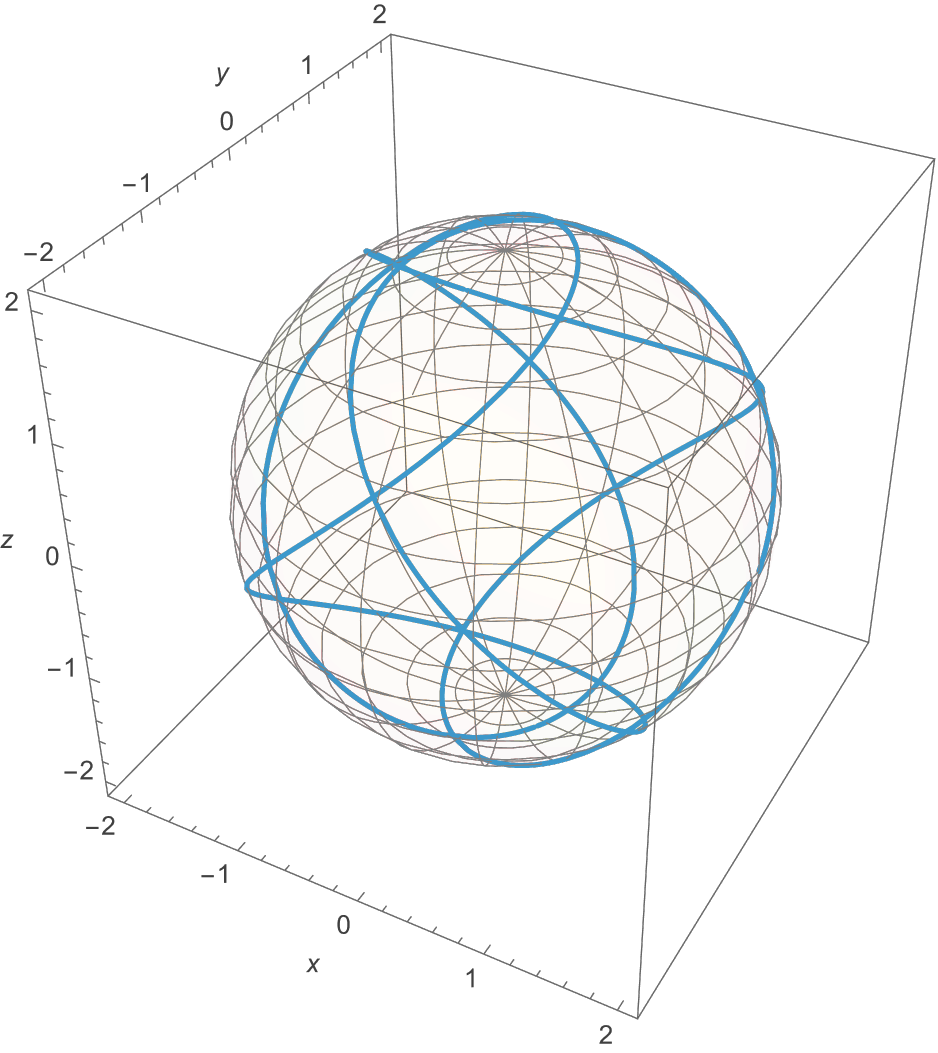}\vspace{0.2cm}
        \includegraphics[width=\linewidth]{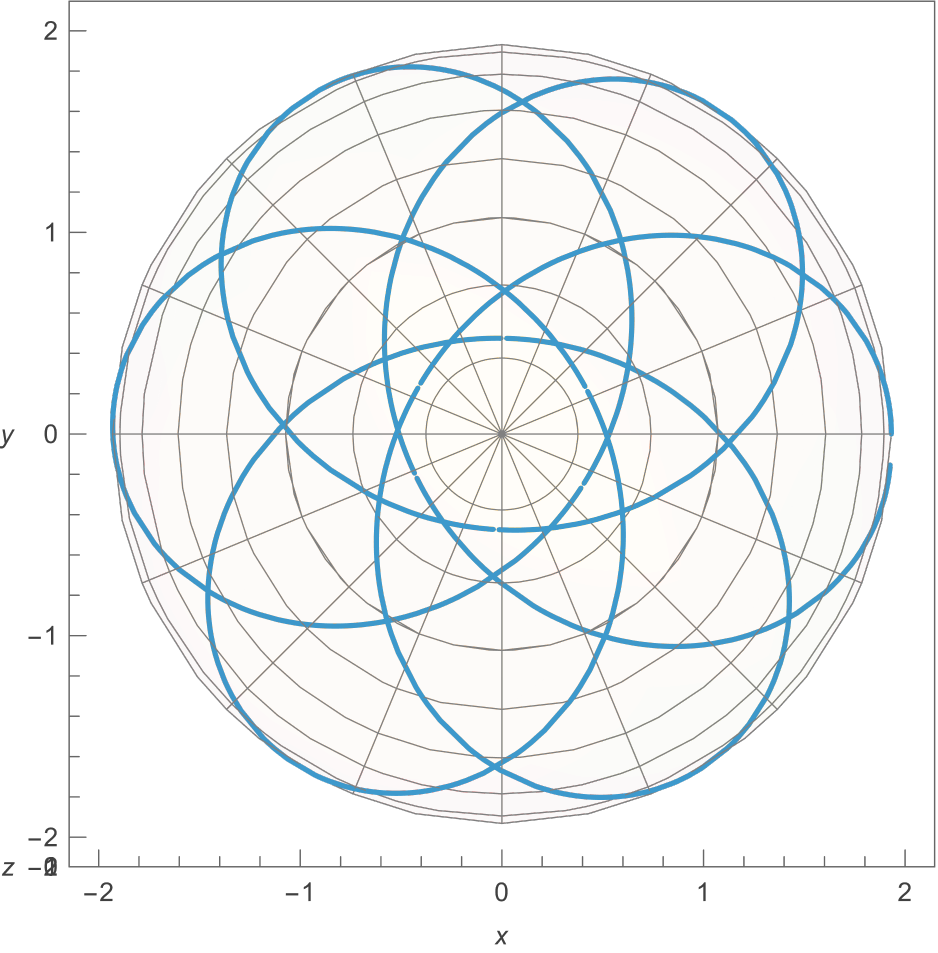}\vspace{0.2cm}
        \includegraphics[width=\linewidth]{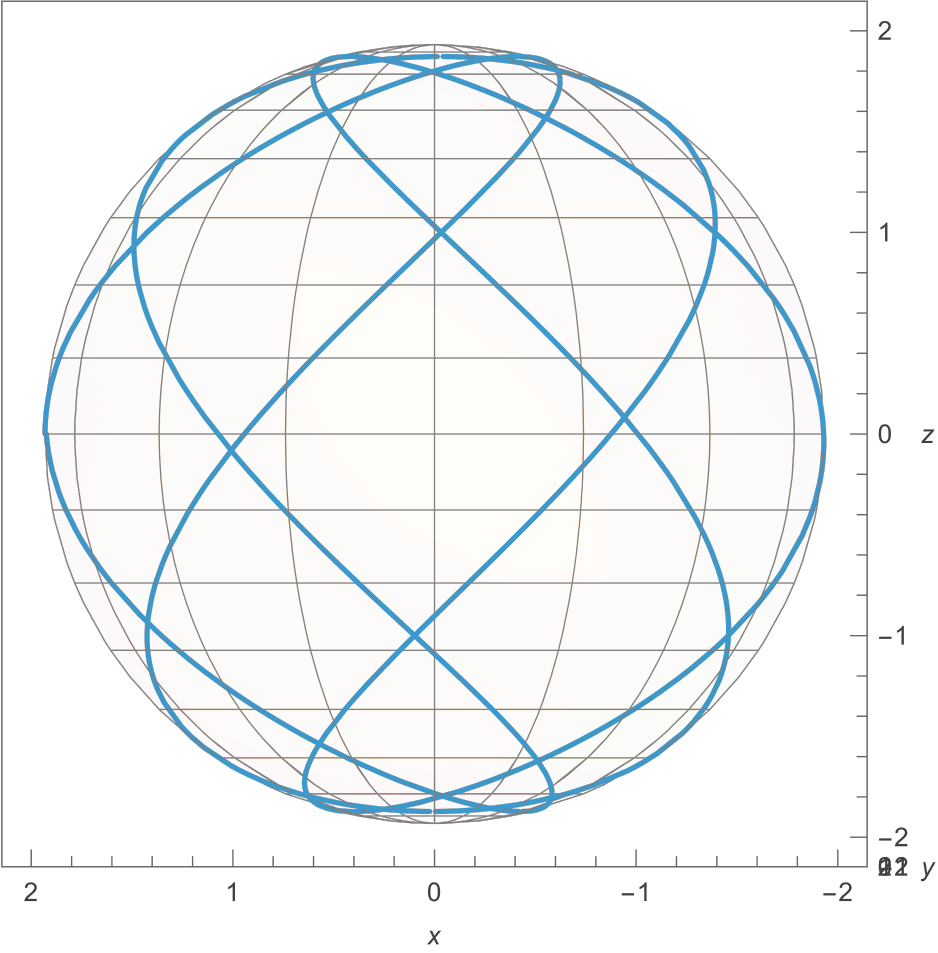}\vspace{0.2cm}
        (c)
    \end{minipage}
    \caption{Extremal SPOs for the parameter sets (a)--(c) in Table~\ref{tab:Extreme}. For each orbit, the upper, middle, and lower panels show the trajectory in the $x$--$y$--$z$ space and its projections onto the $x$--$y$ and $x$--$z$ planes, respectively.}
    \label{Extreme1}
\end{figure}

\begin{figure}[H]
    \centering
    \begin{minipage}{0.32\textwidth}
        \centering
        \includegraphics[width=\linewidth]{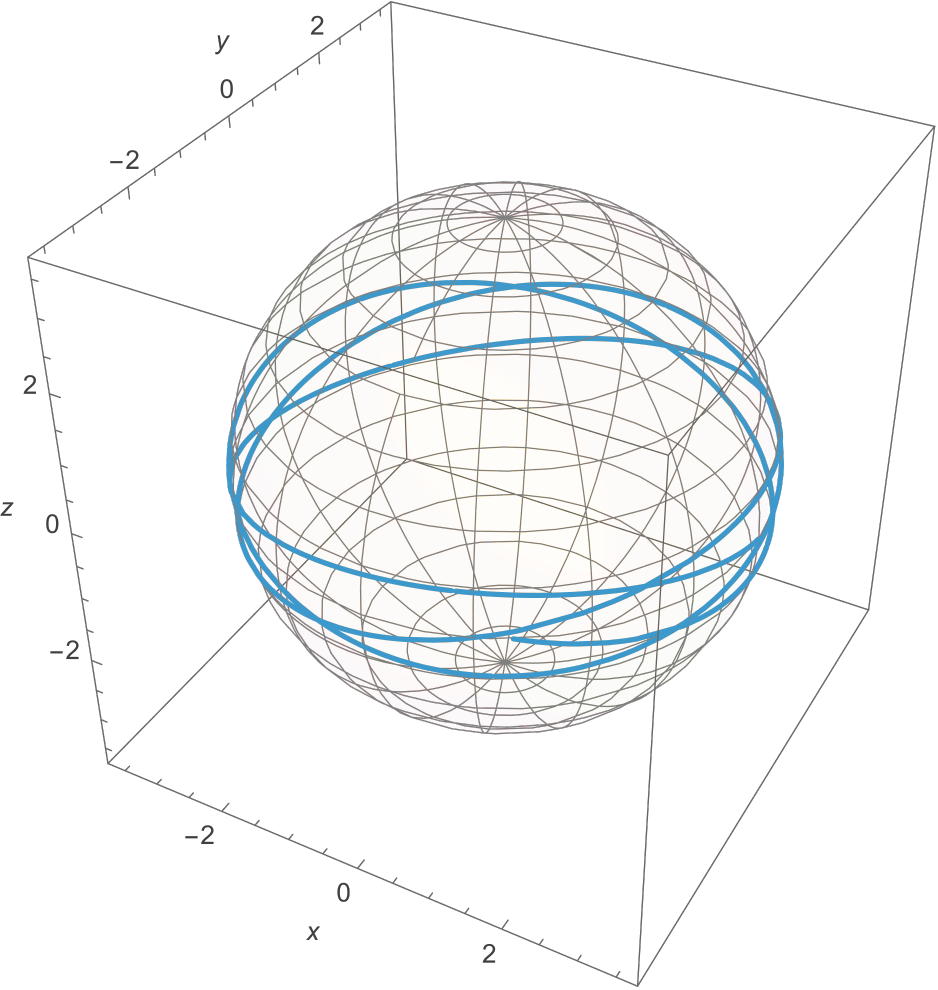}\vspace{0.2cm}
        \includegraphics[width=\linewidth]{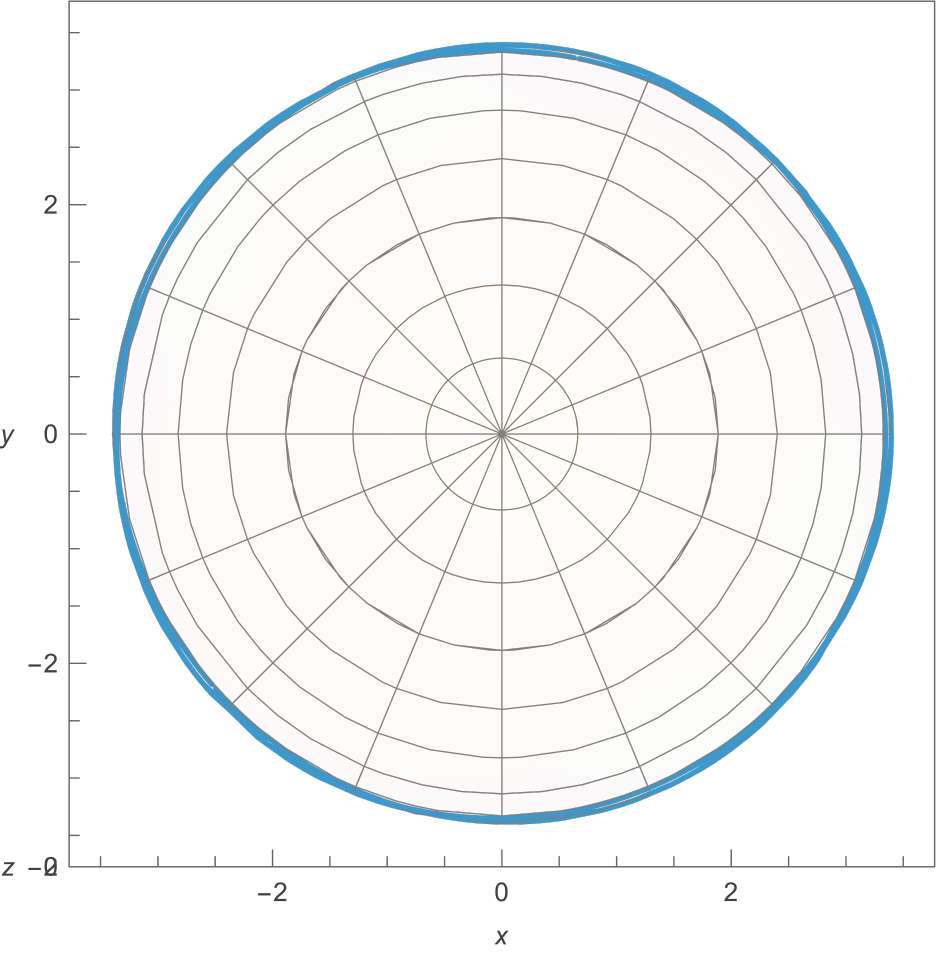}\vspace{0.2cm}
        \includegraphics[width=\linewidth]{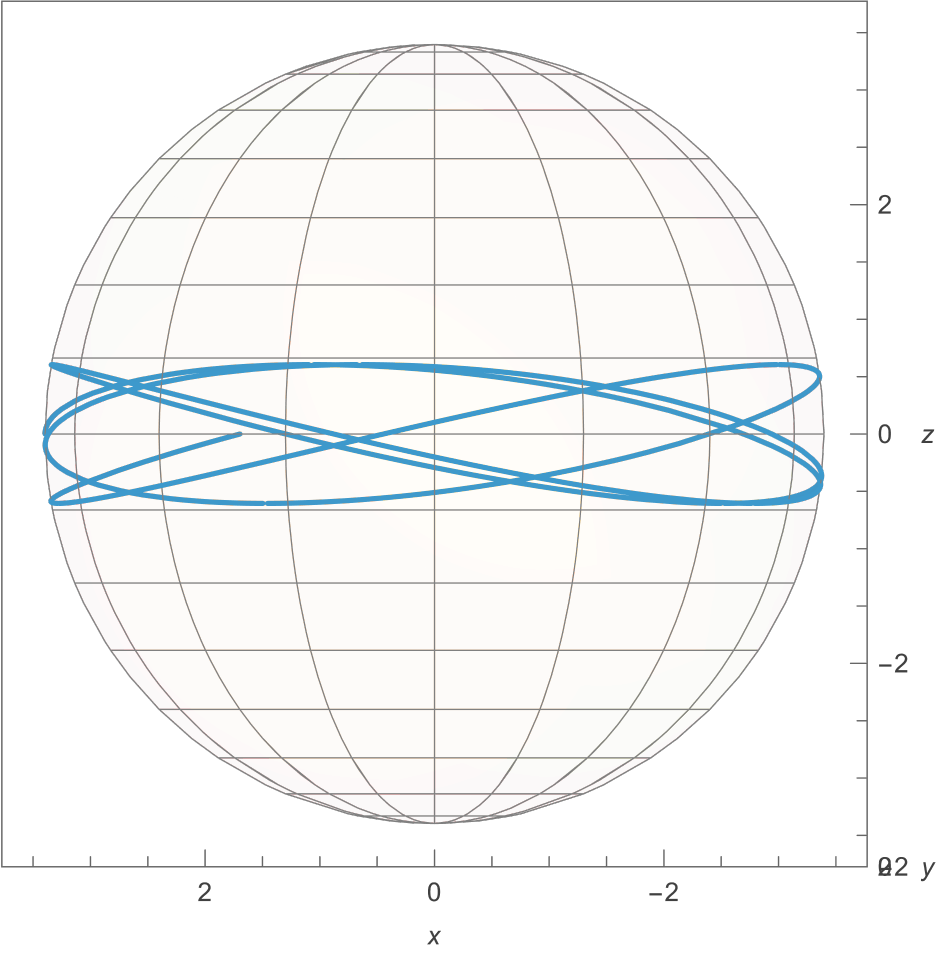}\vspace{0.2cm}
        (d)
    \end{minipage}\hfill
    \begin{minipage}{0.32\textwidth}
        \centering
        \includegraphics[width=\linewidth]{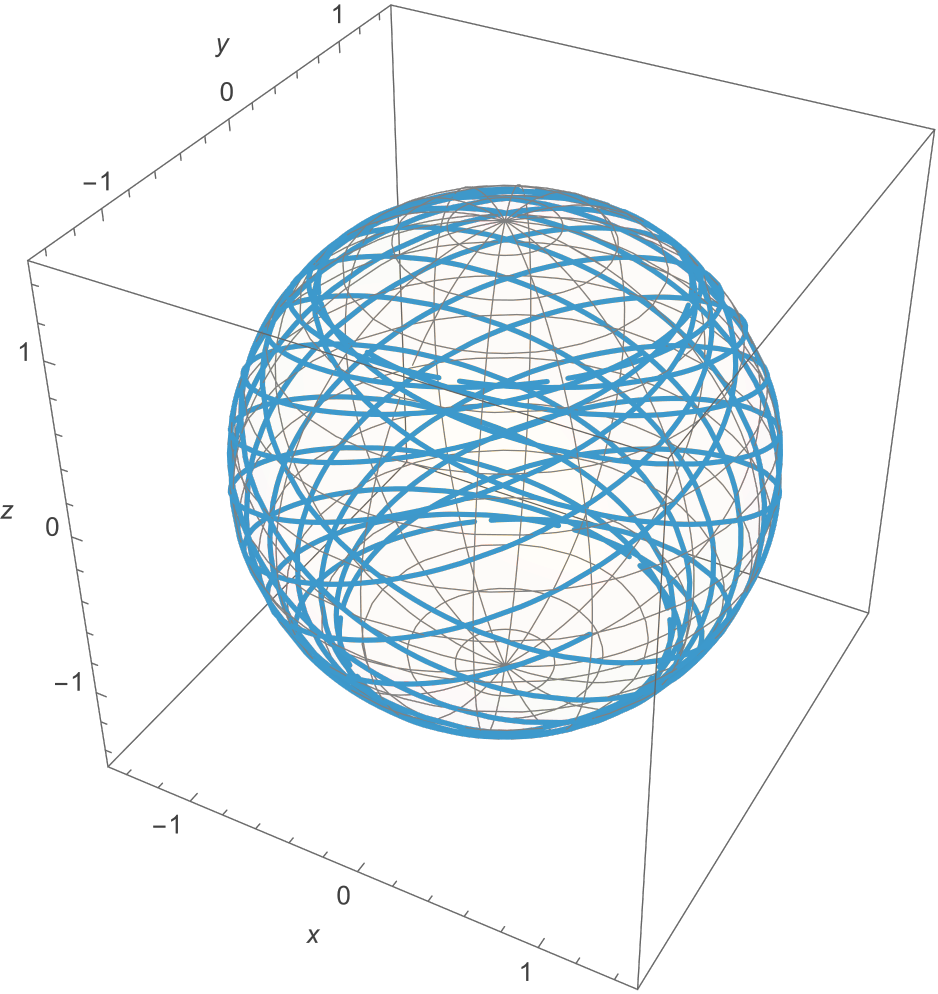}\vspace{0.2cm}
        \includegraphics[width=\linewidth]{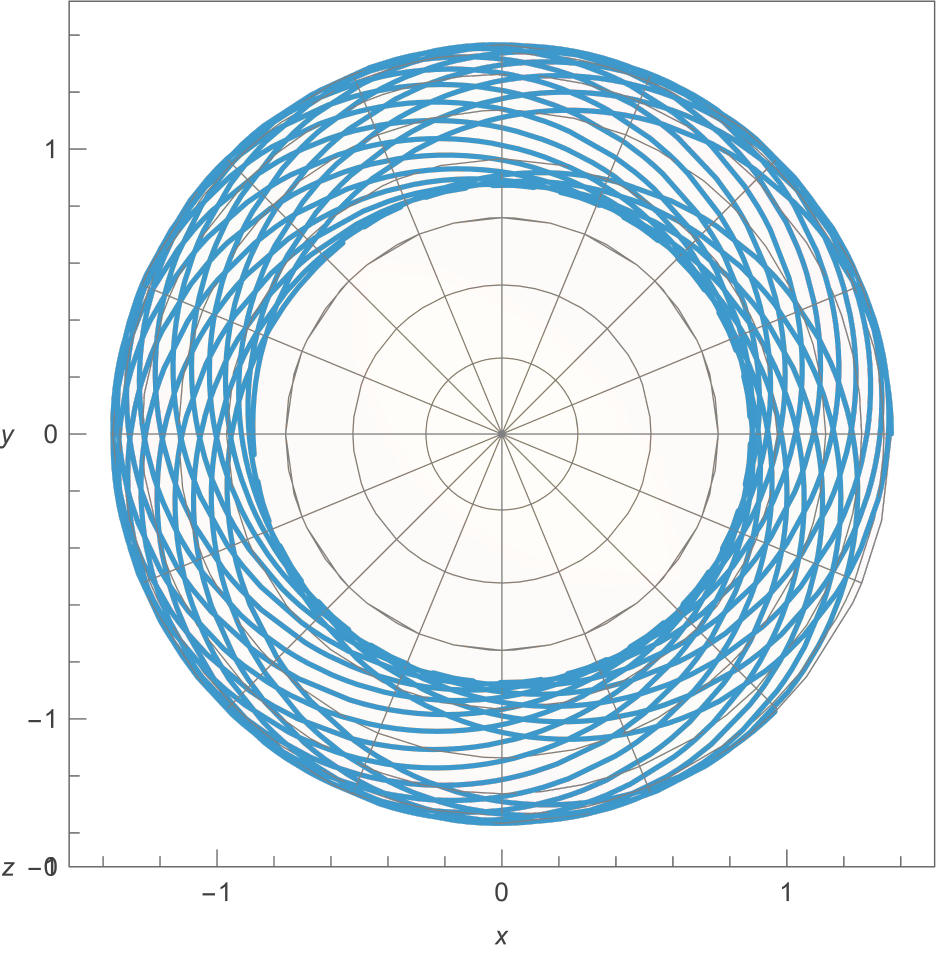}\vspace{0.2cm}
        \includegraphics[width=\linewidth]{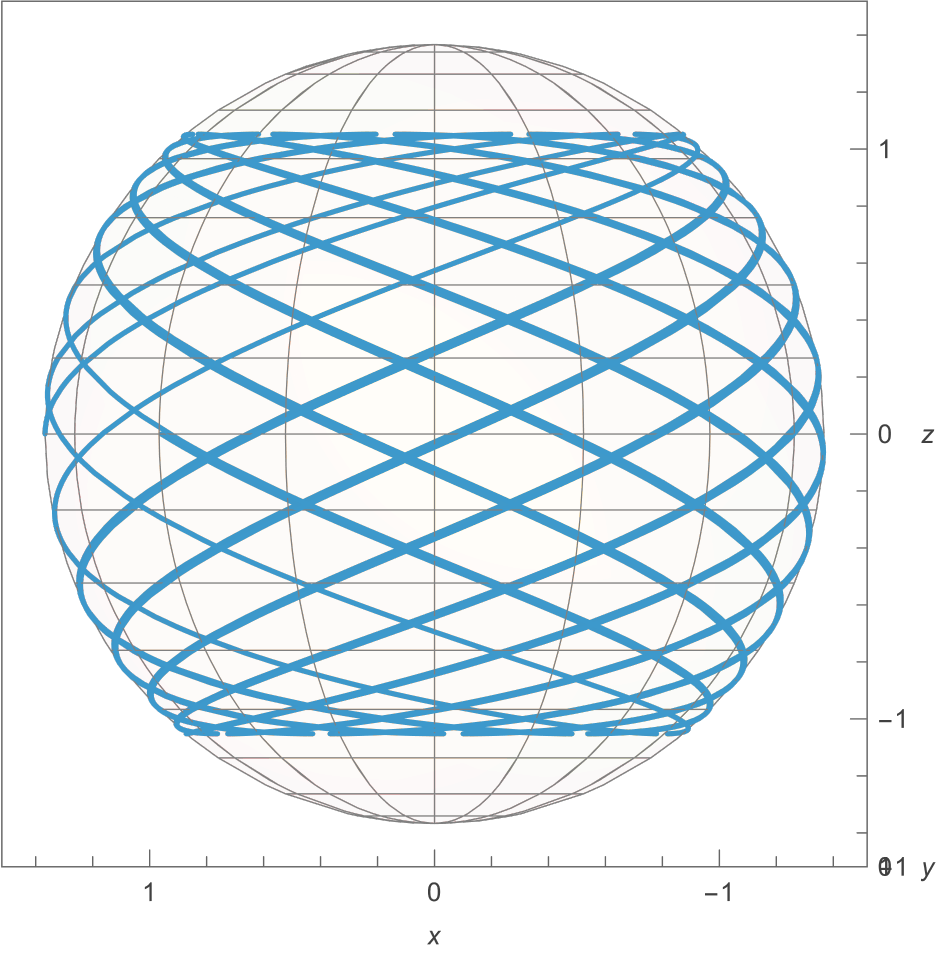}\vspace{0.2cm}
        (e)
    \end{minipage}\hfill
    \begin{minipage}{0.32\textwidth}
        \centering
        \includegraphics[width=\linewidth]{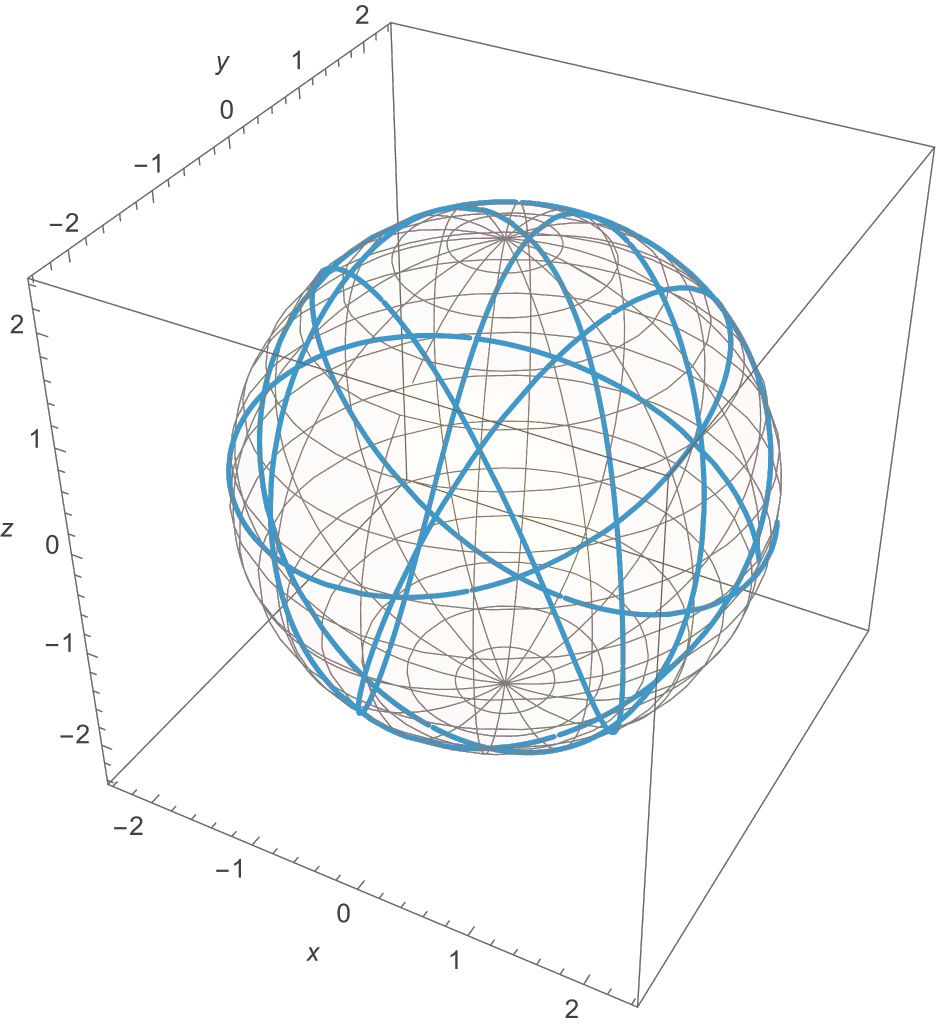}\vspace{0.2cm}
        \includegraphics[width=\linewidth]{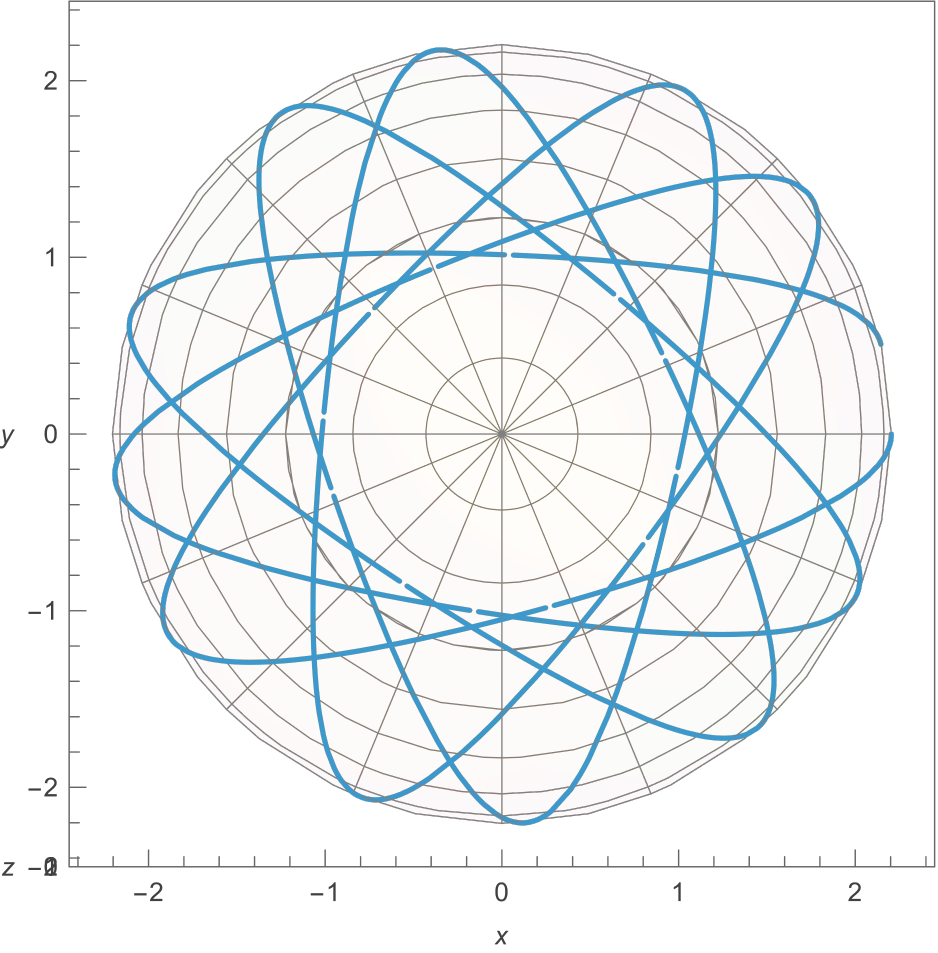}\vspace{0.2cm}
        \includegraphics[width=\linewidth]{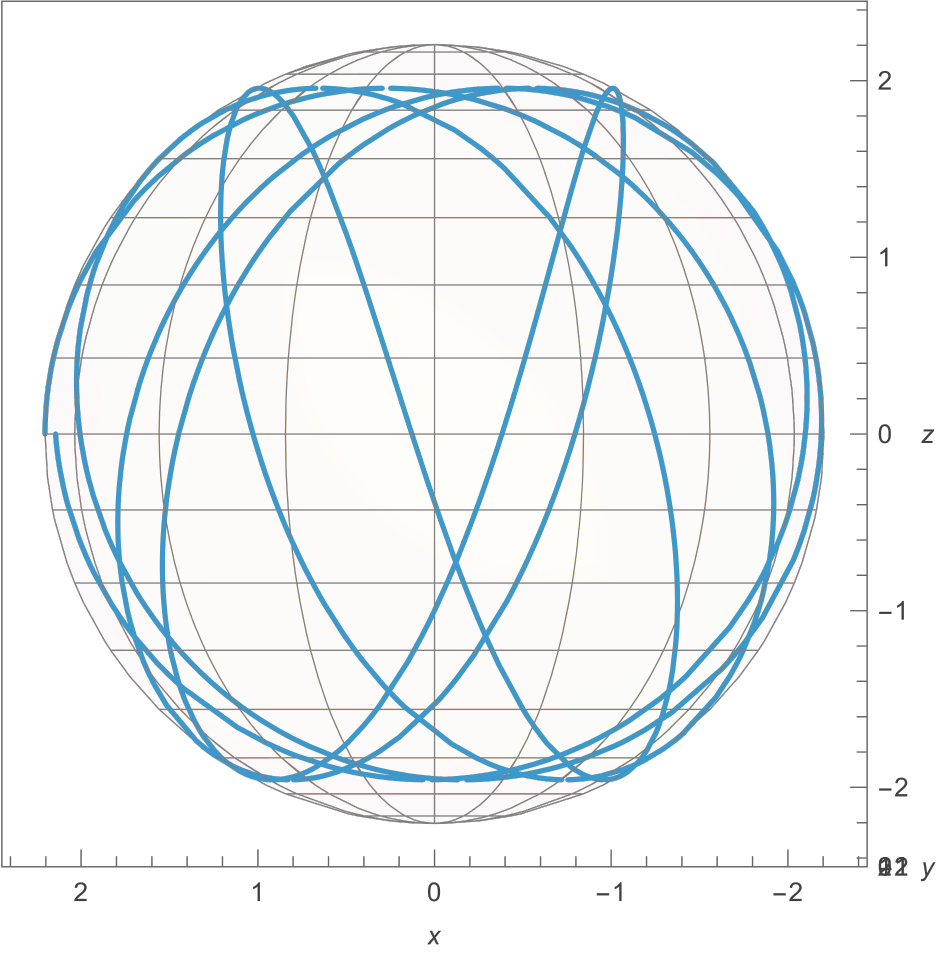}\vspace{0.2cm}
        (f)
    \end{minipage}
    \caption{Extremal SPOs for the parameter sets (d)--(f) in Table~\ref{tab:Extreme}. For each orbit, the upper, middle, and lower panels show the trajectory in the $x$--$y$--$z$ space and its projections onto the $x$--$y$ and $x$--$z$ planes, respectively.}
    \label{Extreme2}
\end{figure}

The corresponding spatial trajectories are displayed in Figs.~\ref{Extreme1} and \ref{Extreme2}, with their parameters summarized in Table~\ref{tab:Extreme}. Each trajectory starts at the equatorial plane and initially moves toward the northern hemisphere. The orbit in Fig.~\ref{Extreme1}(a), corresponding to the Kerr limit $b=0$, exhibits two latitudinal oscillations in the plotted trajectory and reproduces the extremal Kerr orbit discussed by Teo~\cite{Teo2003}. Its accumulated azimuthal angle, $\Delta\phi\simeq\pi$, also makes it resemble the near-photon-boomerang orbit discussed in Ref.~\cite{Alam2024}. 

As the nonlinear parameter $b$ increases, the orbital trajectories exhibit different latitudinal excursions and azimuthal advances. These changes are reflected in the values of $u_0$ and $\Delta\phi$ listed in Table~\ref{tab:Extreme}. The three-dimensional trajectories and their projections onto the $x$--$y$ and $x$--$z$ planes provide complementary views of these orbital features.

\subsubsection{Horizonless Case}

For other combinations of the spin parameter $a$ and the nonlinear parameter $b$, the function $\Delta(r)$ may have no real positive roots. In this case, the spacetime has no event horizon and belongs to the horizonless, or naked, regime. Unlike the black-hole configurations considered above, the absence of an event horizon allows stable SPOs to occur. In terms of the radial potential, a spherical orbit is stable when
\begin{equation}
    \left.
    \frac{d^2R}{dr^2}
    \right|_{r=r_{\rm SO}}<0,
\end{equation}
where the SPO conditions $R(r_{\rm SO})=R'(r_{\rm SO})=0$ are also imposed.

We select representative stable SPOs by fixing $a=M=1$ and varying $b$ and $\Phi$. The resulting orbital parameters are listed in Table~\ref{tab:naked}. In particular, the values of $u_0$ and $\Delta\phi$ characterize the latitudinal excursions and azimuthal advances of the corresponding trajectories.

\begin{table}[H]
    \centering
    \caption{Parameters of representative stable SPOs in the horizonless rotating geometry with $a=M=1$.}
    \label{tab:naked}
    \vspace{0.2cm}
    \begin{tabular}{|c|c|c|c|c|c|c|}
        \hline
        \makebox[2cm]{Orbit} & \makebox[1.5cm]{$b$} & \makebox[1.5cm]{$\Phi$}  & \makebox[2cm]{$r_{\rm SO}$} & \makebox[2cm]{$C$} & \makebox[2cm]{$u_0$} & \makebox[2cm]{$\Delta\phi$} \\
        \hline
        (a) & 0.1   & 0      & 1.13355 & 37.7618  & 1 & 16.2164 \\
        (b) & 0.2   & -1      & 1.247828 & 31.6639  & 0.985022 & 9.84953 \\
        (c) & 0.3   & -1      & 1.611834 & 20.1265  & 0.977094 & 5.73495  \\
        (d) & 0.4   & -2      & 1.68921 & 19.4118  & 0.9138 & 4.93489  \\
        (e) & 0.5   & -3      & 1.730324 & 16.4715  & 0.809647 & 4.45289  \\
        \hline
    \end{tabular}
\end{table}

\begin{figure}[H]
    \centering
    \begin{minipage}{0.32\textwidth}
        \centering
        \includegraphics[width=\linewidth]{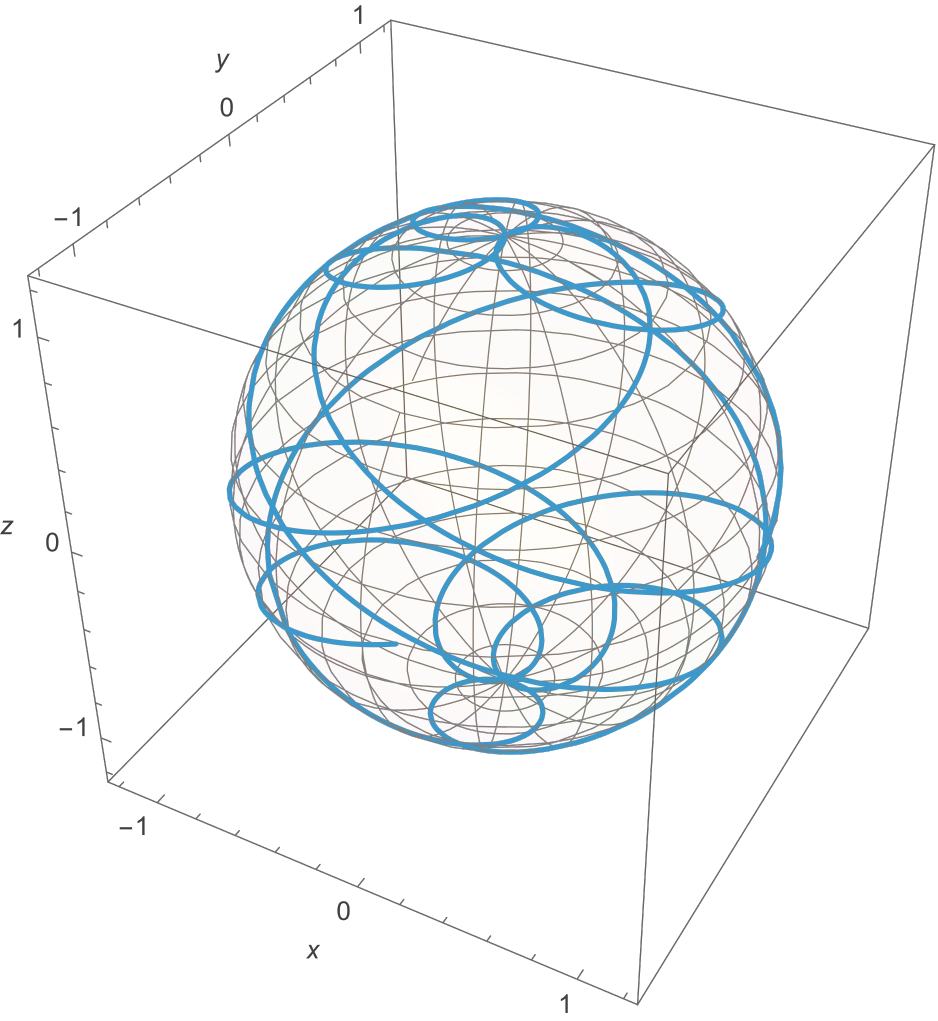}\vspace{0.2cm}
        \includegraphics[width=\linewidth]{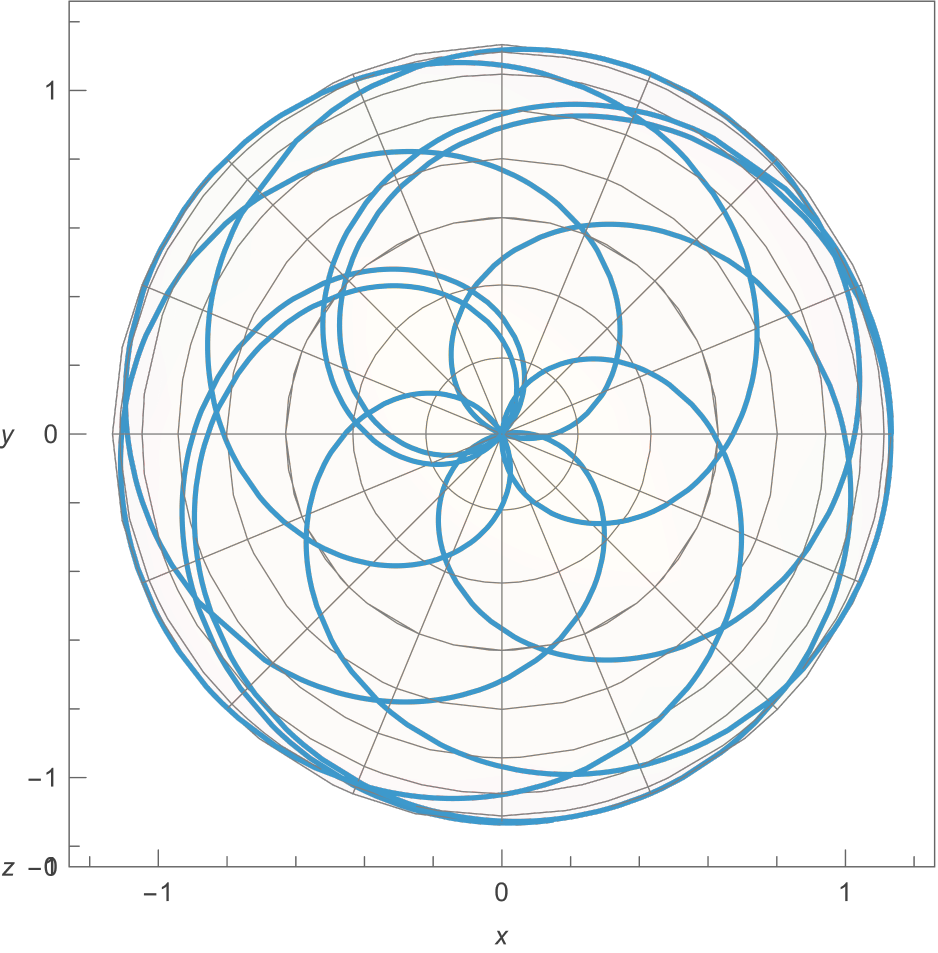}\vspace{0.2cm}
        \includegraphics[width=\linewidth]{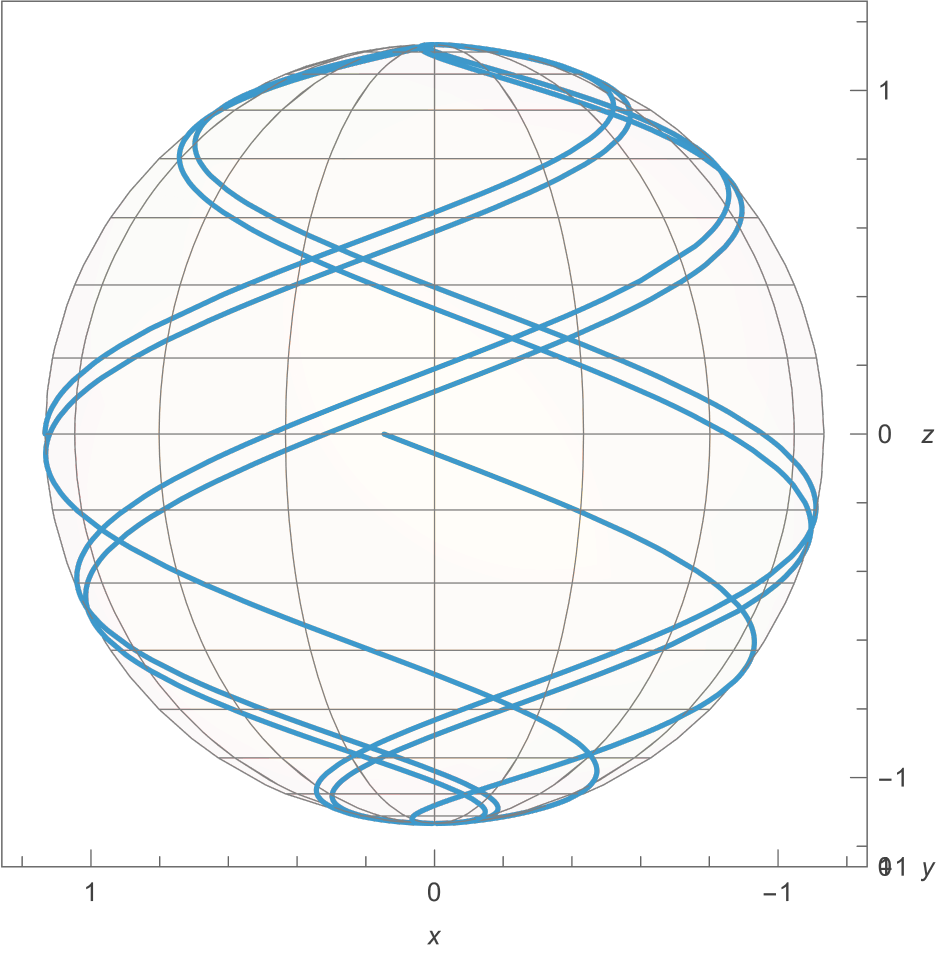}\vspace{0.2cm}
        (a)
    \end{minipage}\hfill
    \begin{minipage}{0.32\textwidth}
        \centering
        \includegraphics[width=\linewidth]{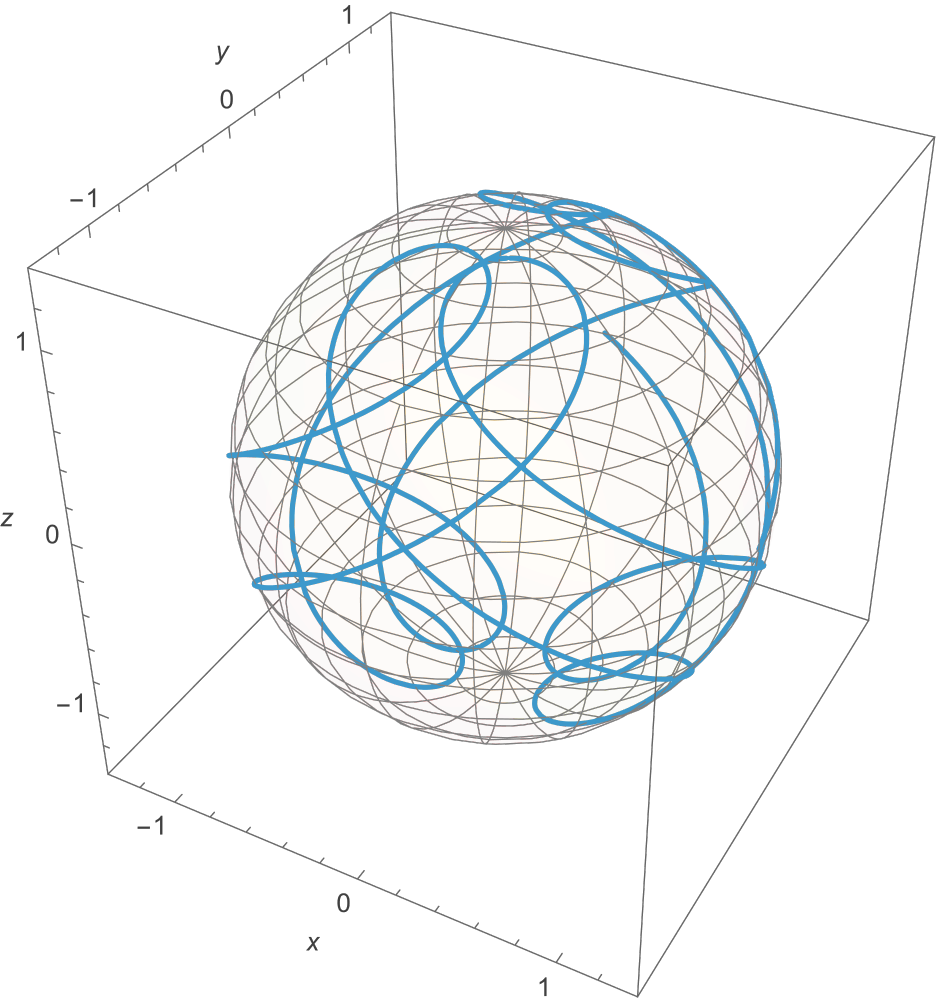}\vspace{0.2cm}
        \includegraphics[width=\linewidth]{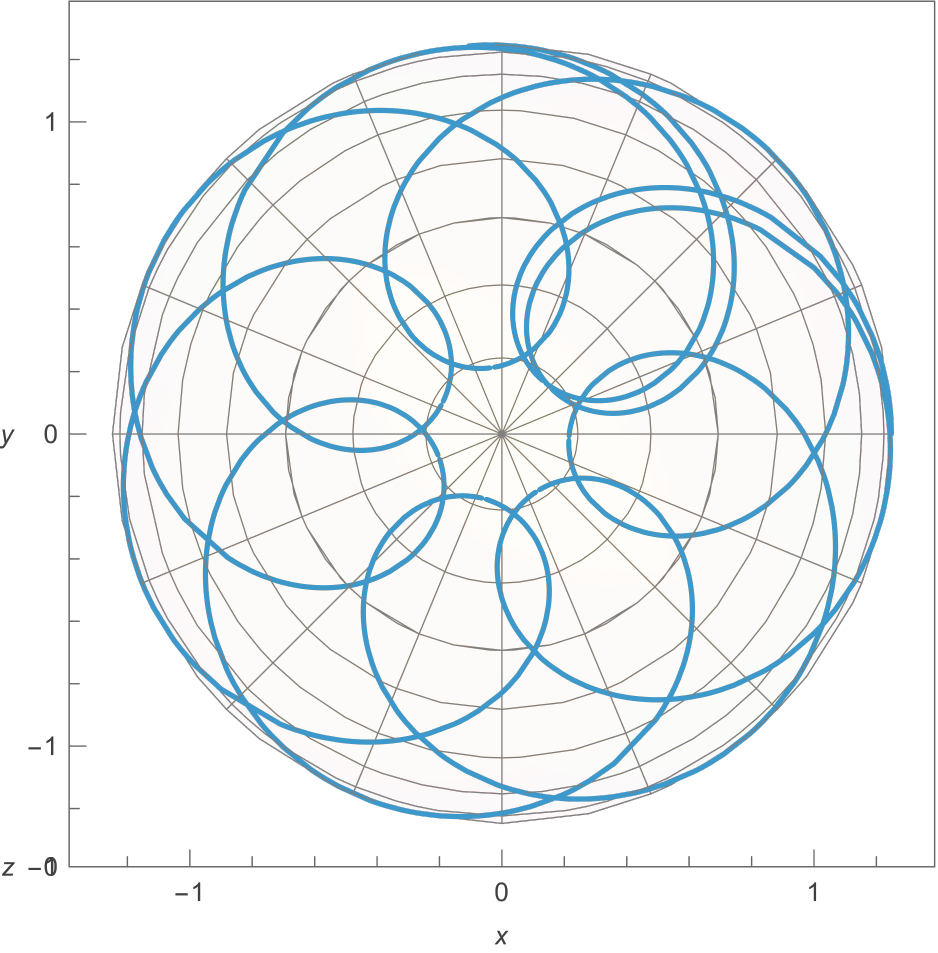}\vspace{0.2cm}
        \includegraphics[width=\linewidth]{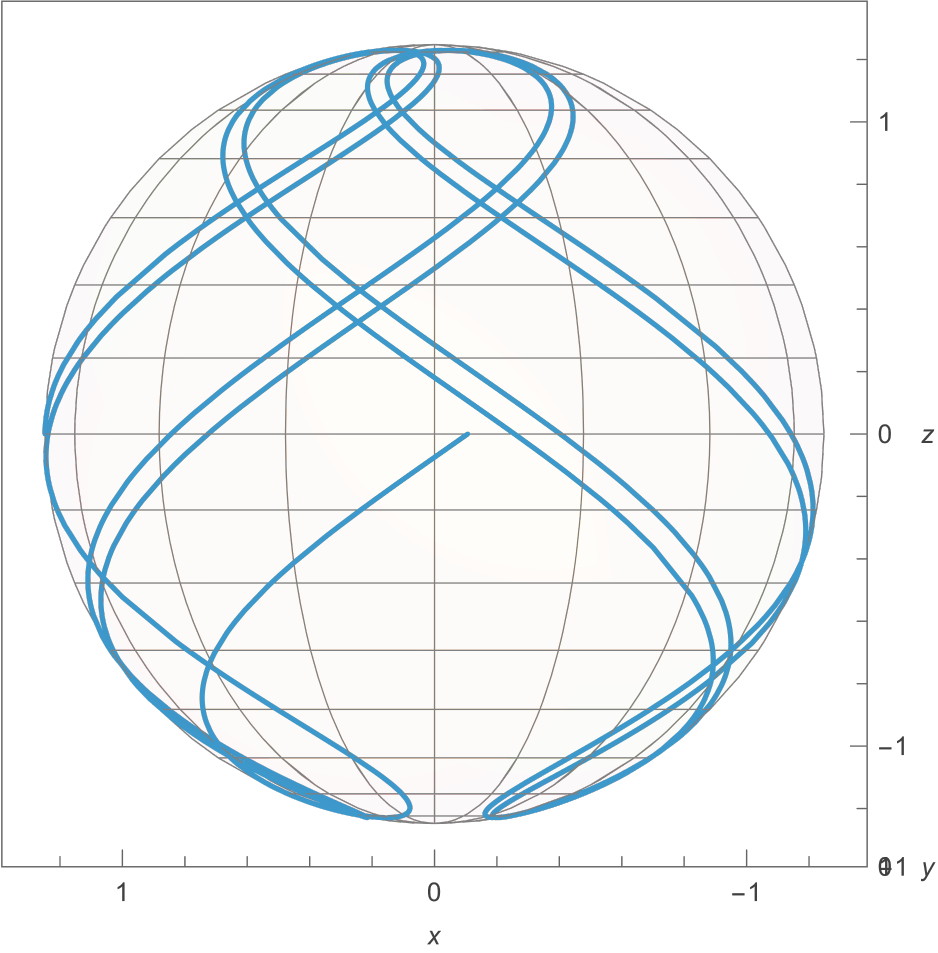}\vspace{0.2cm}
        (b)
    \end{minipage}\hfill
    \begin{minipage}{0.32\textwidth}
        \centering
        \includegraphics[width=\linewidth]{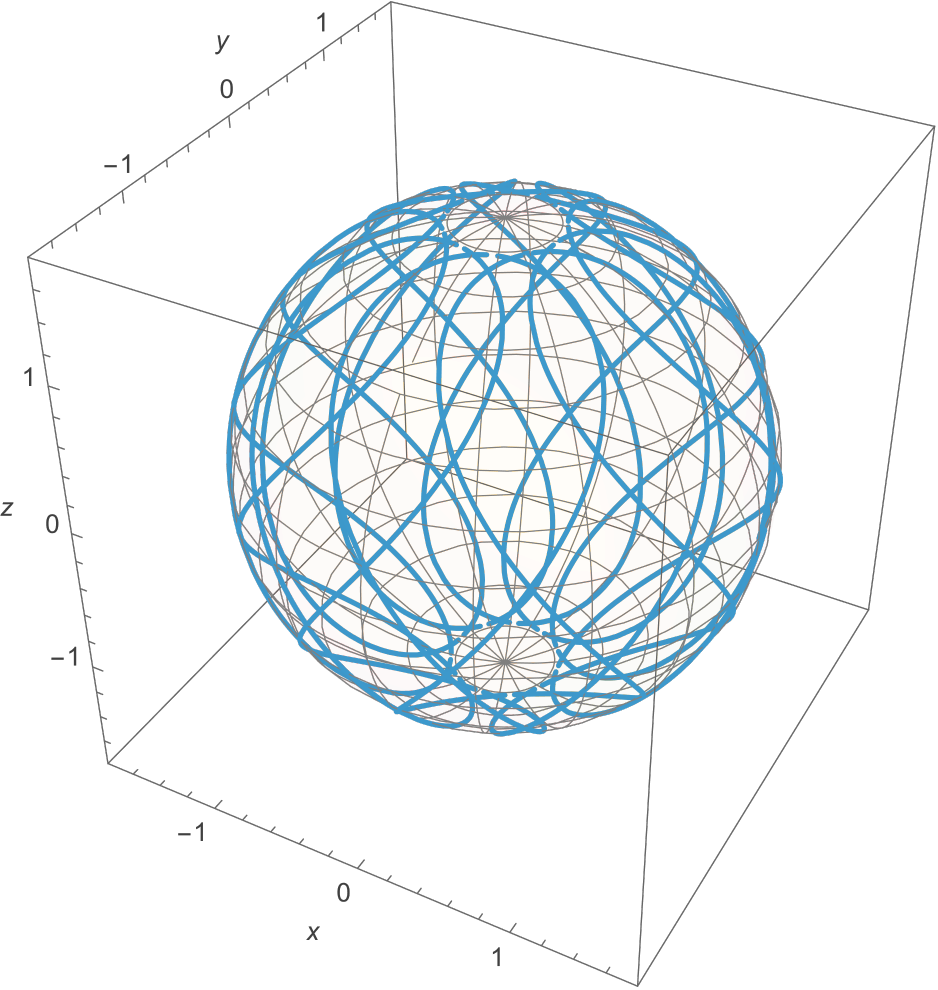}\vspace{0.2cm}
        \includegraphics[width=\linewidth]{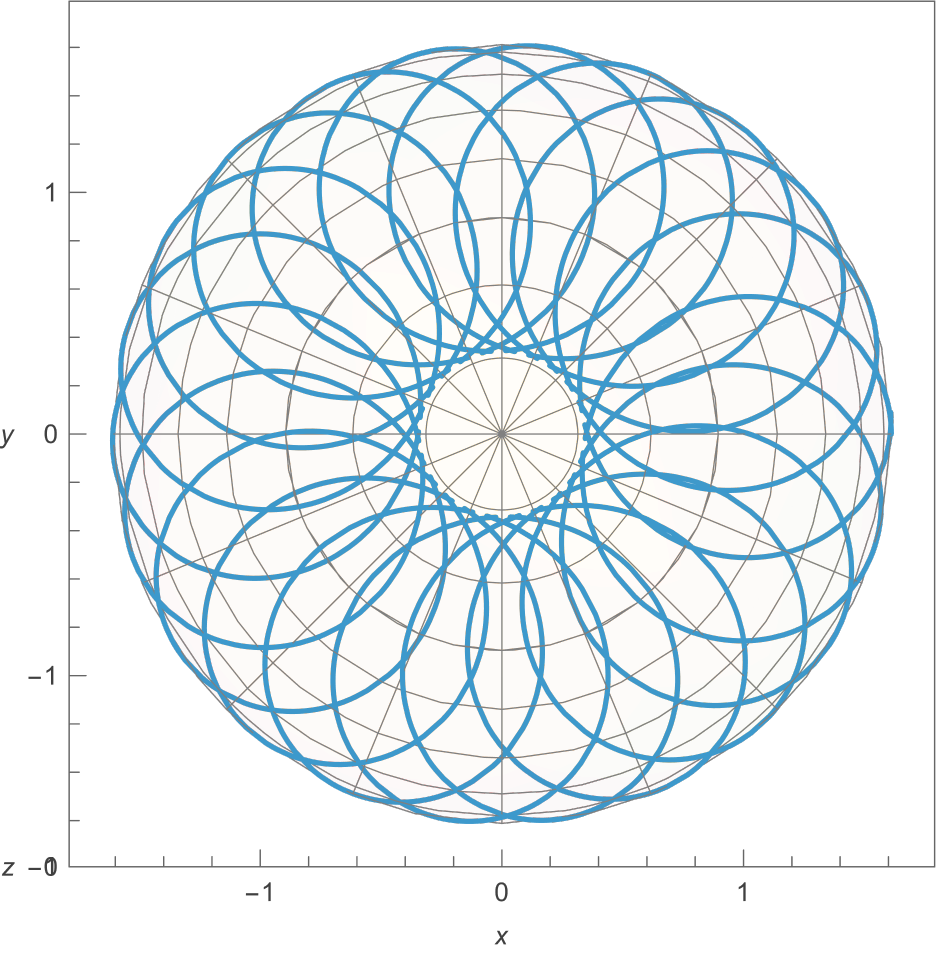}\vspace{0.2cm}
        \includegraphics[width=\linewidth]{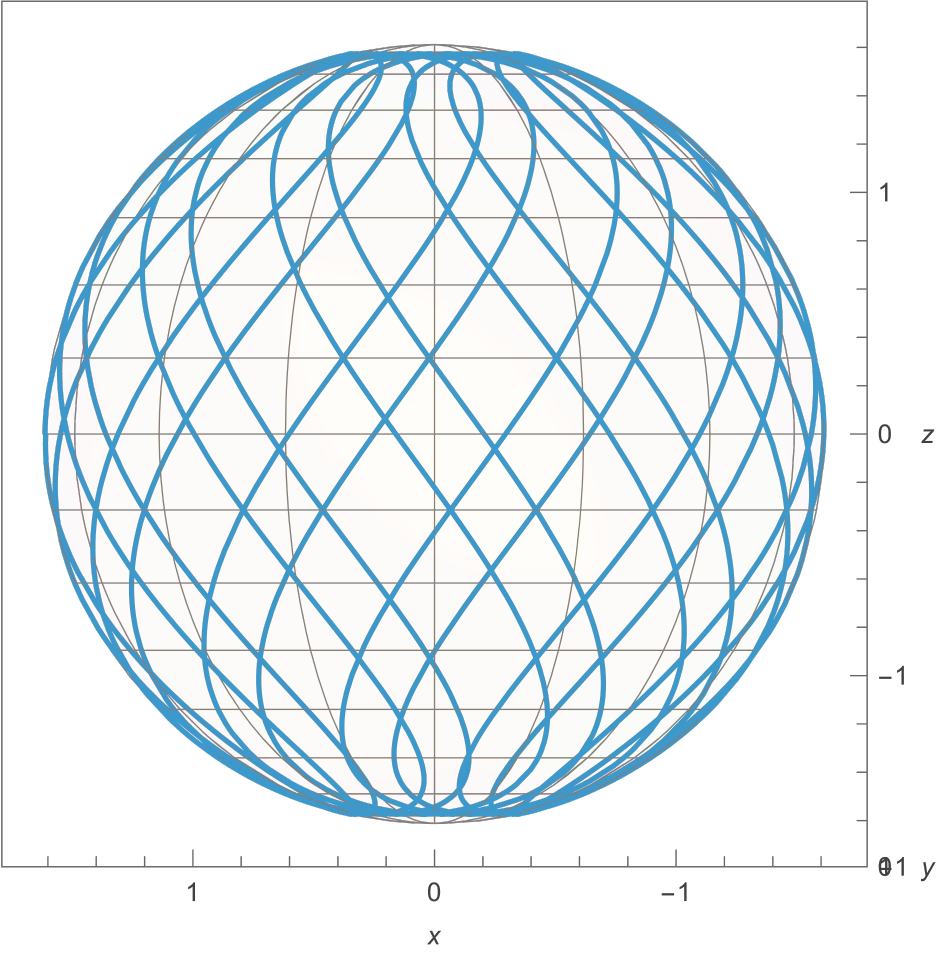}\vspace{0.2cm}
        (c)
    \end{minipage}
    \caption{Stable SPOs in the horizonless geometry for the parameter sets (a)--(c) in Table~\ref{tab:naked}. For each orbit, the upper, middle, and lower panels show the trajectory in the $x$--$y$--$z$ space and its projections onto the $x$--$y$ and $x$--$z$ planes, respectively.}
    \label{naked1}
\end{figure}

\begin{figure}[H]
    \centering
    \begin{minipage}{0.32\textwidth}
        \centering
        \includegraphics[width=\linewidth]{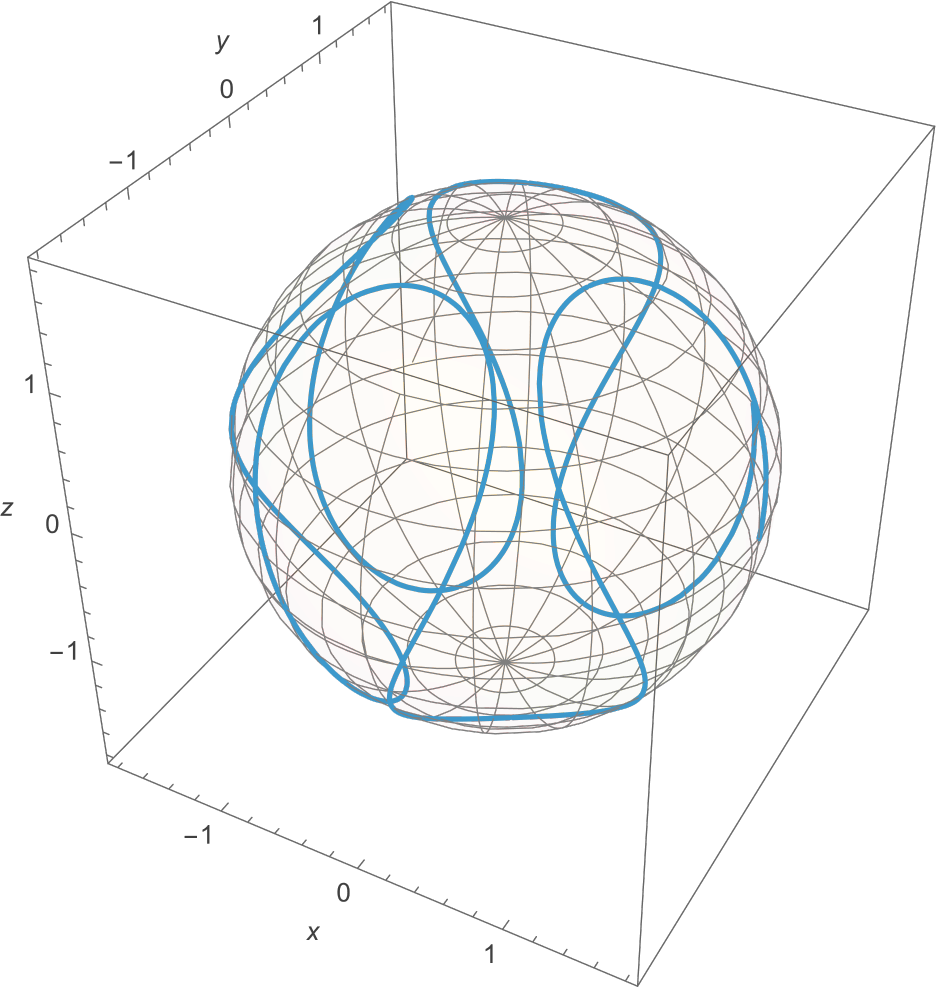}\vspace{0.2cm}
        \includegraphics[width=\linewidth]{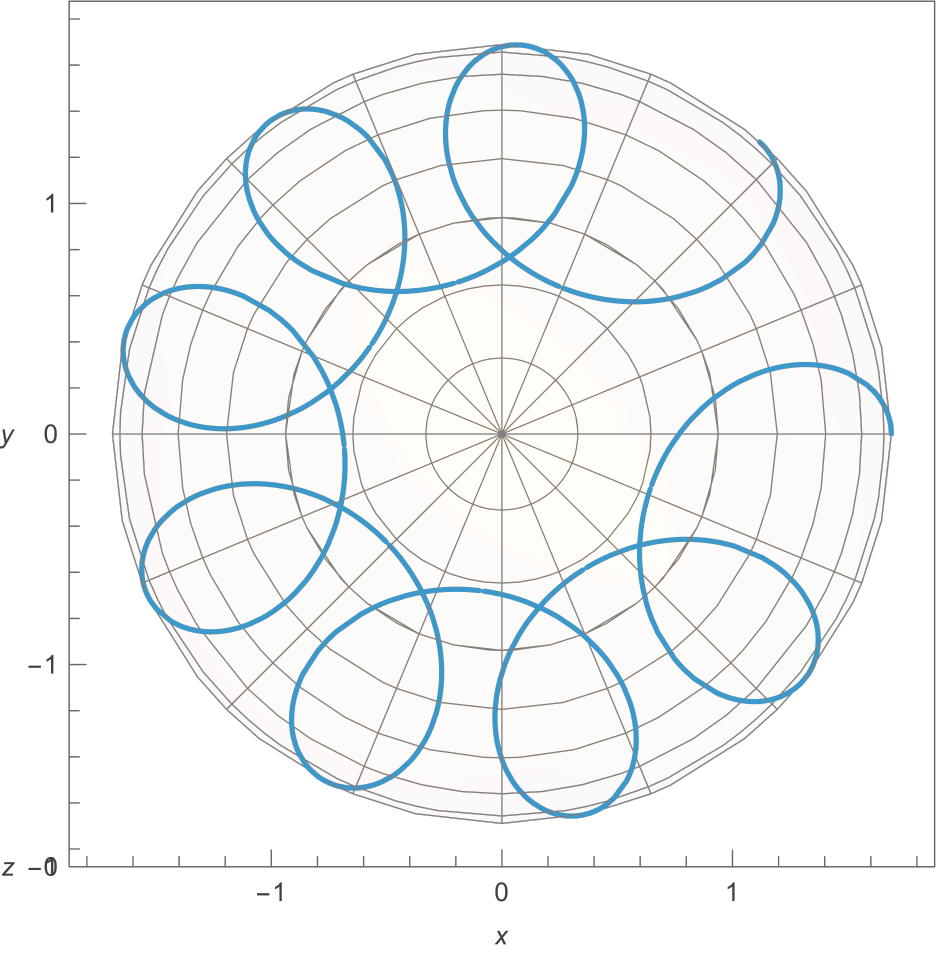}\vspace{0.2cm}
        \includegraphics[width=\linewidth]{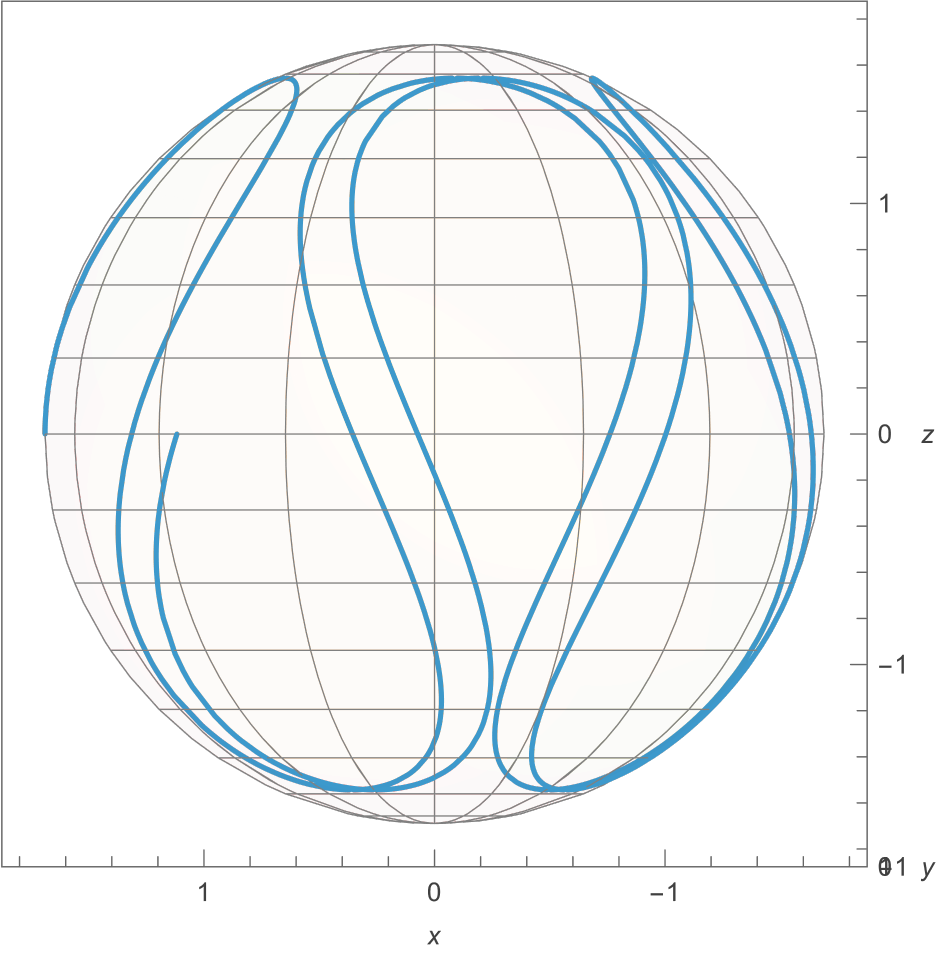}\vspace{0.2cm}
        (d)
    \end{minipage}\hspace{0.03\textwidth}
    \begin{minipage}{0.32\textwidth}
        \centering
        \includegraphics[width=\linewidth]{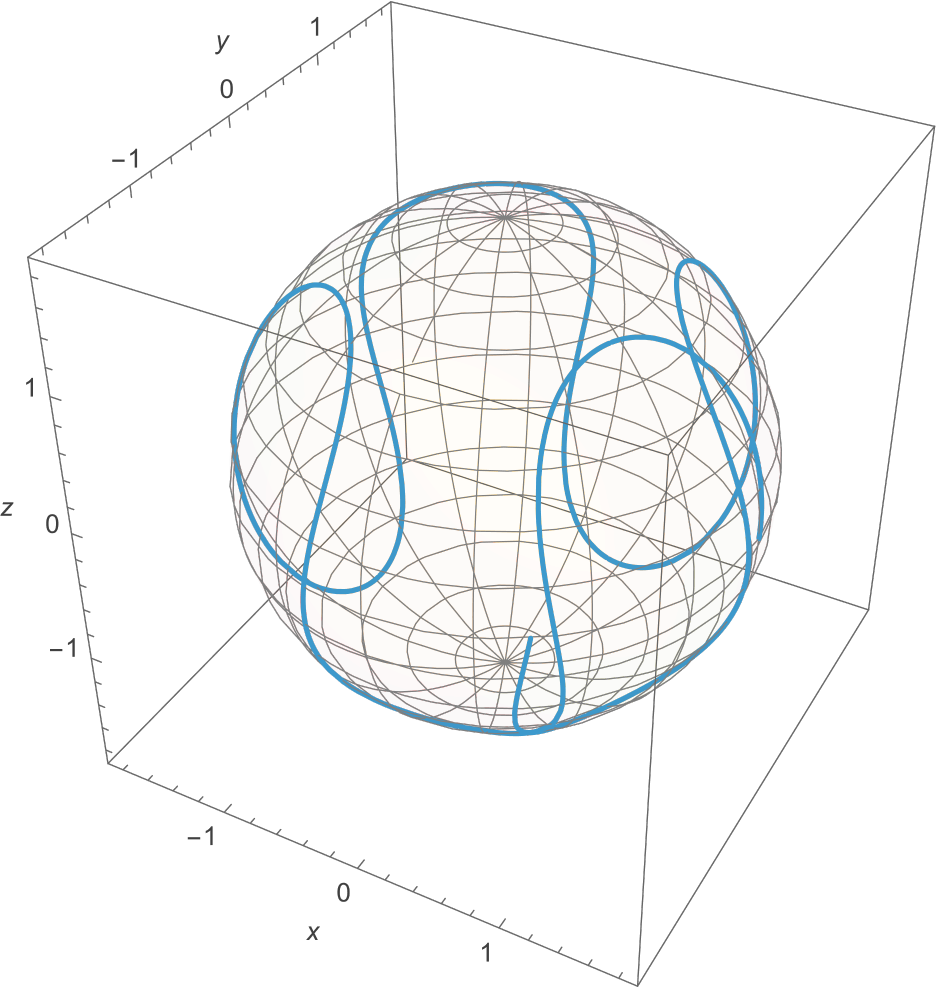}\vspace{0.2cm}
        \includegraphics[width=\linewidth]{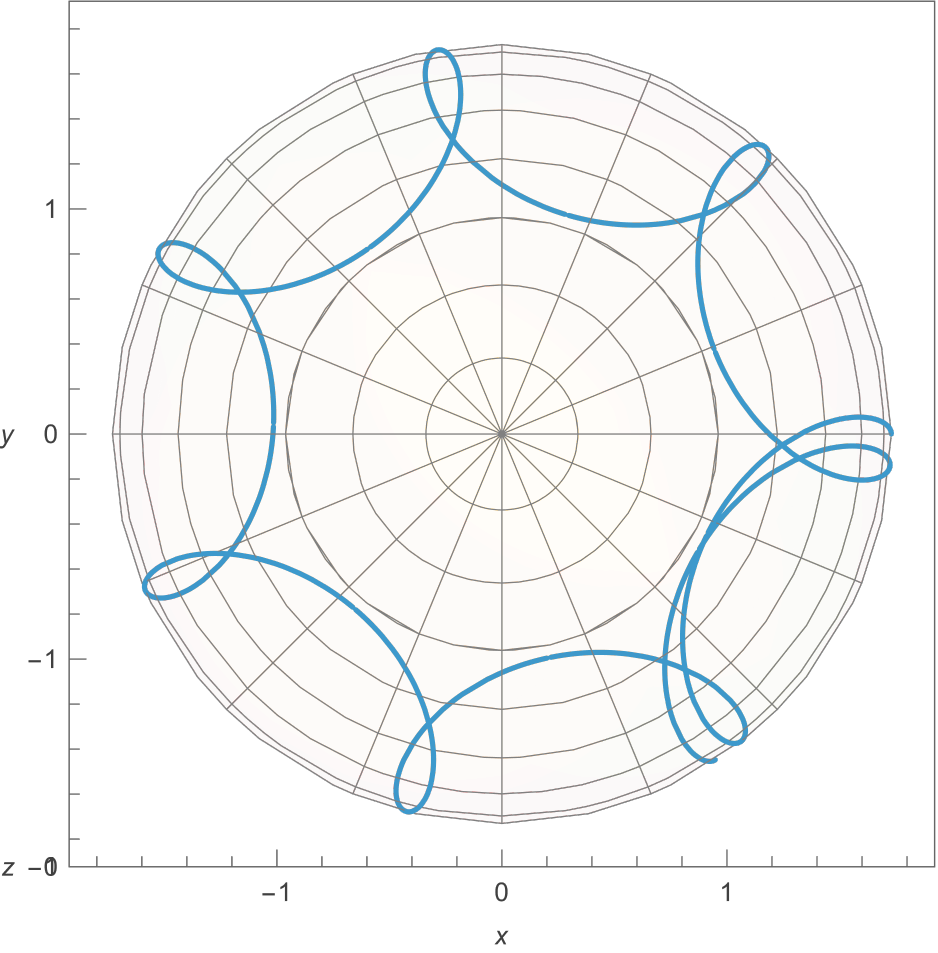}\vspace{0.2cm}
        \includegraphics[width=\linewidth]{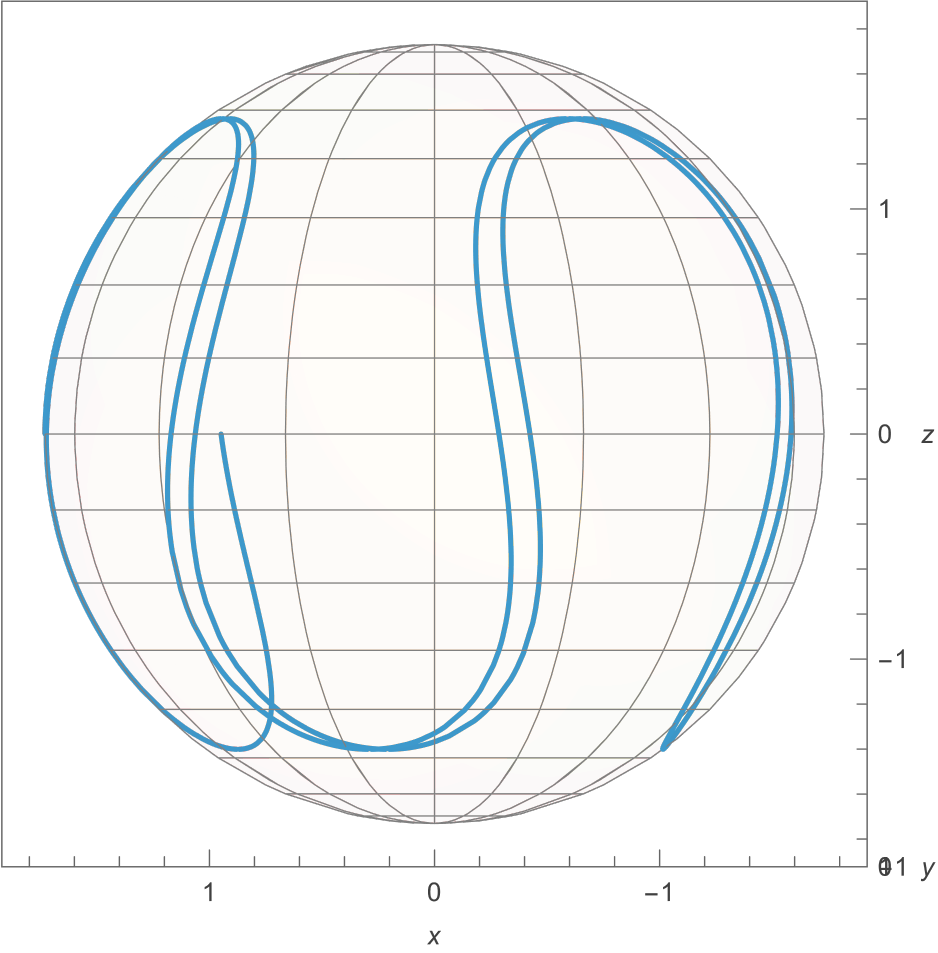}\vspace{0.2cm}
        (e)
    \end{minipage}\hspace{0.03\textwidth}
    \caption{Stable SPOs in the horizonless geometry for the parameter sets (d) and (e) in Table~\ref{tab:naked}. For each orbit, the upper, middle, and lower panels show the trajectory in the $x$--$y$--$z$ space and its projections onto the $x$--$y$ and $x$--$z$ planes, respectively.}
    \label{naked2}
\end{figure}

The corresponding stable SPOs are shown in Figs.~\ref{naked1} and \ref{naked2}. The trajectories display bounded motion in the polar direction while remaining at a constant radial coordinate. Their azimuthal advances, listed in Table~\ref{tab:naked}, are larger than those of several extremal examples. In particular, the orbit in Fig.~\ref{naked1}(a) has $\Delta\phi\simeq16.22$, indicating a substantial azimuthal advance over one complete latitudinal oscillation. As $b$ increases across the selected examples, the values of $\Delta\phi$ decrease, while the polar turning point $u_0$ also changes.

\subsection{Black-Hole Shadow}
\label{sec:shadow}

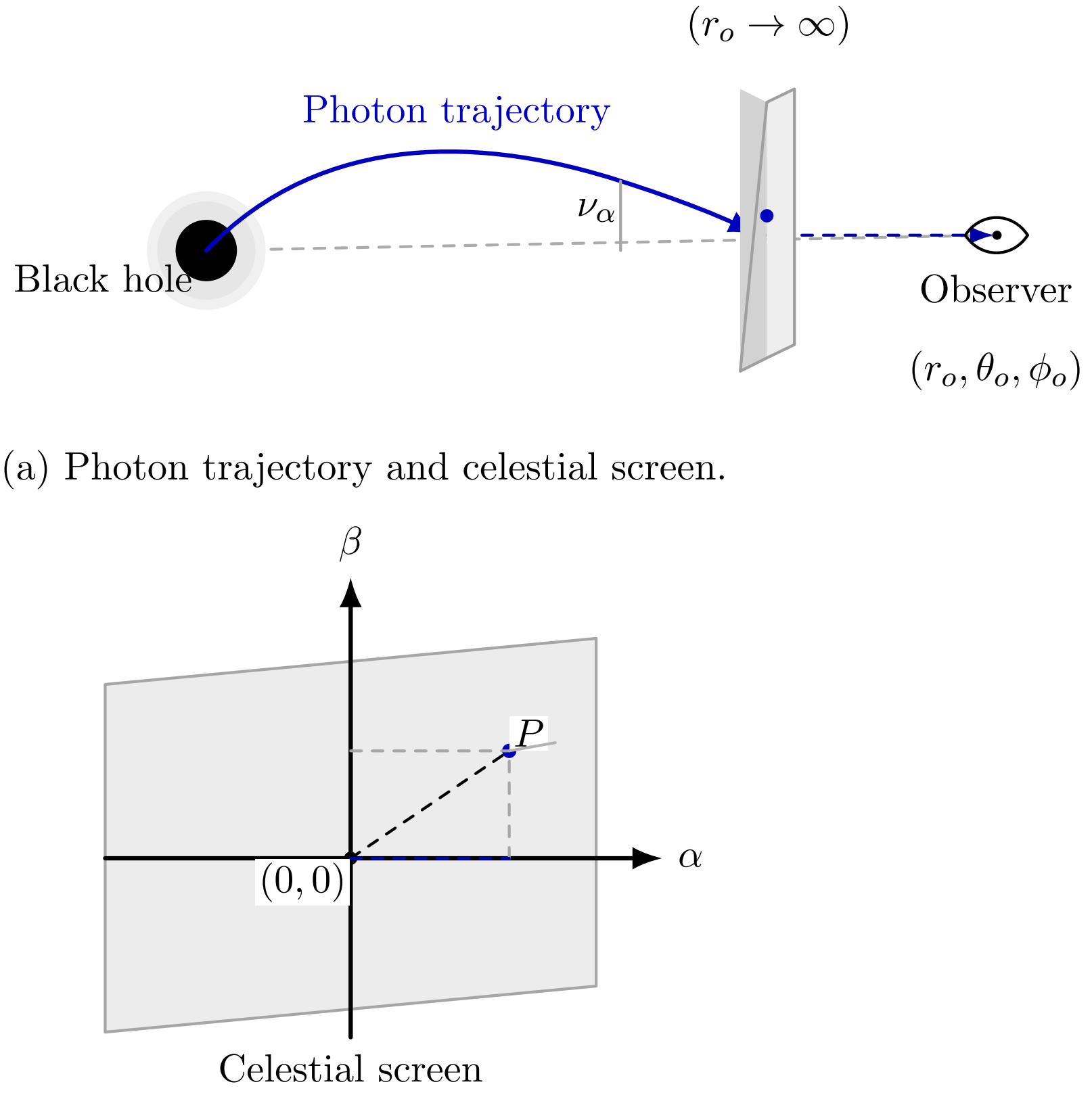
\begin{figure}[h]
\centering

\begin{subfigure}[t]{0.5\textwidth}
\centering

\begin{tikzpicture}[
    >=Latex,
    line cap=round,
    line join=round,
    font=\small,
    scale=1
]




\coordinate (BH)  at (0,0);
\coordinate (Scr) at (5.4,0.15);
\coordinate (Obs) at (7.7,0.15);

\draw[
    dashed,
    gray!65,
    thick
]
    (BH) -- (Obs);

\fill[gray!12] (BH) circle (0.58);
\fill[gray!20] (BH) circle (0.48);
\fill[black]   (BH) circle (0.3);

\node[
    below left
] at (BH)
    {Black hole};

\draw[
    blue!75!black,
    very thick,
    ->,
    >=Latex
]
    (BH)
    .. controls (1.15,1.20)
              and (2.75,1.10)
              .. (4.05,0.68)
    .. controls (4.65,0.49)
              and (5.05,0.32)
              .. (Scr);

\draw[
    blue!65!black,
    dashed,
    thick,
    ->,
    >=Latex
]
    (Scr) -- (Obs);

\node[
    blue!75!black
] at (2.45,1.35)
    {Photon trajectory};

\draw[
    gray!70,
    thick
]
    (4.05,0) -- (4.05,0.68);

\node at (3.82,0.39)
    {$\nu_\alpha$};

\fill[gray!35]
    (5.22,-1.18)
    -- (5.48,-1.05)
    -- (5.48,1.45)
    -- (5.22,1.58)
    -- cycle;

\fill[gray!13]
    (5.48,-1.05)
    -- (5.75,-0.92)
    -- (5.75,1.58)
    -- (5.48,1.45)
    -- cycle;

\draw[
    gray!75,
    thick
]
    (5.22,-1.18)
    -- (5.48,-1.05)
    -- (5.75,-0.92)
    -- (5.75,1.58)
    -- (5.48,1.45)
    -- cycle;

\node[
    align=center
] at (5.50,2.2)
    {
     $(r_o\rightarrow\infty)$};

\coordinate (P) at (5.48,0.34);

\fill[
    blue!75!black
]
    (P) circle (0.065);


\draw[
    thick
]
    (7.42,0.15)
    .. controls (7.57,0.38)
              and (7.88,0.38)
              .. (8.03,0.15)
    .. controls (7.88,-0.08)
              and (7.57,-0.08)
              .. (7.42,0.15);

\fill (7.73,0.15) circle (0.045);

\node[
    align=center
] at (7.72,-0.8)
    {Observer\\
     $(r_o,\theta_o,\phi_o)$};

\end{tikzpicture}

\caption{Photon trajectory and celestial screen.}
\label{fig:photon_celestial_screen}

\end{subfigure}
\hfill
\begin{subfigure}[t]{0.5\textwidth}
\centering

\begin{tikzpicture}[
    >=Latex,
    line cap=round,
    line join=round,
    font=\small,
    scale=1
]



\fill[gray!15]
    (-2.4,-1.7)
    -- (2.4,-1.25)
    -- (2.4,2.15)
    -- (-2.4,1.70)
    -- cycle;

\draw[
    gray!70,
    thick
]
    (-2.4,-1.7)
    -- (2.4,-1.25)
    -- (2.4,2.15)
    -- (-2.4,1.70)
    -- cycle;

\node[
    align=center
] at (0,-2.05)
    {Celestial screen};

\draw[
    ->,
    very thick,
    >=Latex
]
    (-2.4,0)
    -- (3.05,0)
    node[right] {$\alpha$};

\draw[
    ->,
    very thick,
    >=Latex
]
    (0,-1.75)
    -- (0,2.75)
    node[above] {$\beta$};

\fill[black] (0,0) circle (0.065);

\node[
    below left,
    fill=white,
    inner sep=1pt
]
at (0,0)
    {$(0,0)$};

\coordinate (P2) at (1.55,1.05);

\fill[blue!75!black]
    (P2) circle (0.07);

\node[
    above right,
    fill=white,
    inner sep=1pt
]
at (P2)
    {$P$};

\draw[
    dashed,
    gray!70,
    thick
]
    (0,1.05) -- (P2);

\draw[
    dashed,
    gray!70,
    thick
]
    (1.55,0) -- (P2);

\draw[
    dashed,
    thick
]
    (0,0) -- (P2);

\draw[
    blue!70!black,
    dashed,
    thick
]
    (0,0) -- (1.55,0);

\draw[
    gray!60,
    thick
]
    (P2) -- ++(0.45,0.08);

\end{tikzpicture}

\caption{Celestial screen and coordinates.}
\label{fig:celestial_coordinates}

\end{subfigure}

\caption{
Schematic illustration of a shadow formation. Panel (a) shows the physical path of the photon intersecting the celestial screen at an asymptotic angle $\nu_\alpha$. Panel (b) illustrates the celestial screen and the corresponding coordinates $(\alpha,\beta)$.}
\label{fig:celestial_screen}

\end{figure}

The unstable spherical photon orbits determine the boundary
of the black-hole shadow as observed from infinity
\cite{Teo2003,Tsukamoto:2017fxq}. For an observer located at
$r_o$ with inclination angle $\theta_o$, the celestial coordinates
$(\alpha,\beta)$ are defined on the observer's screen, as illustrated
in Fig.~\ref{fig:celestial_screen}~\cite{Tsukamoto:2017fxq,Bardeen:1972fi}. They are given by~\cite{Vazquez:2003zm,Shaikh:2019fpu}
\begin{align}
\alpha
&=
\lim_{r_o\to\infty}
\left(
-r_o^2\sin\theta_o
\frac{d\phi}{dr}
\right),
\label{eq:alpha_def}
\\
\beta
&=
\lim_{r_o\to\infty}
\left(
r_o^2
\frac{d\theta}{dr}
\right).
\label{eq:beta_def}
\end{align}

The shadow boundary is obtained by evaluating these expressions along
the family of physically admissible unstable spherical photon orbits.
Using the parametrization $\Phi(r_{\rm SO})$ and
$C(r_{\rm SO})$ obtained in Eq.~\eqref{35}, the shadow can therefore be
represented parametrically as
\begin{align}
\alpha(r_{\rm SO})=
-\frac{\Phi(r_{\rm SO})}{\sin\theta_o}, \qquad \beta_\pm(r_{\rm SO})
=
\pm
\sqrt{
C(r_{\rm SO})
+a^2\cos^2\theta_o
-\Phi^2(r_{\rm SO})\cot^2\theta_o
}.
\label{eq:beta_parametric_shadow}
\end{align}
Only those values of $r_{\rm SO}$ satisfying the reality condition
\begin{equation}
C(r_{\rm SO})
+a^2\cos^2\theta_o
-\Phi^2(r_{\rm SO})\cot^2\theta_o
\geq 0
\label{eq:shadow_reality}
\end{equation}
and the instability condition~\eqref{unstable} contribute to the shadow boundary.

The endpoints of the relevant spherical-orbit family are associated
with equatorial circular photon orbits. Since
$\Theta(\pi/2)=C$, the corresponding photon-orbit radii are determined by $C(r_{\rm ph}^{\pm})=0$,
where the two branches correspond to the prograde and retrograde
equatorial photon orbits. Varying $r_{\rm SO}$ between the corresponding
admissible endpoints and substituting into
Eqs.~\eqref{eq:beta_parametric_shadow} yields the shadow contour.

For an edge-on observer, $\theta_o=\pi/2$, these expressions simplify to 
\[\alpha=-\Phi,
\quad
\beta_\pm=\pm\sqrt{C}.
\]
In this case, the equatorial photon orbits with $C=0$ determine the
intersections of the shadow boundary with the horizontal
$\alpha$-axis.

\begin{figure}
    \centering
    \includegraphics[width=0.5\linewidth]{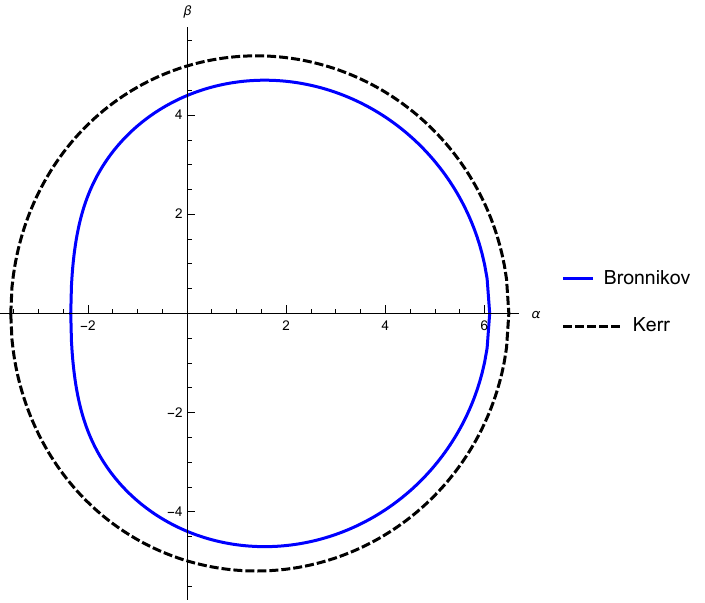}
    \caption{
    Comparison between the shadow of the Newman--Janis-generated
    rotating Bronnikov geometry and that of a Kerr black hole
    with the same ADM mass and spin. Here, $b=0.3$ and $a/M=0.7$.
    }
    \label{shadowrotbronkerr}
\end{figure}

As a consistency check, we recover the Kerr limit by taking
$b\to0$ while keeping the ADM mass~\eqref{qsubs} fixed. This requires $q\to0$, and consequently
$m(r)\to M$. Accordingly, the parametrization of the spherical photon orbits
in terms of $\Phi(r_{\rm SO})$ and $C(r_{\rm SO})$ reduces to Kerr's. The resulting shadow contour therefore approaches the standard Kerr shadow, providing a consistency check of our construction~\cite{Bardeen:1972fi,Hioki:2009na}. The comparison between the shadow of the rotating regular geometry and that of Kerr is shown in Fig.~\ref{shadowrotbronkerr}.

\begin{figure}[htbp]
    \centering
    \includegraphics[width=1.0\linewidth]{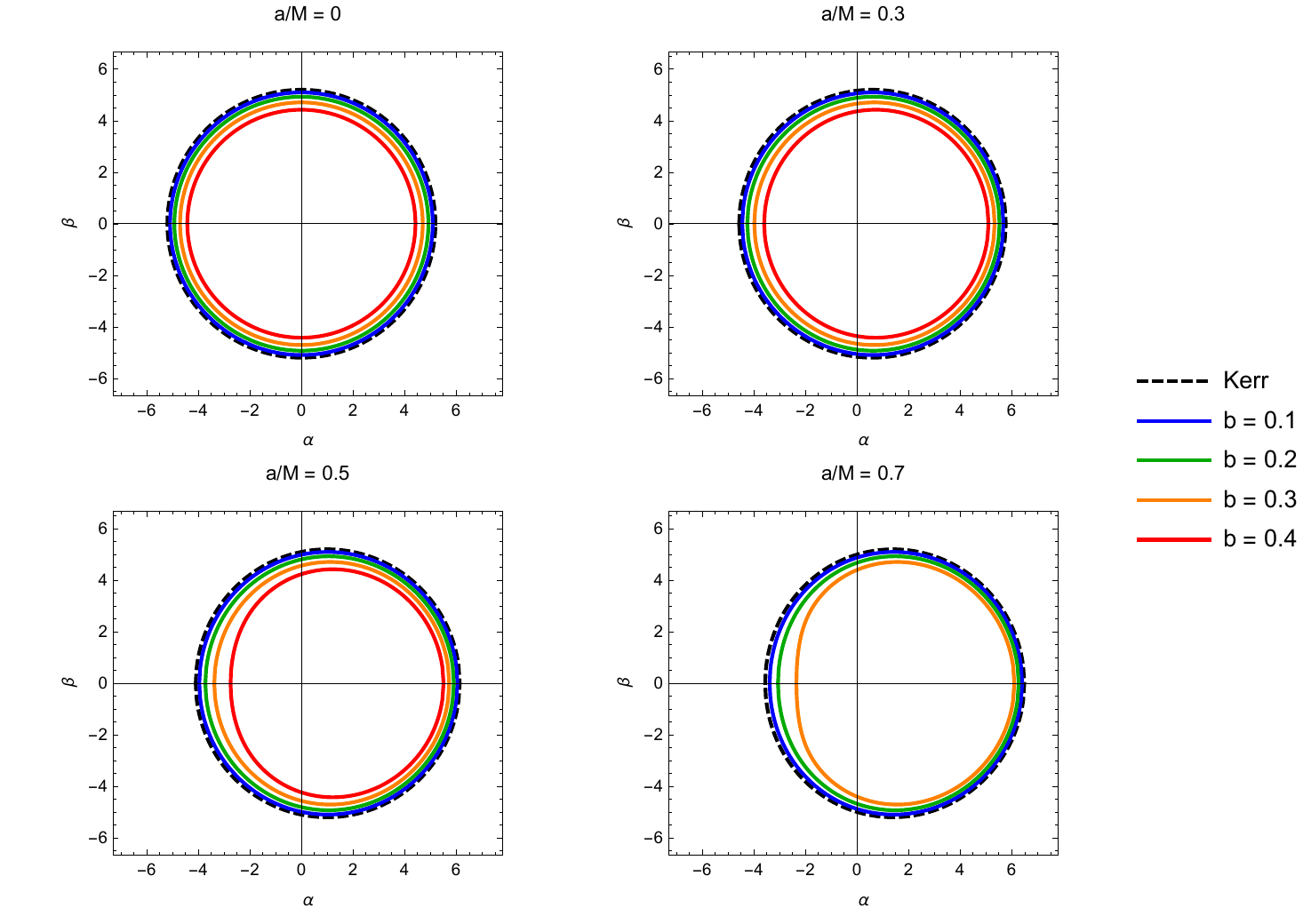}
    \caption{
    Shadow contours of the Newman--Janis-generated rotating
    Bronnikov geometry for different values of $a/M$ and $b$. The charge parameter is determined by $q=(2bM)^{2/3}$ to keep the ADM mass
    fixed. The $b=0.4$ configuration is omitted from the
    $a/M=0.7$ panel because it is horizonless. The
    configuration with $a/M=0.7$ and $b=0.3$ is close
    to extremality, with $a_{\rm ext}/M\simeq 0.71$.}
    \label{fig:shadow_parameter_sweep}
\end{figure}

 Fig.~\ref{fig:shadow_parameter_sweep}
compares the shadow contours for several values of the
spin parameter $a/M$ and nonlinear parameter $b$, while
keeping the ADM mass fixed at $M=1$. The Kerr shadow is
included in each panel as a reference. For a given spin,
increasing $b$ generally reduces the shadow size, although
the magnitude of this effect depends on the spin parameter.
A qualitatively similar dependence on the regularization
parameter and spin has been reported for rotating regular
black holes~\cite{Abdujabbarov:2016hnw}. In particular, the differences
between the Bronnikov and Kerr contours become more
pronounced for the configurations shown at lower and
intermediate spins. At $a/M=0.7$, the configuration with
$b=0.3$ lies close to the extremal limit, with
$a_{\rm ext}/M\simeq 0.71$, and exhibits the most
pronounced deformation among the configurations considered.
The $b=0.4$ configuration is not shown in this panel because
it belongs to the horizonless regime. These results illustrate
how the nonlinear mass profile modifies both the size and
shape of the shadow, with the resulting deformation depending
on the interplay between $b$ and the spin parameter.

\section{Spherical Timelike Orbits}
\label{sec:sto}

Having discussed spherical photon orbits and their role in determining
the black-hole shadow, we now turn to spherical timelike orbits.
For a massive test particle, we set $\mu=1$ in Eqs.~\eqref{timegeod}--\eqref{31}. Imposing the spherical-orbit
conditions in Eq.~\eqref{SOcond}, we obtain four solution branches
for the specific energy $E$ and the impact parameter $\Phi$.

To express these solutions in a compact form, we first introduce
the following auxiliary quantities:
\begin{equation}
    p\equiv b(r-2m(r)),
    \qquad
    s\equiv 2rm'(r),
\end{equation}
In terms of these quantities, two of the solution branches are
given by
\begin{align}
    E_{a,b}
    &=
    \sqrt{\frac{X_{\pm}}{bpr^4\mathcal{D}}},
    \nonumber\\
    \Phi_{a,b}
    &=
    \frac{E_{a,b}}{2ap\mathcal{W}}
    \left(
        \mathcal{K}
        +\frac{p^2r^2\mathcal{D}\mathcal{B}}{X_{\pm}}
    \right),
    \label{EandPhiSTO}
\end{align}
where
\begin{equation}
\begin{aligned}
    X_{\pm} &\equiv -\mathcal{E}
        \pm 4ap\mathcal{W}\mathcal{J}\mathcal{Y}\sqrt{b},
    &\qquad
    \mathcal{W} &\equiv p+b(s-r),
    \\
    \mathcal{Z} &\equiv 2p+\mathcal{W},
    &
    \mathcal{J} &\equiv a^2b+pr,
    \\
    \mathcal{D} &\equiv r\mathcal{Z}^2+8a^2b\mathcal{W},
    &
    \mathcal{U} &\equiv pr\mathcal{Z}+2a^2b\mathcal{W},
    \\
    \mathcal{B} &\equiv
        a^2b\left[(2p-\mathcal{W})\mathcal{C}
        -r^2\mathcal{W}\right]-2p^2r^3,
    &
    \mathcal{Y} &\equiv
        \sqrt{a^2b\mathcal{C}^2-pr^3\mathcal{C}
        -\frac{1}{2}r^3\mathcal{W}(r^2+\mathcal{C})},
    \\
    \mathcal{E} &\equiv
        \mathcal{U}\mathcal{B}
        +2a^2b(r^2+\mathcal{C})\mathcal{J}\mathcal{W}^2,
    &
    \mathcal{K} &\equiv
        r\mathcal{U}+2a^2\mathcal{W}(p-br).
\end{aligned}
\end{equation}

The remaining two solution branches are related to the first pair
by an overall sign reversal:
\begin{equation}
    (E_{c,d},\Phi_{c,d})
    =
    -(E_{a,b},\Phi_{a,b}).
\end{equation}
This structure is also encountered in the analyses of spherical
timelike orbits in Kerr~\cite{Teo2021Spherical} and
Kerr--Newman spacetimes~\cite{Alam2024}. Since we consider
future-directed timelike geodesics with positive specific energy,
only the positive-energy branches are physically relevant.

\subsection{Analytical Solutions}

As in the photon case, to obtain analytical expressions we define the variable $w\equiv u^2$. The roots of $V(u)=0$ are
\begin{equation}
    w_{1,2}
    =
    \frac{
        a^2(1-E^2)-C-\Phi^2
        \mp
        \sqrt{
            \left[C+\Phi^2+a^2(1-E^2)\right]^2
            -4a^2C(1-E^2)
        }
    }{
        2a^2(1-E^2)
    }.
\end{equation}
In the following, we restrict our attention to bound timelike
orbits, for which $E^2<1$. The physically admissible polar motion
is characterized by $0\leq w\leq w_1\leq 1\leq w_2$.

The polar equation can be expressed as
\begin{equation}
    \frac{dw}{d\lambda}
    =
    \pm 2Y(w),
    \qquad
    Y(w)
    =
    \sqrt{
        a^2(1-E^2)w(w_1-w)(w_2-w)
    }.
    \label{MinoTimelike}
\end{equation}
Integrating Eq.~\eqref{MinoTimelike}, we obtain
\begin{equation}
    \lambda
    =
    \frac{1}{2}\int_0^w\frac{dw}{Y(w)}
    =
    \frac{1}{a\sqrt{(1-E^2)w_2}}F(\psi,k),
    \label{lambdatimelike}
\end{equation}
where
\begin{equation}
    \psi
    =
    \arcsin\sqrt{\frac{w}{w_1}},
    \qquad
    k
    =
    \sqrt{\frac{w_1}{w_2}},
\end{equation}
and $F(\psi,k)$ denotes the incomplete elliptic integral of
the first kind. At the polar turning point $w=w_1$, the Mino
parameter becomes
\begin{equation}
    \lambda_0
    =
    \frac{K(k)}{a\sqrt{(1-E^2)w_2}},
\end{equation}
where $K(k)$ is the complete elliptic integral of the first
kind. As in the photon case, $\lambda_0$ corresponds to one
quarter of a complete latitudinal oscillation.

Inverting Eq.~\eqref{lambdatimelike}, the polar motion can be
expressed in terms of the Jacobi elliptic sine function as
\begin{equation}
    u(\lambda)
    =
    \sqrt{w_1}\,
    \operatorname{sn}
    \left(
        a\sqrt{(1-E^2)w_2}\,\lambda,k
    \right).
    \label{ueqtimelike}
\end{equation}

Having obtained the polar solution, we can proceed to the
azimuthal motion. Following the same procedure as in the photon
case, we integrate the azimuthal geodesic equation with respect
to the Mino parameter. The resulting azimuthal coordinate is
\begin{equation}
\begin{aligned}
    \phi(\lambda)
    ={}&
    \frac{a}{\Delta}
    \left(2m(r)rE-a\Phi\right)\lambda
    \\
    &+
    \frac{\Phi}{a\sqrt{(1-E^2)w_2}}
    \Pi(\psi,w_1,k),
\end{aligned}
\label{phitimelike}
\end{equation}
where $\Pi(\psi,w_1,k)$ denotes the incomplete elliptic
integral of the third kind, with characteristic $w_1$.
Consequently, the total azimuthal change over one complete
latitudinal oscillation is
\begin{equation}
    \Delta\phi
    =
    \frac{4}{\sqrt{(1-E^2)w_2}}
    \left[
        \frac{\Phi}{a}\Pi(w_1,k)
        +
        \frac{2m(r)rE-a\Phi}{\Delta}K(k)
    \right],
\end{equation}
where $\Pi(w_1,k)$ is the corresponding complete elliptic
integral of the third kind.

\subsection{Representative Orbits}

As in the null case, we employ the analytical solutions, Eqs.~\eqref{ueqtimelike} and \eqref{phitimelike}, to construct the spherical
timelike orbits (STOs) for various configurations, including extremal
and horizonless geometries. In this subsection, we focus on bound
orbits with $E^2<1$ around extremal configurations. Furthermore, we
consider marginally stable orbits, characterized by the condition
$R''(r_{\rm SO})=0$. The representative orbital parameters are listed in
Table~\ref{tab:TimelikeExtreme}.

\begin{table}[H]
    \centering
    \caption{Parameters of representative marginally stable STOs
    around extremal Newman--Janis-generated rotating regular geometries
    with $M=1$ and $\Phi=0$.}
    \label{tab:TimelikeExtreme}
    \vspace{0.2cm}
    \begin{tabular}{|c|c|c|c|c|c|c|c|}
        \hline
        \makebox[2cm]{Orbit} & \makebox[1.5cm]{$b$} & \makebox[1.5cm]{$a$} & \makebox[1.5cm]{$E$} & \makebox[2cm]{$r_{SO}$} & \makebox[2cm]{$C$} & \makebox[2cm]{$u_0$} & \makebox[2cm]{$\Delta\phi$} \\
        \hline
        (a) & 0   & 1      & 0.937 & 5.28  & 11.24 & 1 & -5.268 \\
        (b) & 0.3   & 0.71      & 0.93 & 4.67  & 9.68 & 1 & -5.408 \\
        (c) & 0.5   & 0.285      & 0.917 & 3.92  & 7.91 & 1 & 6.748  \\
        \hline
    \end{tabular}
\end{table}

\begin{figure}[H]
    \centering
    \begin{minipage}{0.32\textwidth}
        \centering
        \includegraphics[width=\linewidth]{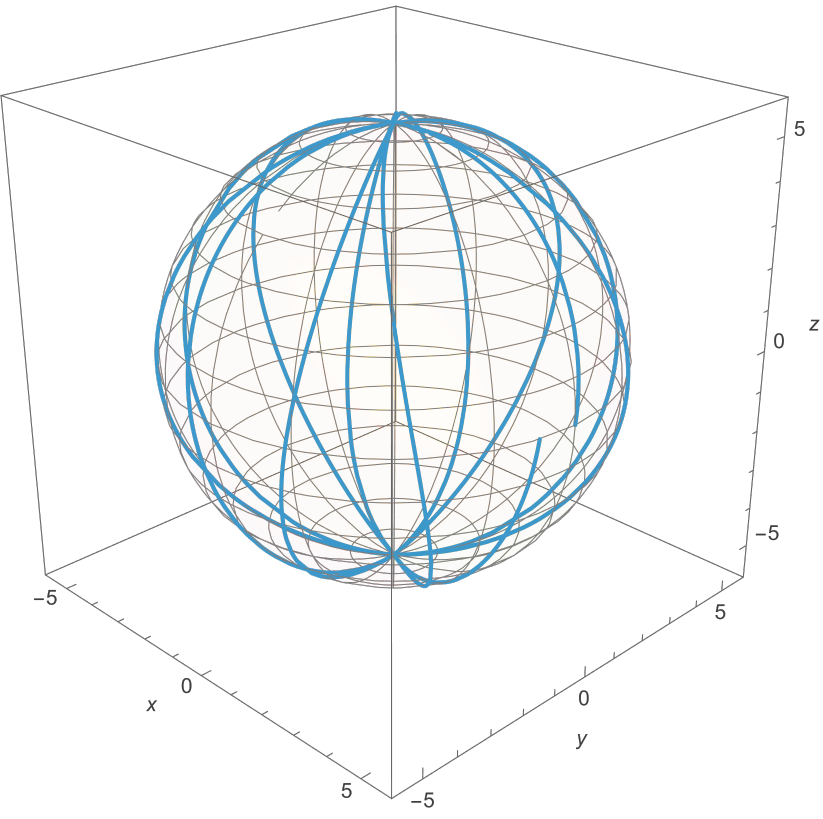}\vspace{0.2cm}
        \includegraphics[width=\linewidth]{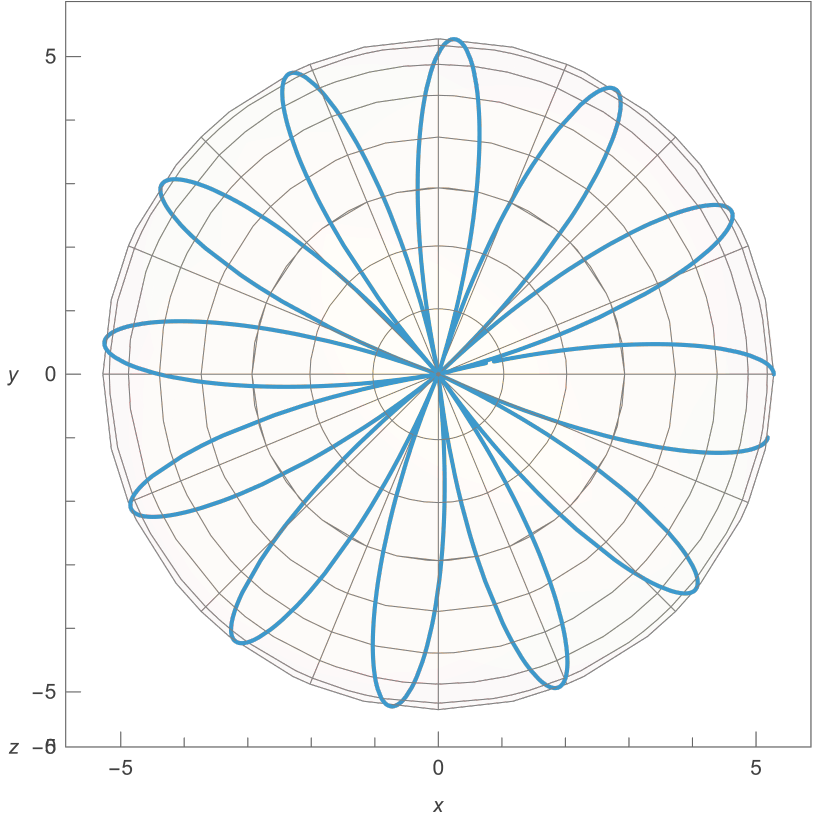}\vspace{0.2cm}
        \includegraphics[width=\linewidth]{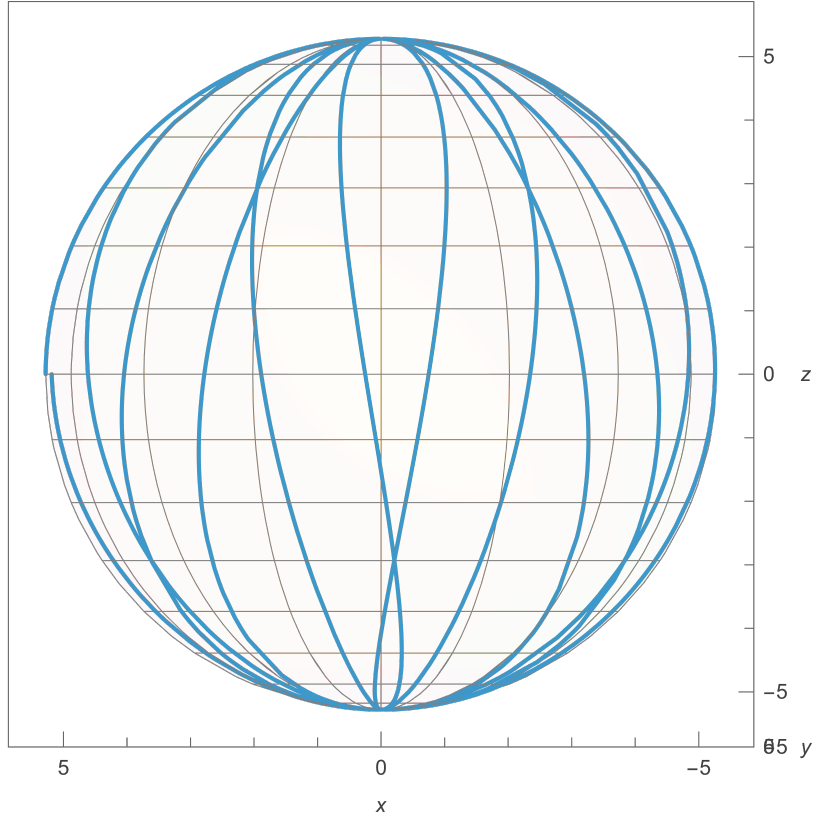}\vspace{0.2cm}
        (a)
    \end{minipage}\hfill
    \begin{minipage}{0.32\textwidth}
        \centering
        \includegraphics[width=\linewidth]{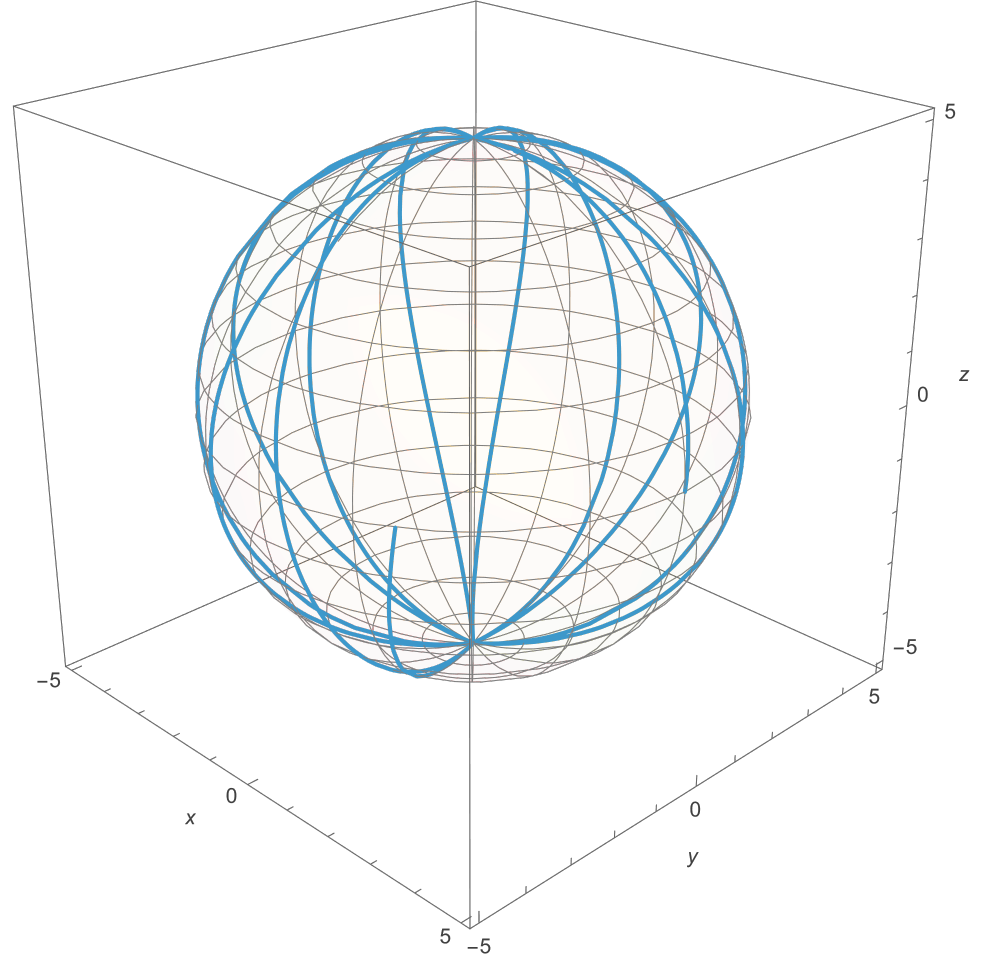}\vspace{0.2cm}
        \includegraphics[width=\linewidth]{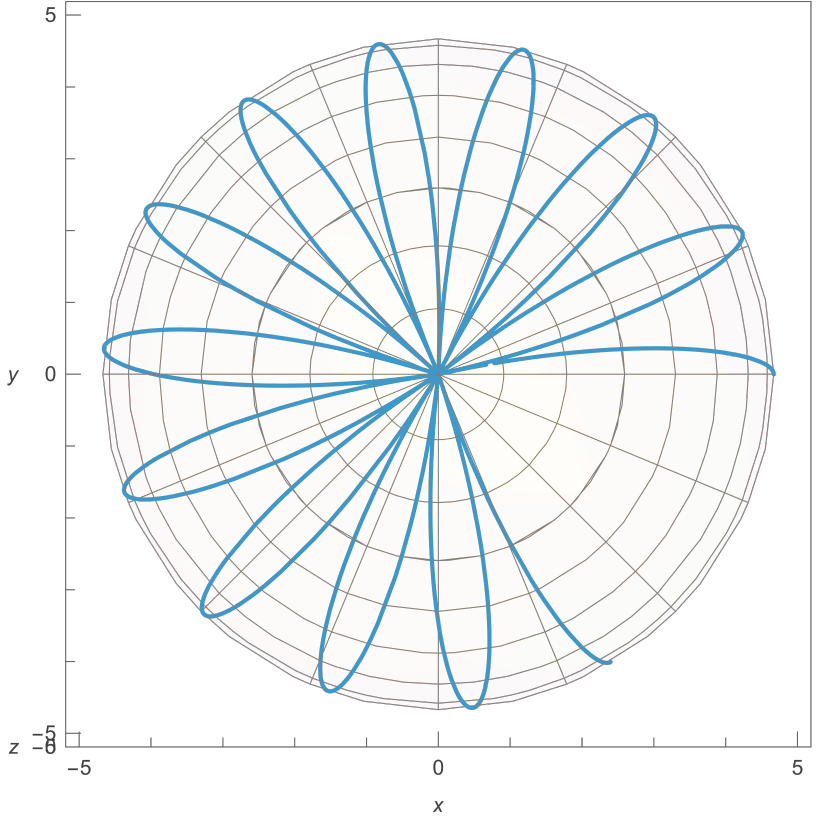}\vspace{0.2cm}
        \includegraphics[width=\linewidth]{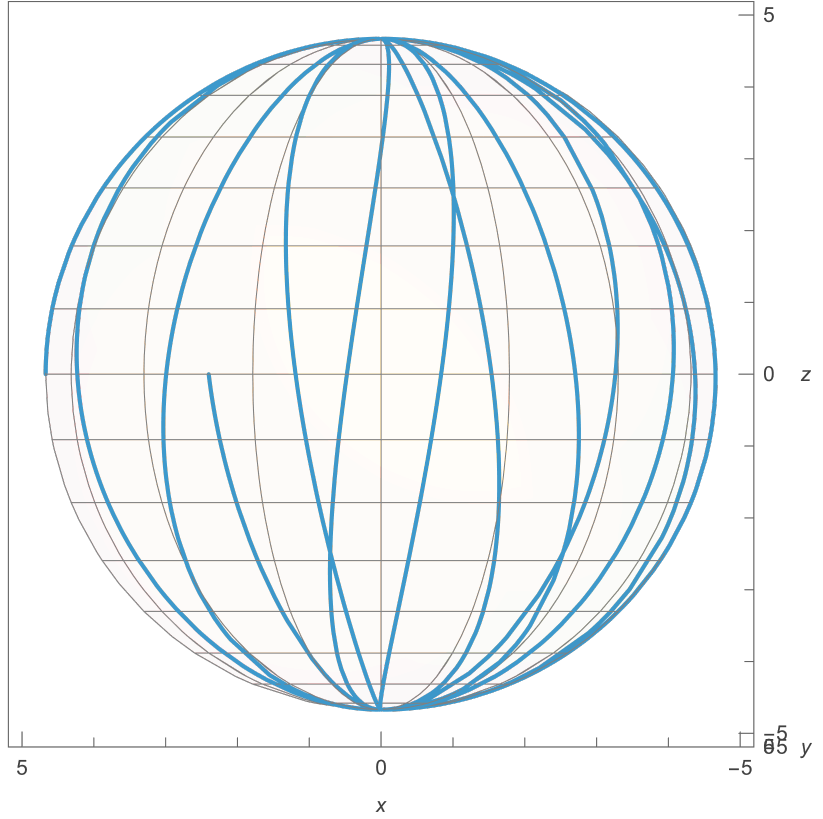}\vspace{0.2cm}
        (b)
    \end{minipage}\hfill
    \begin{minipage}{0.32\textwidth}
        \centering
        \includegraphics[width=\linewidth]{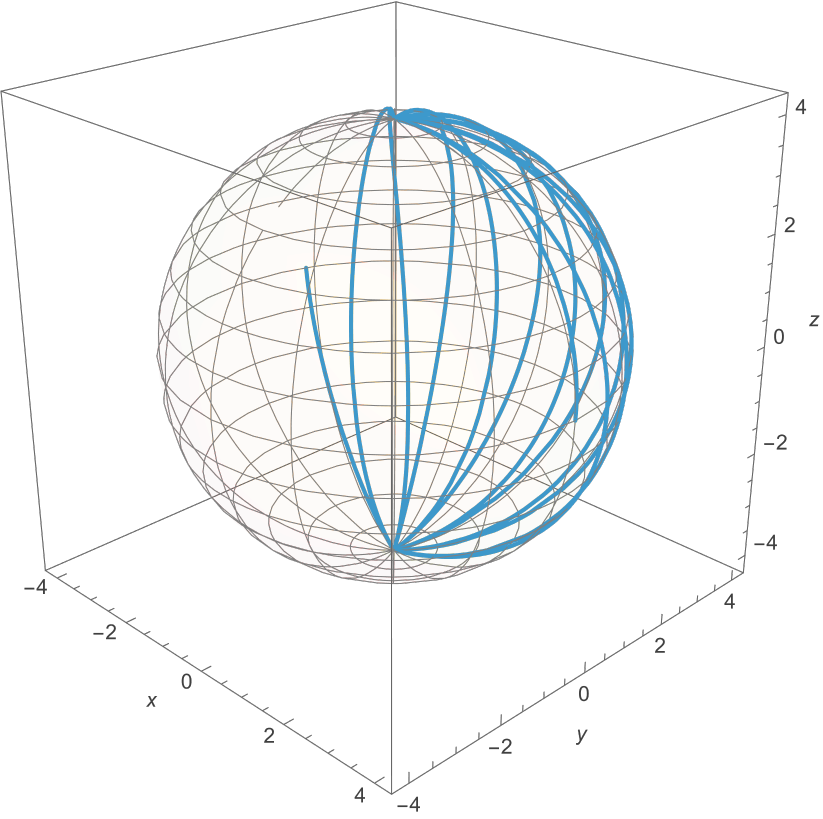}\vspace{0.2cm}
        \includegraphics[width=\linewidth]{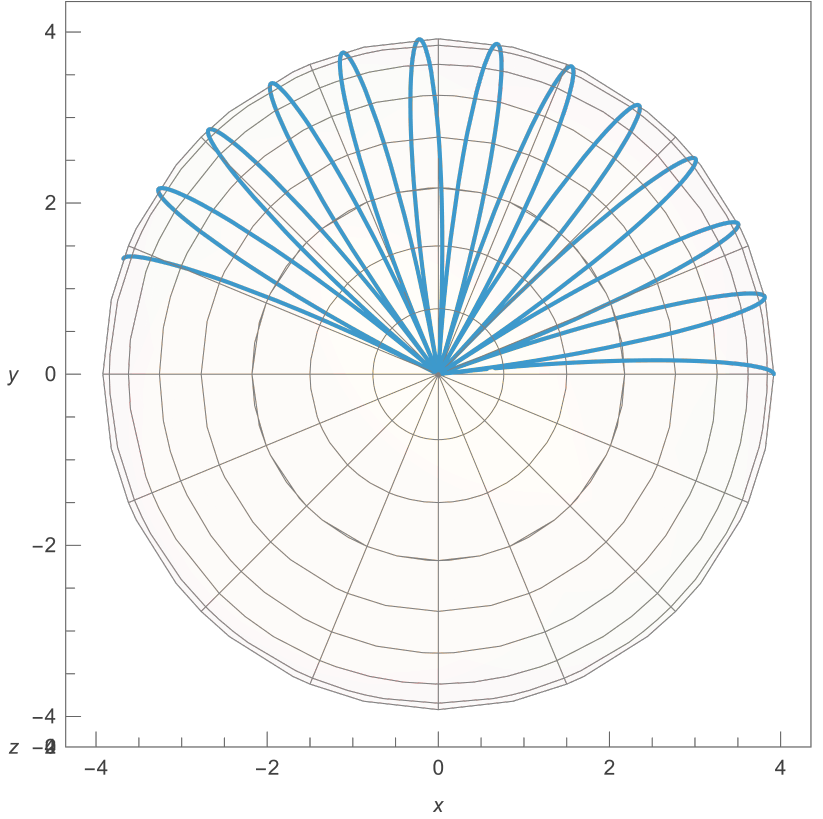}\vspace{0.2cm}
        \includegraphics[width=\linewidth]{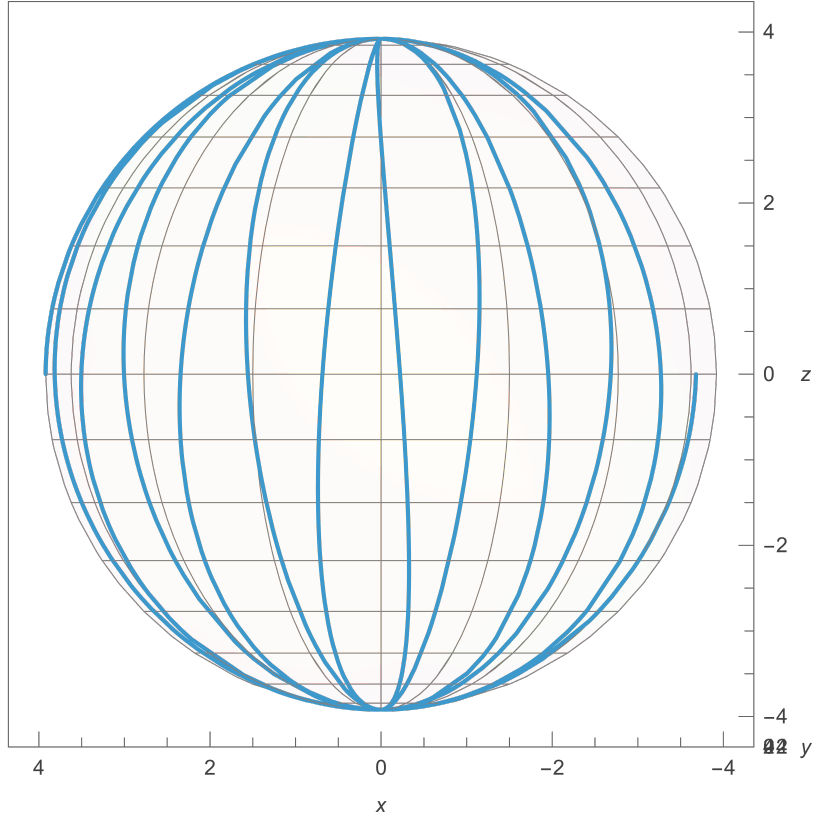}\vspace{0.2cm}
        (c)
    \end{minipage}
    \caption{Four complete latitudinal oscillations of the extremal
    STOs, shown in three-dimensional space (top), together with their
    projections onto the $x$-$y$ (middle) and $x$-$z$ (bottom) planes.
    The parameter sets corresponding to panels (a)--(c) are listed in
    Table~\ref{tab:TimelikeExtreme}.}
    \label{Timelike}
\end{figure}

Fig.~\ref{Timelike} illustrates the corresponding STOs in three
dimensions, together with their projections onto the $x$-$y$ and
$x$-$z$ planes. Each trajectory is displayed over six complete
latitudinal oscillations. The three configurations are distinguished
by their values of the regularization parameter $b$, as specified in
Table~\ref{tab:TimelikeExtreme}.

\subsection{Innermost Stable Spherical Orbits}

The function $R(r)$ in Eq.~\eqref{eq:radial_null_shadow}
determines the radial motion of a test particle and can be interpreted
in terms of an effective potential, $V_{\rm eff}(r)\equiv -R(r)$.
Its radial stability is determined by the sign of $R''(r)$:
the orbit is stable for $R''(r)<0$ and unstable for $R''(r)>0$.
The marginally stable orbit marks the boundary between these two
regimes. On the equatorial plane, where $C=0$, this orbit reduces to
the familiar innermost stable circular orbit (ISCO). For non-equatorial
orbits with $C\neq 0$, it is referred to as the innermost stable
spherical orbit (ISSO)~\cite{Druart:2023}. It is therefore determined by the three conditions
\begin{equation}
  R(r_{\rm ISSO}) = R'(r_{\rm ISSO}) = R''(r_{\rm ISSO}) = 0 ,
  \label{eq:isso}
\end{equation}
where primes denote derivatives with respect to $r$, evaluated at
fixed constants of motion $(E,\Phi,C)$.

\begin{figure}[H]
    {\centering
    \includegraphics[width=0.8\linewidth]{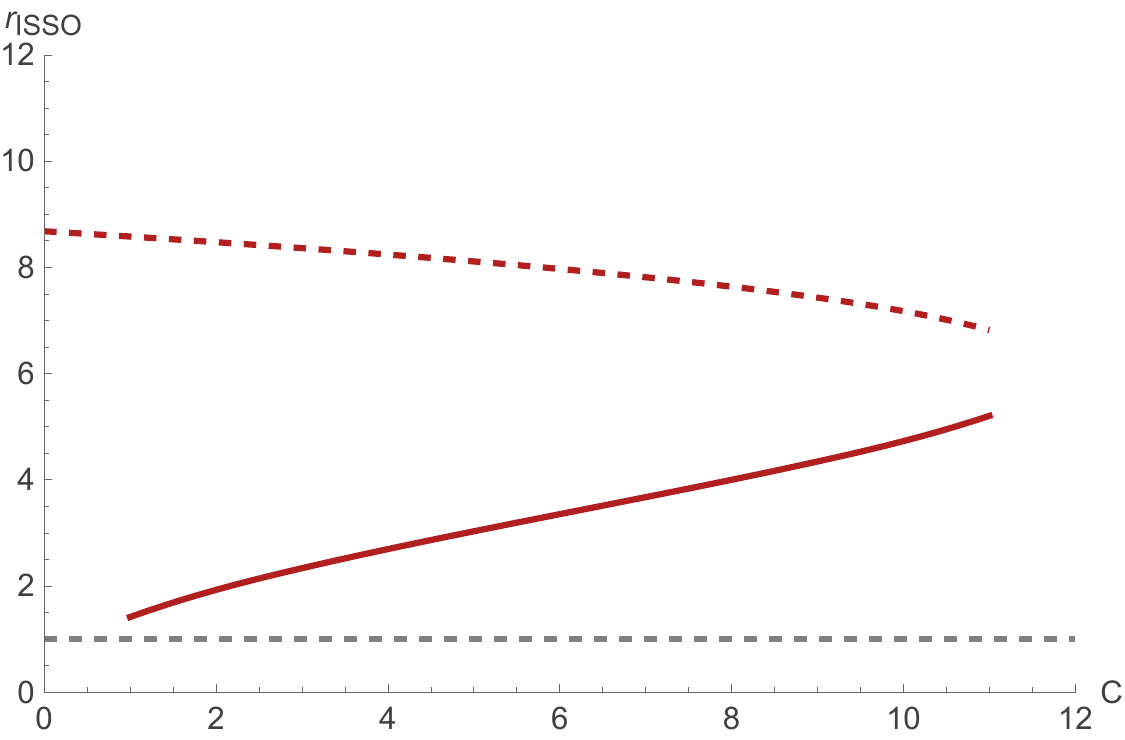}\par}
    \caption{ISSO radius $r_{\rm ISSO}$ as a function of the Carter
    constant $C$ for an extremal configuration with $b=0.1$ and
    $a=0.94$. The gray dashed line indicates the outer horizon,
    while the red dashed and solid curves represent the positive
    and negative roots of Eq.~\eqref{eq:isso_reduced}, respectively.}
    \label{ISSOCplot}
\end{figure}

The first two conditions in Eq.~\eqref{eq:isso} define the STOs and
are solved by the expressions for $E_{a,b}$ and $\Phi_{a,b}$ given
in Eq.~\eqref{EandPhiSTO}. Substituting these expressions into the
third condition reduces the ISSO determination to a single equation
for $r$ at a given value of $C$:
\begin{equation}
  R''(r)\Big|_{E=E_{a,b}(r,C),\;\Phi=\Phi_{a,b}(r,C)} = 0 .
  \label{eq:isso_reduced}
\end{equation}
Fig.~\ref{ISSOCplot} shows the resulting ISSO radii as functions
of the Carter constant $C$ for an extremal configuration with
$b=0.1$ and $a=0.94$. The two curves correspond to the positive
and negative roots of Eq.~\eqref{eq:isso_reduced}. The outer horizon is also shown for reference.

\begin{figure}[H]
    \centering
    \begin{minipage}{0.32\textwidth}
        \centering
        \includegraphics[width=\linewidth]{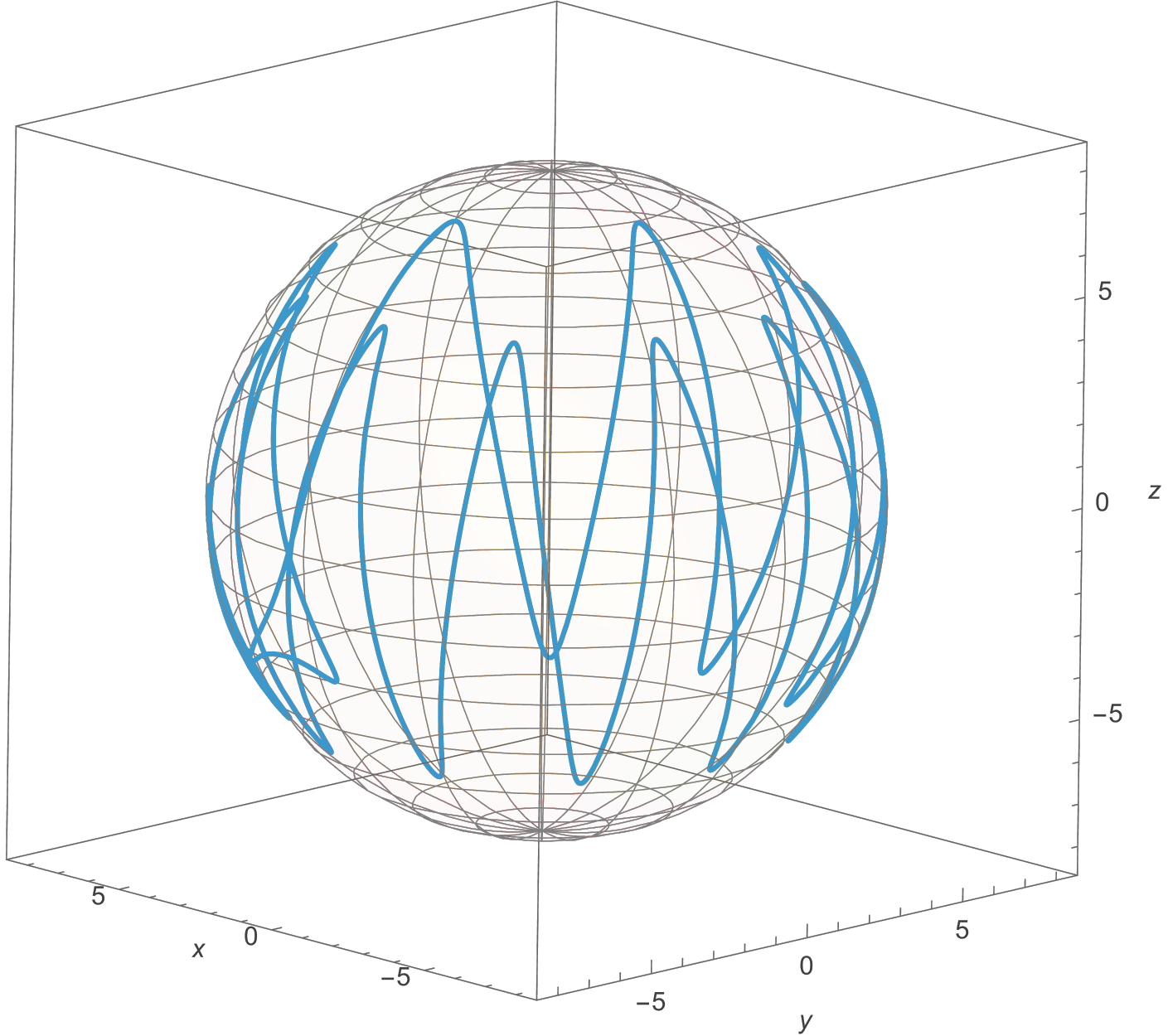}\vspace{0.2cm}
        (a)
    \end{minipage}\hfill
    \begin{minipage}{0.32\textwidth}
        \centering
        \includegraphics[width=\linewidth]{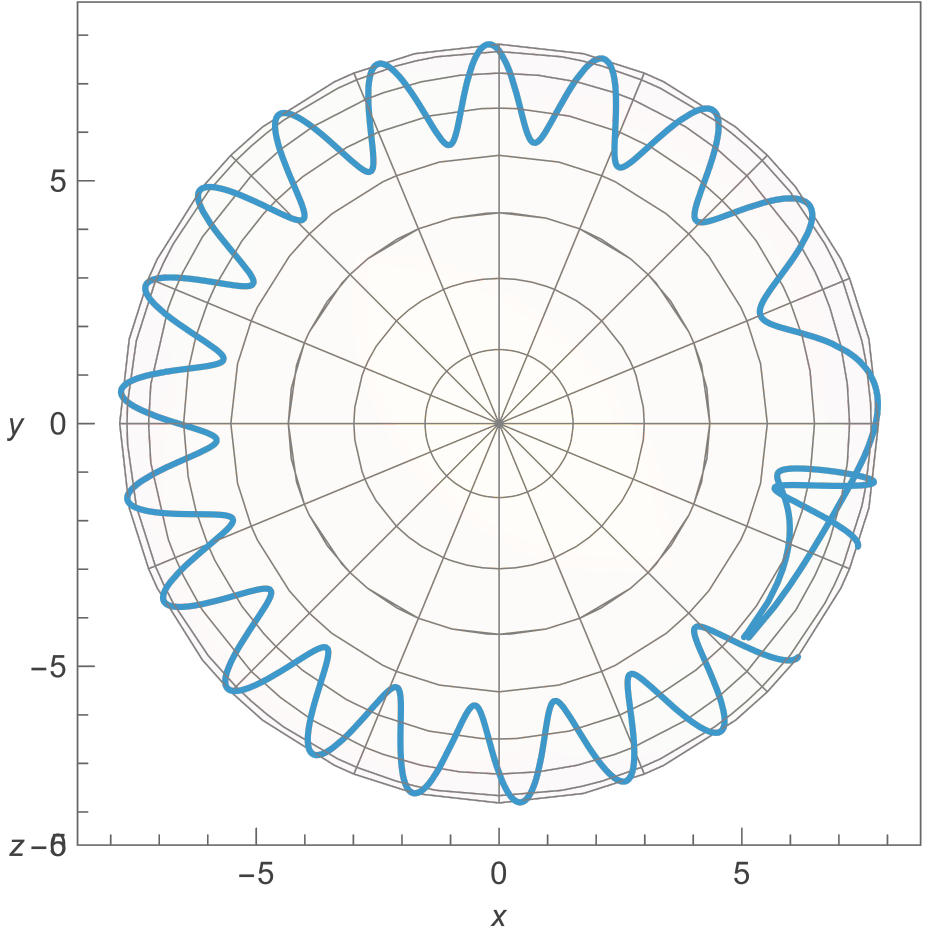}\vspace{0.2cm}
        (b)
    \end{minipage}\hfill
    \begin{minipage}{0.32\textwidth}
        \centering
        \includegraphics[width=\linewidth]{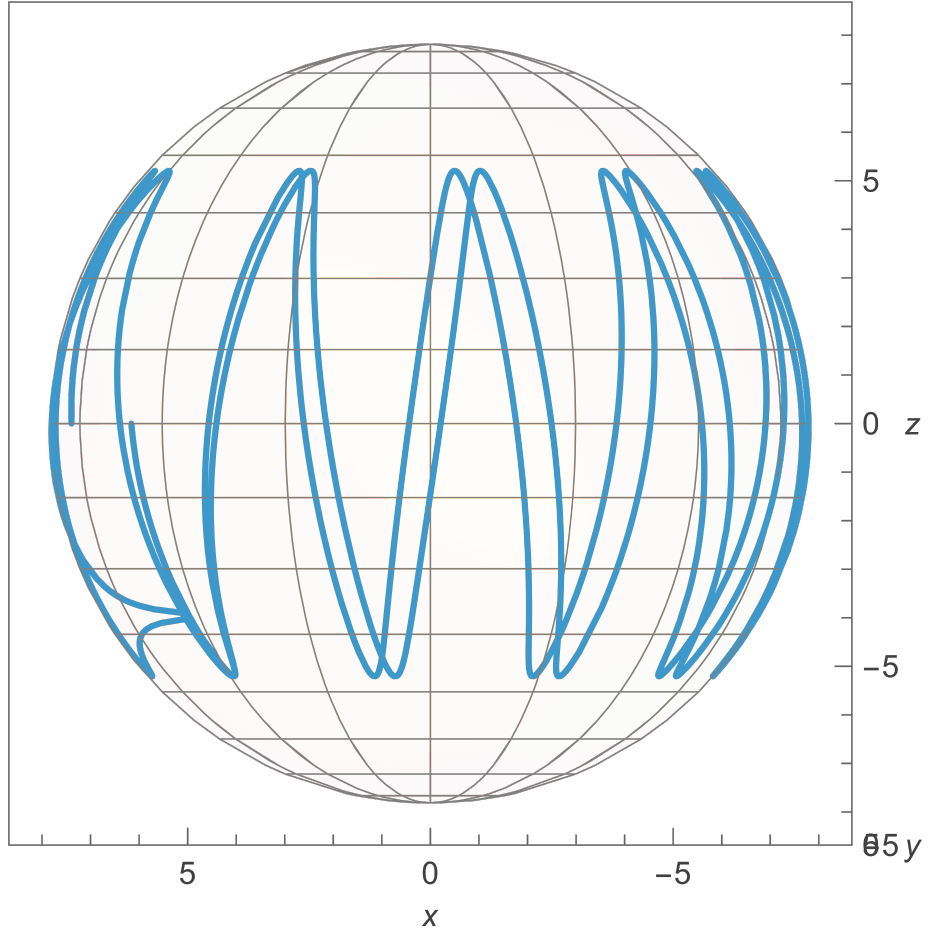}\vspace{0.2cm}
        (c)
    \end{minipage}
    \caption{Ten complete latitudinal oscillations of the ISSO for
    an extremal configuration with $b=0.1$ and $a=0.94$, shown in
    three-dimensional space (a) and projected onto the $x$-$y$ (b)
    and $x$-$z$ (c) planes. The corresponding orbital parameters
    are $r_{\rm ISSO}=7.814$ and $C=7$.}
    \label{ISSOSTOs}
\end{figure}

The corresponding marginally stable spherical orbit is illustrated
in Fig.~\ref{ISSOSTOs}, showing its three-dimensional trajectory and
its projections onto the $x$-$y$ and $x$-$z$ planes. The trajectory
is displayed over ten complete latitudinal oscillations.

\section{Energy Conditions}
\label{sec:energy}

The construction of regular black holes is often accompanied by
nontrivial violations of the standard energy conditions. The strong energy condition (SEC), for instance, is necessarily violated in static regions beneath the event horizon~\cite{Zaslavskii:2010qz}. For rotating case, the particular rotating Bardeen and Hayward geometries have been shown to violate the weak energy condition (WEC) for nonvanishing spin~\cite{Bambi2013}. Such
violations exist generically within a broad class of rotating regular
geometries satisfying physically reasonable conditions
\cite{Neves:2014aba}.

An important qualification was emphasized by
Maeda~\cite{Maeda:2021jdc}, who proposed a set of physical criteria
for assessing regular black-hole geometries (see Sec.~\ref{sec:metric}). In particular, one of these criteria, C3, requires the effective energy-momentum tensor to satisfy all the standard energy conditions in the
asymptotically flat regions. This requirement distinguishes
violations occurring in the strong-field region from those
persisting into the asymptotic regime. Another criterion, C4, requires the energy conditions to hold on the event horizon of a sufficiently large black hole~\cite{Maeda:2021jdc}, so that any violation is hidden behind it. It therefore provides
a useful criterion for examining the physical properties of
regular black-hole geometries. Here, we employ the effective energy-momentum tensor associated with the
considered geometry, defined through the Einstein tensor, to
investigate this criterion. Our analysis also allows us to establish whether both criteria are satisfied by the present geometry.

Throughout this section, $\rho$ and $P_i$ denote the effective
energy density and principal pressures, respectively, in an
orthonormal frame
\begin{equation}
T^{\mathrm{eff}}_{\hat{\mu}\hat{\nu}}=\operatorname{diag}(\rho,P_1,P_2,P_3).
\end{equation}
For the geometries considered here, $P_2=P_3$. The standard
energy conditions can then be expressed as
\begin{align}
    &\text{NEC:}
    &&\rho+P_1\geq 0,\qquad
      \rho+P_2\geq 0,
    \label{eq:NEC}\\
    &\text{WEC:}
    &&\rho\geq 0,\qquad
      \rho+P_1\geq 0,\qquad
      \rho+P_2\geq 0,
    \label{eq:WEC}\\
    &\text{DEC:}
    &&\rho\geq 0,\qquad
      \rho\geq |P_1|,\qquad
      \rho\geq |P_2|,
    \label{eq:DEC}\\
    &\text{SEC:}
    &&\rho+P_1\geq 0,\qquad
      \rho+P_2\geq 0,\qquad
      \rho+P_1+2P_2\geq 0.
    \label{eq:SEC}
\end{align}

\subsection{Static Case}

As a warm-up, we first consider the static Bronnikov. It is worth mentioning that the original Bronnikov solution violates the energy conditions. 

The effective energy density and principal pressures of the
static geometry are given by
\begin{align}
    \rho &=
    \frac{q^2}{8\pi r^4}\operatorname{sech}^2 \left(\frac{b \sqrt{q}}{r} \right),
    \label{eq:rho_static}\\
    P_1 &=-\rho,
    \label{eq:P1_static}\\
    P_2=P_3 &=
    \frac{q^2}{8\pi r^4}\operatorname{sech}^2 \left(\frac{b \sqrt{q}}{r} \right)
    \left(1-\left(\frac{b \sqrt{q}}{r} \right)\tanh \left(\frac{b \sqrt{q}}{r} \right)\right).
    \label{eq:P2_static}
\end{align}
The radial NEC is identically saturated,
$\rho+P_1=0$, whereas the transverse NEC and SEC combinations
are
\begin{align}
    \rho+P_2 &=
    \frac{q^2}{8\pi r^4}\operatorname{sech}^2 \left(\frac{b \sqrt{q}}{r} \right)
    \left(2-\left(\frac{b \sqrt{q}}{r} \right)\tanh \left(\frac{b \sqrt{q}}{r} \right)\right),
    \label{eq:NEC_static}\\
    \rho+P_1+2P_2 &=
    \frac{q^2}{4\pi r^4}\operatorname{sech}^2 \left(\frac{b \sqrt{q}}{r} \right)
    \left(1-\left(\frac{b \sqrt{q}}{r} \right)\tanh \left(\frac{b \sqrt{q}}{r} \right)\right).
    \label{eq:SEC_static}
\end{align}

\begin{figure}[H]
    \centering
    \begin{minipage}{0.48\textwidth}
        \centering
        \includegraphics[width=\linewidth]{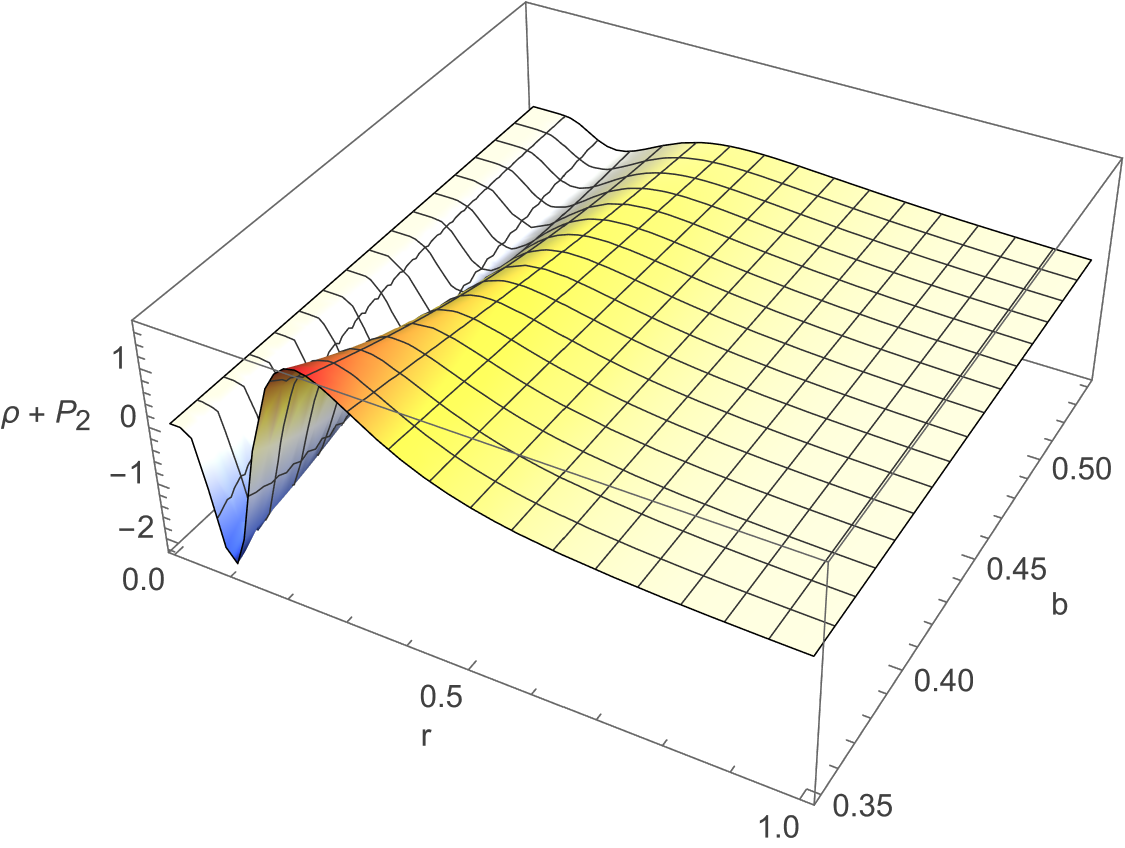}\vspace{0.2cm}
    \end{minipage}\hfill
    \begin{minipage}{0.48\textwidth}
        \centering
        \includegraphics[width=\linewidth]{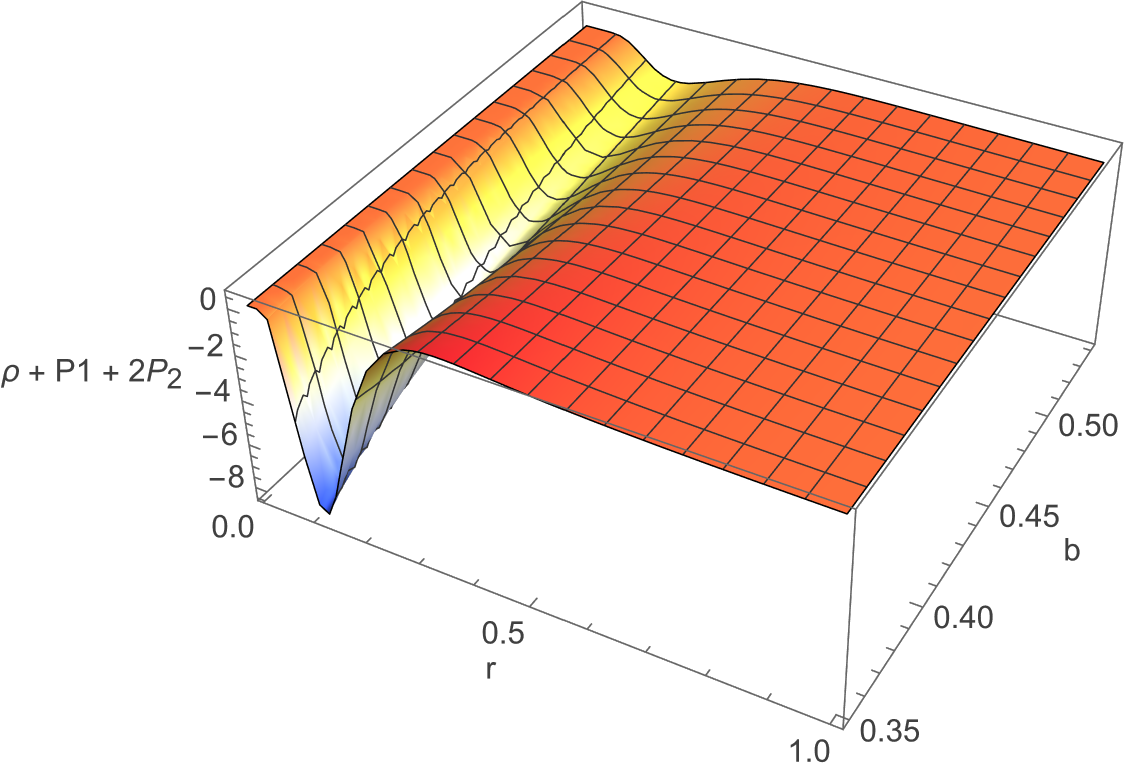}\vspace{0.2cm}
    \end{minipage}\hfill
    \caption{(Left) Plot of $\rho + P_2$ (NEC) vs $r$ with several $b$ values for static Bronnikov black hole with $M=1$. (Right) Plot of $\rho + P_1+ 2P_2$ (SEC) vs $r$ with several $b$ values for static Bronnikov black hole with $M=1$.}
    \label{3DECStaticBronnikov}
    \end{figure}

Since the function $x\tanh x$, with $x\equiv b\sqrt{q}/r$, is monotonically increasing in $x$, the transverse NEC
is violated when $x\tanh x>2$, whereas the SEC is
violated when $x\tanh x>1$. Moreover, since $\rho>0$
and $P_1=-\rho$, the NEC, WEC, and DEC share the same
violation domain. Consequently, violations of the energy
conditions are confined to the inner region and disappear
at sufficiently large radii. Fig.~\ref{3DECStaticBronnikov}
displays the radial profiles of $\rho+P_2$, corresponding
to the transverse NEC, and $\rho+P_1+2P_2$, corresponding
to the SEC, for several values of the nonlinear parameter
$b$ at fixed ADM mass $M=1$.

\begin{figure}[H]
    \centering
    \includegraphics[width=0.7\linewidth]{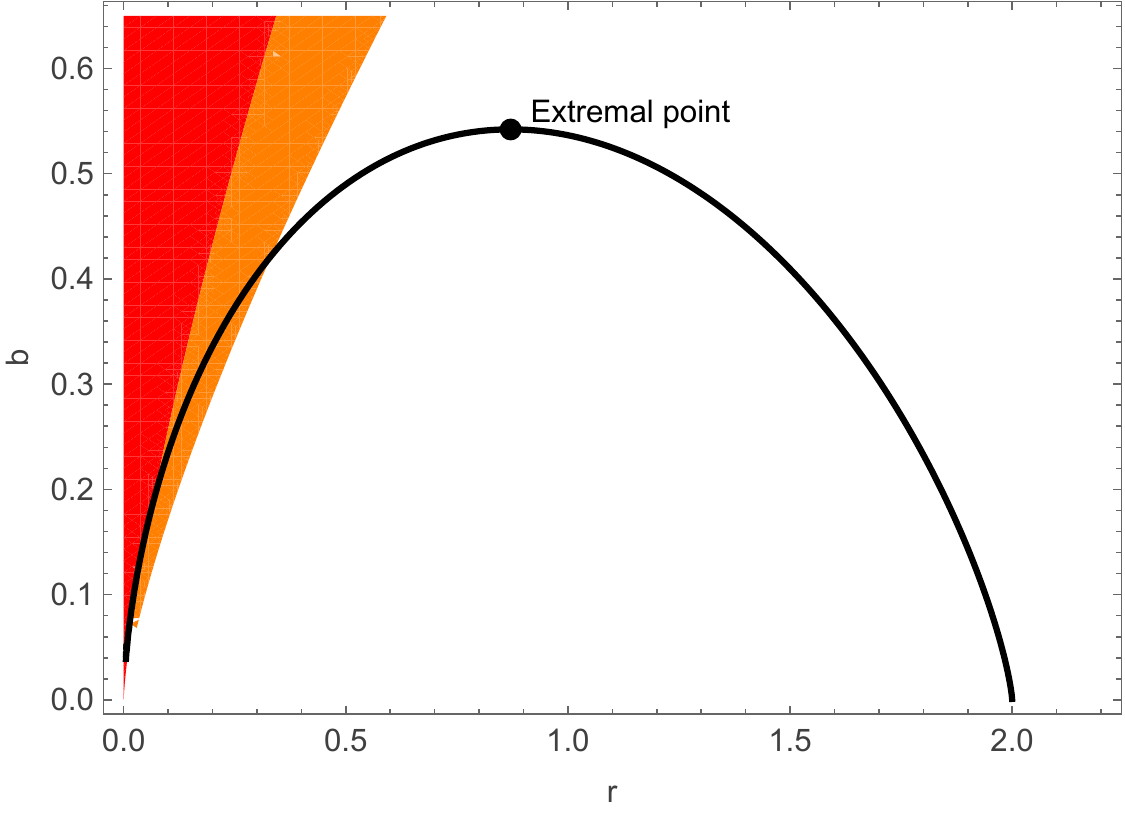}
    \caption{Region of the $(r,b)$ plane in which the NEC, WEC, and DEC is violated (Red region) and SEC is violated (Orange region). The solid curve is the horizon.}
    \label{Regionplot}
\end{figure}

The corresponding violation domains in the $(r,b)$ plane
are presented in Fig.~\ref{Regionplot}. The red region denotes
violations of the NEC, WEC, and DEC, whereas the orange region
indicates violation of the SEC. The solid curve represents
the event horizon, with the extremal configuration marked
by the corresponding point.


\subsection{Rotating Case}

\begin{figure}[H]
    \centering
    \begin{minipage}{0.49\textwidth}
        \centering
        \includegraphics[width=\linewidth]{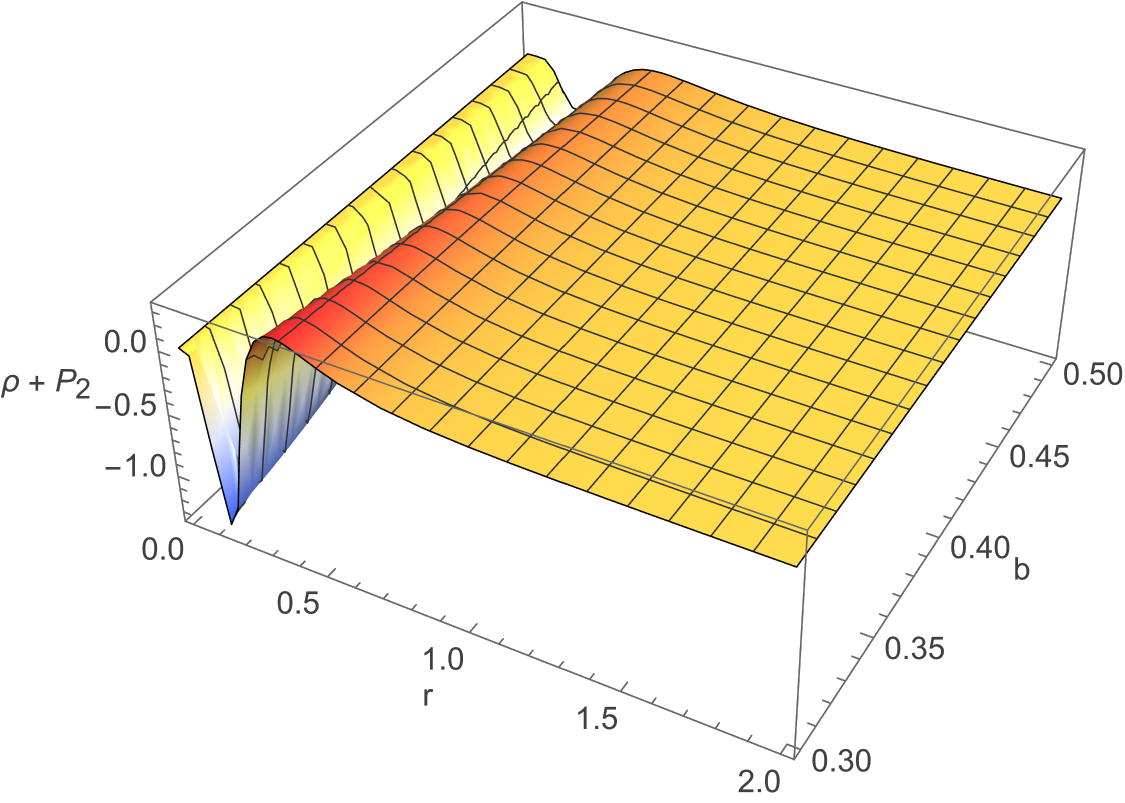}\vspace{0.2cm}
        \includegraphics[width=\linewidth]{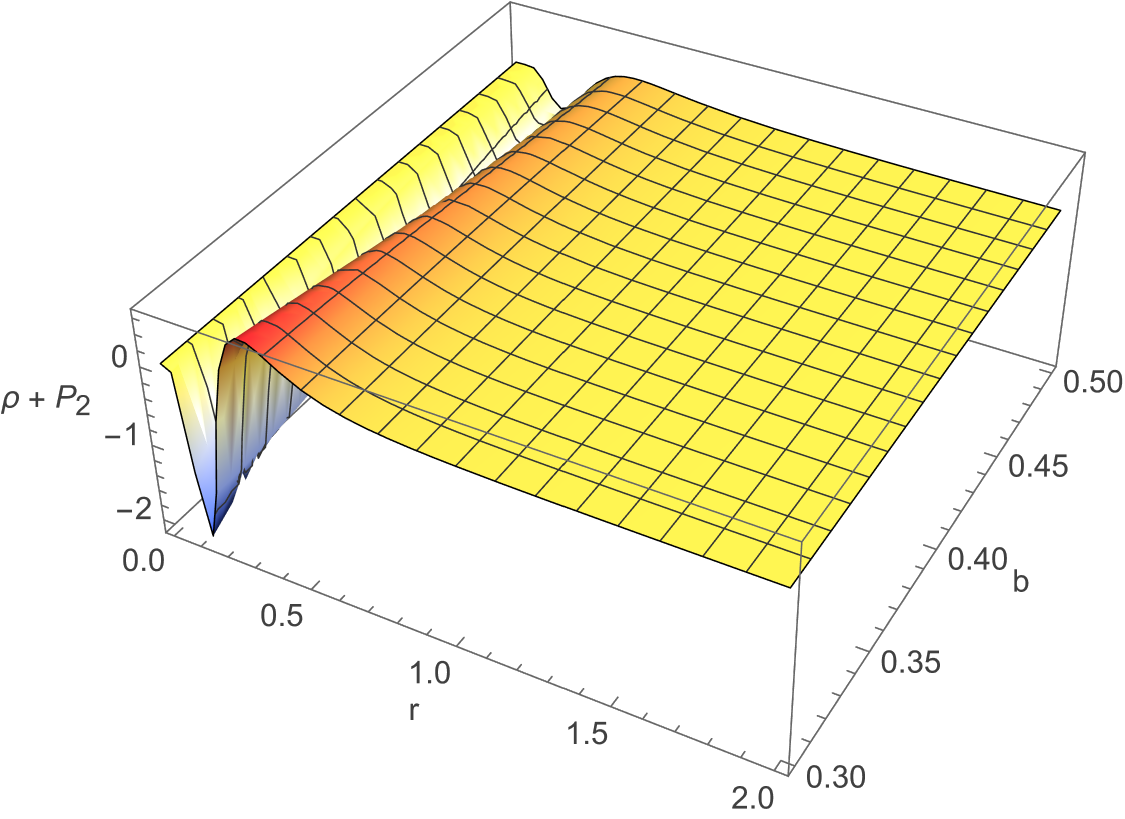}\vspace{0.2cm}
        \vspace{0.2cm}
    \end{minipage}\hfill
    \begin{minipage}{0.49\textwidth}
        \centering
        \includegraphics[width=\linewidth]{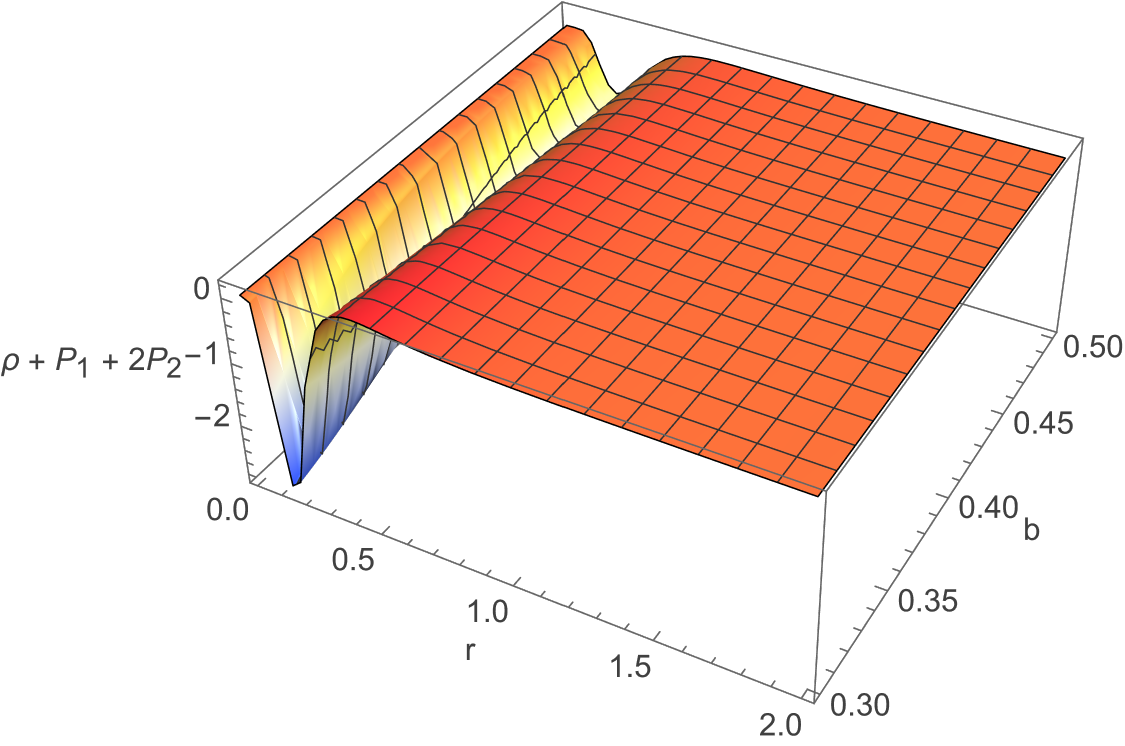}\vspace{0.2cm}
        \includegraphics[width=\linewidth]{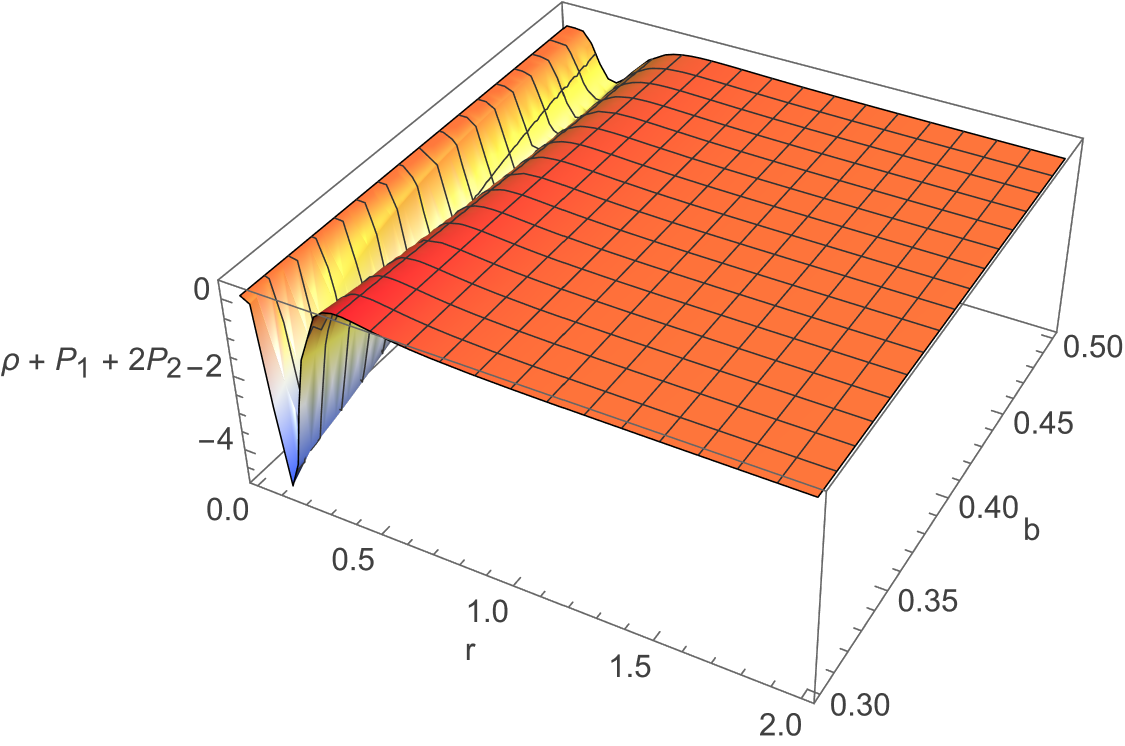}\vspace{0.2cm}
        \vspace{0.2cm}
    \end{minipage}\hfill
    \caption{(Top Left) Plot of $\rho + P_2$ (NEC) vs $r$ with several $b$ values for rotating geometry of Bronnikov black hole with $M=1, \quad \theta = 0$. (Bottom Left) Plot of $\rho + P_2$ (NEC) vs $r$ with several $b$ values for rotating geometry of Bronnikov black hole with $M=1, \quad \theta = \pi/4$. (Top Right) Plot of $\rho + P_1+ 2P_2$ (SEC) vs $r$ with several $b$ values for rotating geometry of Bronnikov black hole with $M=1, \quad \theta = 0$. (Bottom Right) Plot of $\rho + P_1+ 2P_2$ (SEC) vs $r$ with several $b$ values for rotating geometry of Bronnikov black hole with $M=1, \quad \theta = \pi/4$.)}
    \label{3DrotatingEC}
\end{figure}

\begin{figure}[H]
    \centering
    \includegraphics[width=1\linewidth]{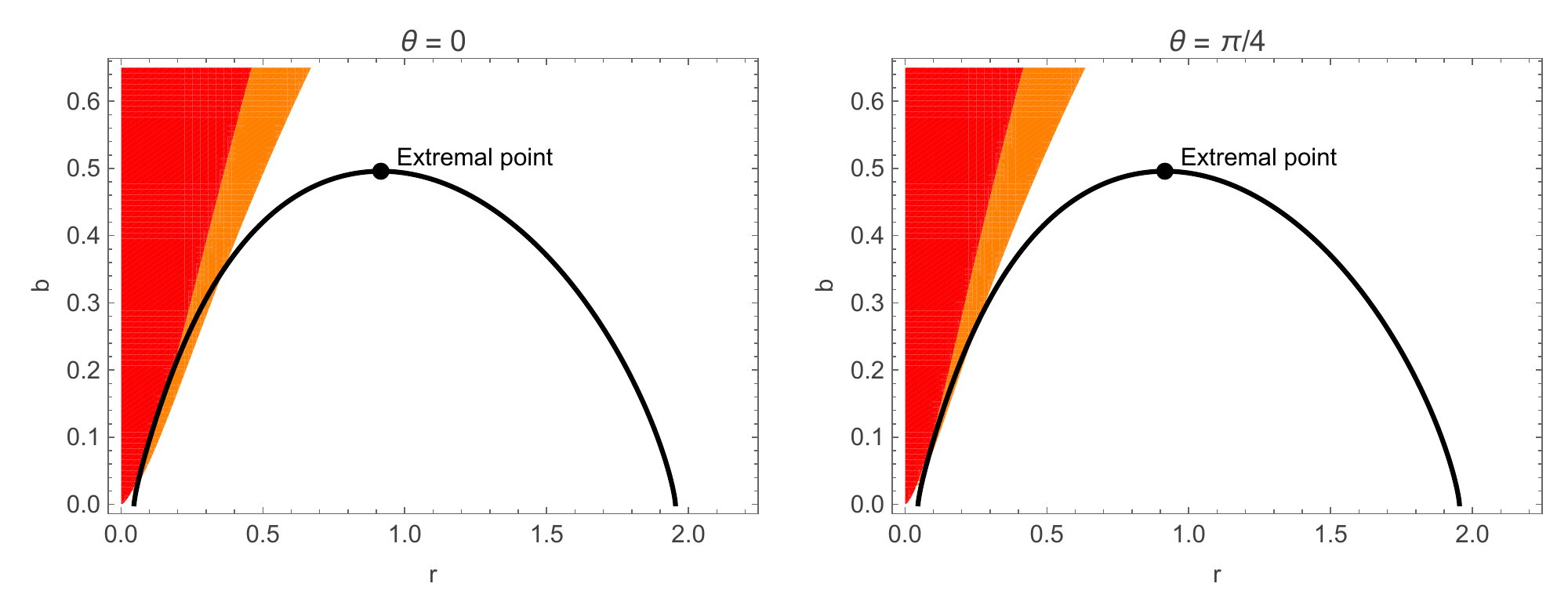}
    \caption{Region of the $(r,b)$ plane in rotating geometry of Bronnikov black hole for two $\theta$-values in which the NEC, WEC, and DEC is violated (Red region) and SEC is violated (Orange region). The solid curve is the horizon.}
    \label{Regionplotrotating}
\end{figure}

We now turn to the Newman--Janis-generated rotating geometry.
The energy conditions
acquire an angular dependence. For the rotating metric~\eqref{eq:rotating_regular_metric}, the effective energy density and pressures obtained from
the Einstein tensor are~\cite{Burinskii2002}
\begin{align}
    \rho =
    \frac{r^2m'}{4\pi\Sigma^2}, \qquad P_1 =-\rho, \qquad  P_2=P_3 =
    \rho-\frac{(rm)''}{8\pi\Sigma}.
    \label{eq:rho_rotating}
\end{align}
These expressions are used to evaluate the effective energy
conditions throughout this section.
These combinations take the exact forms
\begin{align}
    \rho &= \frac{q^2}{8\pi\Sigma^2}\operatorname{sech}^2\!\left(\frac{b\sqrt{q}}{r}\right)\ \ge0,
    \label{eq:rot_rho}\\
    \rho-P_2 &= \frac{b\,q^{5/2}}{8\pi r^3\Sigma}\tanh\!\left(\frac{b\sqrt{q}}{r}\right)\operatorname{sech}^2\!\left(\frac{b\sqrt{q}}{r}\right)\ \ge0,
    \label{eq:rot_exact1}\\
    \rho+P_2 &= \frac{q^2}{8\pi r^2\Sigma^2}\operatorname{sech}^2\!\left(\frac{b\sqrt{q}}{r}\right)
    \left[2r^2-\frac{b\sqrt{q}}{r}\,\Sigma\tanh\!\left(\frac{b\sqrt{q}}{r}\right)\right],
    \label{eq:rot_exact2}\\
    \rho+P_1+2P_2 &= \frac{q^2}{4\pi r^2\Sigma^2}\operatorname{sech}^2\!\left(\frac{b\sqrt{q}}{r}\right)
    \left[r^2-\frac{b\sqrt{q}}{r}\,\Sigma\tanh\!\left(\frac{b\sqrt{q}}{r}\right)\right],
    \label{eq:rot_exact3}
\end{align}
which reduce to Eqs.~\eqref{eq:rho_static}--\eqref{eq:SEC_static}
in the static limit $a=0$. Since $\rho\ge0$ and
$\rho-P_2\ge0$ everywhere, the NEC, WEC, and DEC share
the same violation domain, which is determined by the
transverse NEC, $\rho+P_2\ge0$. The angular dependence
of these conditions enters through
$\Sigma=r^2+a^2\cos^2\theta$. For a given radius, $\Sigma$
is largest on the rotation axis ($\theta=0$), making the
transverse NEC and SEC more restrictive there than
at other polar angles. This behavior is consistent with
the comparison between $\theta=0$ and $\theta=\pi/4$
in Figs.~\ref{3DrotatingEC} and~\ref{Regionplotrotating}.

Fig.~\ref{3DrotatingEC} shows the radial profiles of $\rho+P_2$ (NEC) and $\rho+P_1+2P_2$ (SEC) for several
values of $b$, with the rotation parameter fixed at $a=0.3$.
The upper and lower panels correspond to $\theta=0$ and
$\theta=\pi/4$, respectively. The profiles exhibit violations
in the inner radial region, while their angular dependence
reveals the influence of rotation on the effective matter
distribution. In particular, the radial extent and magnitude
of the negative regions vary with the polar angle.

The corresponding violation domains in the $(r,b)$ plane
are shown in Fig.~\ref{Regionplotrotating} for
$\theta=0$ and $\theta=\pi/4$. As in the static case, the
red and orange regions denote violations of the NEC, WEC,
and DEC, and of the SEC, respectively. The solid curve
marks the event horizon. The comparison between the two
angular sections shows that rotation modifies the boundaries
of the violation domains, introducing an angular dependence
into the distribution of energy-condition violations.

To further characterize the physical behavior of the
rotating geometry, we examine the energy conditions in
the asymptotically flat region. The relevant asymptotic
behavior follows from the expansion of the mass function
in Eq.~\eqref{eq:seed_mass_profile},
\begin{equation}
    m(r)=M-\frac{q^2}{2r}
    +\frac{b^2q^3}{6r^3}
    -\frac{b^4q^4}{15r^5}
    +\mathcal{O}(r^{-7})\textcolor{blue}{.}
    \label{eq:m_asymptotic}
\end{equation}
Substitution into
Eqs.~\eqref{eq:rho_rotating} gives
\begin{align}
    \rho={}&
    \frac{q^2}{8\pi r^4}
    -\frac{b^2q^3+2a^2q^2\cos^2\theta}
    {8\pi r^6}
    +\mathcal{O}(r^{-8}),
    \label{eq:rho_asymptotic}\\
    P_1={}&-\rho,
    \label{eq:P1_asymptotic}\\
    P_2={}&
    \frac{q^2}{8\pi r^4}
    -\frac{2b^2q^3+2a^2q^2\cos^2\theta}
    {8\pi r^6}
    +\mathcal{O}(r^{-8}).
    \label{eq:P2_asymptotic}
\end{align}
These expressions allow us to examine the four standard
energy conditions independently.

First, the radial NEC is identically saturated, $\rho+P_1=0$.
The transverse NEC, on the other hand, behaves asymptotically as
\begin{equation}
    \rho+P_2=
    \frac{q^2}{4\pi r^4}
    -\frac{3b^2q^3+4a^2q^2\cos^2\theta}
    {8\pi r^6}
    +\mathcal{O}(r^{-8}).
\end{equation}
For $q>0$ the leading contribution is positive and independent of $a$ and $\theta$, so the transverse NEC holds at sufficiently large radii for fixed $b$, $q>0$, and $a$.

The WEC additionally requires a nonnegative effective
energy density. From Eq.~\eqref{eq:rho_asymptotic},
\begin{equation}
    \rho=
    \frac{q^2}{8\pi r^4}
    +\mathcal{O}(r^{-6})>0.
\end{equation} 
Together with the asymptotic
NEC result, this establishes that the WEC is also satisfied
in the asymptotically flat region.

For the DEC, we must further require
$\rho\geq |P_1|$ and $\rho\geq |P_2|$. The radial condition
is identically saturated because $P_1=-\rho$. In the
asymptotic region, both $\rho$ and $P_2$ are positive,
so the remaining condition reduces to $\rho-P_2\geq0$.
Using Eqs.~\eqref{eq:rho_rotating} 
we obtain
\begin{equation}
    \rho-P_2
    =\frac{(rm(r))''}{8\pi\Sigma}
    =\frac{b^2q^3}{8\pi r^6}
    +\mathcal{O}(r^{-8}).
\end{equation}
Thus, for $b>0$ and $q>0$, the leading nonvanishing
contribution is positive, and the DEC is satisfied at
sufficiently large radii. It is worth noting that this
difference first appears at order $r^{-6}$, whereas
$\rho$ and $P_2$ share the same leading term of order
$r^{-4}$.

Finally, since $P_1=-\rho$, the SEC combination reduces to $\rho+P_1+2P_2=2P_2$.
Its asymptotic behavior is therefore
\begin{equation}
    \rho+P_1+2P_2
    =
    \frac{q^2}{4\pi r^4}
    -\frac{2b^2q^3+2a^2q^2\cos^2\theta}
    {4\pi r^6}
    +\mathcal{O}(r^{-8}).
\end{equation}
The leading contribution is again positive for $q>0$,
showing that the SEC is satisfied in the asymptotically
flat region.

Fig.~\ref{3DrotatingEC} illustrates how the radial profiles of the NEC and SEC depend on the polar angle, whereas Fig.~\ref{Regionplotrotating}
displays the corresponding violation domains in the
$(r,b)$ plane. The comparison between $\theta=0$ and
$\theta=\pi/4$ reveals that rotation modifies the
boundaries of these domains through its angular dependence. The effect of rotation is not merely to alter the radial profiles, but also to change the spatial distribution of the regions in which the energy
conditions are violated. Nevertheless, all four standard energy conditions are satisfied at sufficiently large radii. Consequently, the rotating geometry satisfies the asymptotic energy-condition criterion proposed by Maeda~\cite{Maeda:2021jdc}.

The violation domains are, moreover, hidden behind the event horizon for every rotating black hole. Since
$\Sigma\le r^2+a^2$, Eqs.~\eqref{eq:rot_exact2} and
\eqref{eq:rot_exact3} show that the NEC and SEC hold at
a given radius for all $\theta$ if and only if
$a^2\le A_{\rm NEC}(r)$ and $a^2\le A_{\rm SEC}(r)$,
respectively, where
\begin{equation}
    A_{\rm NEC}(r)\equiv r^2\left[\frac{2r}{b\sqrt q}\coth\left(\frac{b\sqrt q}{r}\right)-1\right],
    \qquad
    A_{\rm SEC}(r)\equiv r^2\left[\frac{r}{b\sqrt q}\coth\left(\frac{b\sqrt q}{r}\right)-1\right].
    \label{eq:ANEC-ASEC}
\end{equation}
Both functions are monotonically increasing with $r$ in
their positive domains and satisfy
$A_{\rm SEC}<A_{\rm NEC}$. It therefore suffices to show
that $a^2\le A_{\rm SEC}(r_h)$ at the outer horizon $r_h$.
There, $\Delta(r_h)=0$ gives $a^2=2m(r_h)r_h-r_h^2$,
so that this condition becomes
\begin{equation}
    \frac{2b\sqrt q}{r_h}\tanh\left(\frac{b\sqrt q}{r_h}\right)\frac{m(r_h)}{r_h}\le1.
    \label{eq:C4_rot_condition}
\end{equation}

At the outer horizon, we also have $\Delta'(r_h)\ge0$,
which implies
\begin{equation}
    \frac{m(r_h)}{r_h}\le
    \left.\frac{m}{m+rm'}\right|_{r=r_h}.
\end{equation}
For the mass profile~\eqref{eq:seed_mass_profile},
the left-hand side of Eq.~\eqref{eq:C4_rot_condition}
is therefore bounded by
\begin{equation}
    \frac{\dfrac{2b\sqrt q}{r_h}\tanh\left(\dfrac{b\sqrt q}{r_h}\right)}
    {1+\dfrac{b\sqrt q}{r_h}\left[1+\tanh\left(\dfrac{b\sqrt q}{r_h}\right)\right]}<1,
    \label{eq:C4_rot_bound}
\end{equation}
where the last inequality follows from
$\tanh(b\sqrt q/r_h)<1$. Consequently,
\begin{equation}
a^2\le A_{\rm SEC}(r_h)\le A_{\rm SEC}(r)<A_{\rm NEC}(r), \quad \forall r\ge r_h.
\end{equation}
Hence, all four standard energy
conditions are satisfied on and outside the event horizon
for every parameter configuration admitting a rotating
black hole. For horizonless configurations, the violation
region near the center is not shielded by an event horizon,
and criterion C4 is therefore not applicable.

Combined with the results of Sec.~\ref{sec:metric}, these
findings show that the static and rotating geometries
satisfy all five geometric criteria C1--C5 of
Ref.~\cite{Maeda:2021jdc}, as summarized in
Table~\ref{tab:maeda}.

\begin{table}[H]
\centering
\caption{Maeda's criteria~\cite{Maeda:2021jdc} for the static
Bronnikov black hole and its rotating counterpart
~\eqref{eq:rotating_regular_metric}. A dash indicates that
the criterion cannot be assessed or is not studied here.}
\label{tab:maeda}
\vspace{0.2cm}
\begin{tabular}{|c|c|c|}
\hline
Criterion & Static & Rotating \\
\hline
C1 & $\checkmark$ & $\checkmark$ \\
C2 & $\checkmark$ & $\checkmark$ \\
C3 & $\checkmark$ & $\checkmark$ \\
C4 & $\checkmark$ & $\checkmark$ \\
C5 & $\checkmark$ & $\checkmark$ \\
C6 & $\times$ & -- \\
C7 & -- & -- \\
\hline
\end{tabular}
\end{table}

\section{Conclusions}
\label{sec:conc}

The structure of particle orbits provides a direct probe of how modifications to spacetime geometry influence motion in the strong-gravity regime. In this work, we investigate this connection by studying spherical photon and timelike orbits in a Newman--Janis-generated rotating regular black hole constructed from the Bronnikov mass profile. We show how the nonlinear mass distribution influences photon propagation and the black-hole shadow.
The resulting picture reveals a rich orbital structure
that distinguishes the rotating regular geometry from
Kerr, while continuously recovering the latter in the
appropriate limit.

In the black-hole regime, all physically admissible SPOs outside the event horizon are radially
unstable, as determined by the condition $R''(r_{\rm SO})>0$.
The parameter dependence is reflected not only in the
allowed orbital radii or the
Carter constant $C$ and impact parameter $\Phi$, but also
in the polar turning point $u_0$ and the accumulated
azimuthal angle $\Delta\phi$. In particular, for the Kerr limit, the selected extremal orbit with $a/M=1$ and $\Phi/M=0$ has
$r_{\rm SO}/M=2.414211$ and $\Delta\phi=3.17612$, closely approaching the photon-boomerang condition
$\Delta\phi=\pi$~\cite{Page:2021rhx}. For the other representative extremal
configurations, the spin decreases from $a/M=1$ at $b=0$ to $a/M=0.286$ at $b=0.5$. The corresponding orbital parameters exhibit non-monotonic variations.

The orbital instability plays a central role in the formation of a black hole shadow, as the unstable photon orbits determine its boundary
\cite{Cunha:2017eoe}. Using the family of physically admissible unstable SPOs, we construct the shadow boundary on the observer's celestial screen. The parameter $b$ modifies both the size and
shape of the shadow relative to the Kerr case. This deviation is illustrated in Fig.~\ref{shadowrotbronkerr},
which compares the shadow of the rotating regular geometry
with that of a Kerr black hole having the same spin, $a/M=0.7$, for an edge-on observer and $b=0.3$. The resulting contour exhibits a noticeable reduction in size together with a deformation of its shape. 

Unlike the black hole configurations, we find stable SPOs in representative horizonless geometries. For
$a/M=1$, the selected examples with $b=0.1$--$0.5$ have
$r_{\rm SO}/M$ ranging from $1.13355$ to $1.730324$.
Their azimuthal changes range from $\Delta\phi=16.2164$
to $4.45289$, respectively. The large azimuthal change in
the first example corresponds to more than five complete
azimuthal revolutions during one latitudinal oscillation.
As in our previous case~\cite{Alam2024}, the absence of
an event horizon in the present geometry allows for a distinct spherical photon-orbit structure, including stable orbits that are not available outside the horizon
in the black hole configurations considered here~\cite{Cunha:2017qtt}.

Stable SPOs can also be found for massive test particles. In the selected extremal configurations with $\Phi=0$, the marginally stable orbits have
$r_{\rm SO}/M=3.73071$, $3.0196$, and $2.67439$ for
$b=0$, $0.3$, and $0.5$, respectively. Their corresponding
spin parameters are $a/M=1$, $0.71$, and $0.285$.
All three examples are bound, with $E<1$, and satisfy
the marginal-stability condition
$R''(r_{\rm SO})=0$. These results illustrate the
modification of the marginally stable spherical-orbit
configurations in the extremal family.

The ISSO is obtained by imposing the simultaneous conditions
$R=R'=R''=0$. For the representative
extremal configuration with $b=0.1$ and $a/M=0.94$, the
orbit shown in our analysis has
$r_{\rm ISSO}/M=7.814$ and $C/M^2=7$. Its trajectory
exhibits repeated latitudinal oscillations while maintaining
a constant radial coordinate. This provides an explicit
example of an ISSO in the rotating regular geometry.

We further investigated the null, weak, strong, and
dominant energy conditions for both the static and rotating
geometries. The results show that regularization is
accompanied by violations of the energy conditions in
certain regions, particularly in the strong-field regime
\cite{Zaslavskii:2010qz,Neves:2014aba}. In the rotating case,
the violation domains are further modified by the spin
and depend on the polar angle $\theta$. Nevertheless, all
four standard energy conditions are satisfied sufficiently
far from the central object and, for every black-hole
configuration, throughout the region on and outside the
event horizon.

We have also assessed the geometries against Maeda's
criteria for physically reasonable nonsingular black holes
\cite{Maeda:2021jdc}. Both the static and rotating geometries
are free of curvature singularities, and the rotating
geometry admits a smooth extension across the disk,
including the ring, within the single-sheeted domain
$r\geq0$. Furthermore, both geometries are stably causal.
The Kretschmann scalar is bounded, independent of $q$ and $a$ at fixed
coupling $b$, establishing the limiting curvature condition. Together with the energy-condition results,
these properties show that both geometries satisfy all
five criteria C1--C5. This combination of properties
distinguishes the present rotating geometry from the
rotating models examined in Ref.~\cite{Maeda:2021jdc}.

Several directions remain open for further investigation.
On the observational side, it would be interesting to
quantify the deviations of the shadow from the Kerr
prediction through the shadow diameter and distortion for different observer
inclinations \cite{Tsukamoto:2017fxq,Chang:2020miq}. The orbital
structure may also be extended to investigate the characteristic frequencies of spherical timelike
motion and their possible implications for astrophysical
phenomena such as quasi-periodic oscillations, which have
been studied in the framework of relativistic precession
and epicyclic resonance models
\cite{Stella:1997tc}. A further direction concerns extreme mass-ratio inspirals (EMRIs), for which
the geodesic motion of a small compact object around a
massive central body provides the leading-order description
in the extreme mass-ratio limit. It would therefore be
interesting to extend the present analysis to generic
bound timelike geodesics and investigate their fundamental
orbital frequencies, before incorporating gravitational
radiation reaction and self-force effects to study the
resulting inspiral and gravitational-wave signatures
\cite{Amaro-Seoane:2012lgq,Barack:2018yvs}. Such an investigation
could provide a framework for assessing possible
deviations from Kerr-based EMRI waveforms and their
relevance to future space-based gravitational-wave
observations such as LISA.

More fundamentally, an important question concerns the
physical origin of the rotating regular geometry itself.
In this work, we do not claim that the rotating metric~\eqref{eq:rotating_regular_metric} is the solution of Einstein's equations with NLED source. Such claim, if exists, remains an open problem. Indeed, the Newman--Janis algorithm does not, in general, guarantee that the generated rotating geometry satisfies the same field equations as its static seed
\cite{Junior:2020lya}. This issue has been examined explicitly in the context of nonlinear electrodynamics:
the standard Newman--Janis prescription has been shown
to be incompatible with certain classes of nonlinear
electrodynamic theories
\cite{Kubiznak:2022vft}, while inconsistencies in specific
Newman--Janis-generated rotating regular black-hole
solutions have also been pointed out
\cite{Rodrigues:2017tfm}. These results motivate a systematic investigation of the matter-field content required to support the rotating geometry considered here. Addressing these questions would provide a more complete understanding of both the observable signatures and the physical foundations of rotating regular
spacetimes.

\section*{Acknowledgement}

We thank Byon Jayawiguna, Hardian Laudza, and Leonardus Putra for enlightening discussions on regular black holes and spherical orbits. HSR is supported by Hibah PUTI Q1 UI No.~PKS196/UN2.RST /HKP.05.00/2025. 


\end{document}